\documentclass[11pt]{article}

\usepackage[latin1]{inputenc}

\usepackage{amstext,amsmath,amssymb,amsfonts,bbm,amsmath}
\usepackage{extarrows}
\usepackage{xcolor} 
\usepackage{hyperref}
\hypersetup{
    colorlinks=false,
    menubordercolor=red,
    linkbordercolor=blue
}
\usepackage{authblk}
\usepackage{caption} 
\usepackage{epsfig} 
\usepackage{color} 
\usepackage{graphicx} 
\usepackage{braket} 
\usepackage{slashed} 
\usepackage{dsfont} 
\usepackage{amsthm}

\usepackage{cite}

\allowdisplaybreaks[4]

\usepackage{psfrag}
\usepackage{tikz} 
\usetikzlibrary{arrows}
\usetikzlibrary{plotmarks}
\usetikzlibrary{external}
\usetikzlibrary{patterns}
\usepackage{pgfplots}
\pgfplotsset{compat=1.9}
\usetikzlibrary{patterns}
\usetikzlibrary{calc}
\usetikzlibrary{automata,positioning}

\usepackage{float}  
\usepackage{subfig}
\usepackage[lmargin=50pt,rmargin=60pt,tmargin=60pt,bmargin=65pt]{geometry}

\DeclareMathOperator{\Tr}{Tr}

\newcommand{\be}{\begin{equation}}
\newcommand{\ee}{\end{equation}} 
\newcommand{\nn}{\nonumber}
\newcommand{\f}{\frac}
\newcommand{\p}{\partial}
\newcommand{\la}{\langle}
\newcommand{\ra}{\rangle}
\newcommand{\dd}{{\rm d}^d}

\let\a=\alpha \let\b=\beta  \let\g=\gamma  \let\d=\delta  
\let\z=\zeta        \let\l=\lambda
\let\m=\mu    \let\n=\nu          \let\r=\rho \let\om=\omega
\let\s=\sigma     \let\ph=\phi 
    
\let\G=\Gamma \let\D=\Delta   \let\L=\Lambda \let\X=F
         
\let\Om=\Omega  \let\eps=\epsilon

\newcommand{\cB}{\mathcal{B}}

\newcommand{\cD}{\mathcal{D}}
\newcommand{\cE}{\mathcal{E}}
\newcommand{\cF}{\mathcal{F}}
\newcommand{\cG}{\mathcal{G}}

\newcommand{\cI}{\mathcal{I}}

\newcommand{\cN}{\mathcal{N}}
\newcommand{\cO}{\mathcal{O}}
\newcommand{\cP}{\mathcal{P}}

\newcommand{\cS}{\mathcal{S}}

\newcommand{\cV}{\mathcal{V}}

\newcommand{\cZ}{\mathcal{Z}}

\newcommand{\htilde}{\tilde{h}}

\newcommand{\wtD}{\widetilde{\Delta}}

\DeclareMathOperator{\im}{\mathrm{i}}

\newcommand{\mba}{\mathbf{a}}
\newcommand{\mbb}{\mathbf{b}}
\newcommand{\mbc}{\mathbf{c}}
\newcommand{\mbd}{\mathbf{d}}

\allowdisplaybreaks[4]

\numberwithin{equation}{section}

\newtheorem{theorem}{Theorem}

\theoremstyle{remark}

\begin{document}

\title{\bf The $F$-theorem in the melonic limit}

\author[1]{Dario Benedetti}
\author[1,2,3]{Razvan Gurau}
\author[1,3]{Sabine Harribey}
\author[3]{Davide Lettera}

\affil[1]{\normalsize \it 
 CPHT, CNRS, Ecole Polytechnique, Institut Polytechnique de Paris, Route de Saclay, \authorcr 91128 PALAISEAU, 
 France
  \authorcr \hfill}

\affil[2]{\normalsize\it 
Perimeter Institute for Theoretical Physics, 31 Caroline St. N, N2L 2Y5, Waterloo, ON,
Canada
 \authorcr \hfill}

\affil[3]{\normalsize\it 
Heidelberg University, Institut f\"ur Theoretische Physik, Philosophenweg 19, 69120 Heidelberg, Germany
 \authorcr \hfill
 \authorcr
emails: dario.benedetti@polytechnique.edu, gurau@thphys.uni-heidelberg.de, sabine.harribey@polytechnique.edu, lettera@thphys.uni-heidelberg.de 
 \authorcr \hfill }
 
\date{}

\maketitle

\hrule\bigskip

\begin{abstract}

The $F$-theorem states that in three dimensions the sphere free energy of a field theory must decrease between ultraviolet and infrared fixed points of the renormalization group flow, and it has been proven for unitary conformal field theories (CFTs). 

We consider here the long-range bosonic $O(N)^3$ model on a spherical background, at next-to-next-to-leading order of the $1/N$ expansion. The model displays four large-$N$ fixed points and we test and confirm the $F$-theorem holds in this case. This is non-trivial as one of the couplings is imaginary, and therefore the model is non-unitary at finite $N$. Despite this, several tests indicating that the large-$N$ CFTs are in fact unitary have been performed: for instance all the OPE coefficients computed so far in the large-$N$ limit are real, and the spectrum of bilinear operators is real and above unitarity bounds. Our result, namely that the F theorem holds at large $N$, can be viewed as further indication that such theories are unitary.

As an added bonus, we show how conformal partial waves expansions in conformal field theory can be used to resum infinite classes of vacuum diagrams.
Non-perturbatively, the jump in the value of the free energy has the interpretation of the inclusion at the ultraviolet fixed point of an extra non-normalizable contribution in the conformal partial wave expansion. This can be seen in perturbation theory as the reversal of the sign of an infinite class of diagrams due to the flow of a coupling constant.

\end{abstract}

\hrule\bigskip

\tableofcontents

\section{Introduction}
\label{sec:introduction}

Among the most intriguing features of quantum field theory in various dimensions are the so called $c$-, $a$- and $F$-theorems \cite{Zamolodchikov:1986gt,Komargodski:2011vj,Jafferis:2011zi,Klebanov:2011gs}. These lettered theorems state that under the RG flow between various fixed points some quantities (aptly denoted $c$, $a$ or $F$) always decrease. Intuitively, these quantities must in some way count the degrees of freedom in the theory, as the RG flow decimates the degrees of freedom when going from one fixed point to another.

The most well known of the lettered theorems, the $c$-theorem in dimension 2 was first proven by Zamolodchikov \cite{Zamolodchikov:1986gt}. The quantity $c$ in this case was defined using the two-point functions of the stress-energy tensor. Interestingly, the obtained $c$-function coincides at the RG fixed point with the Weyl anomaly coefficient $c$, that is the central charge. 

In $d=4$ dimensions, the $a$-theorem was first conjectured by Cardy \cite{Cardy:1988cwa}. In this case, there are two universal Weyl anomaly coefficients, usually denoted $a$ and $c$. Cardy conjectured that the quantity that should decrease along the RG flow is the $a$-coefficient, multiplying the Euler density. In practice, this coefficient can be computed from the expectation value of the trace of the stress-energy tensor in the Euclidean theory on $S^4$. After a long time, the $a$-theorem was finally proven in \cite{Komargodski:2011vj,Komargodski:2011xv}.

The latest addition to this list of monotonicity theorems is the $F$-theorem, concerning field theories in $d=3$.
In this case, $F$ has a relatively straightforward definition as the free energy of the CFT on the sphere. 
The compactness of the sphere regulates the infrared divergences, however ultraviolet divergences persist and need to be regularized properly: $F$ is defined as the finite part of the free energy.
This choice was first proposed in \cite{Jafferis:2010un,Jafferis:2011zi}, where various checks were performed on supersymmetric theories, then extended to non-supersymmetric ones in \cite{Klebanov:2011gs} (see also \cite{Pufu:2016zxm} for a review).
Shortly after, the $F$-theorem was proven in \cite{Casini:2012ei}, using the relation between the free energy and the entanglement entropy across a circle \cite{Casini:2011kv}; so far, this is the only method that works for all three theorems \cite{Casini:2004bw,Casini:2017vbe}.

One common feature of the proofs of these theorems is that unitarity plays a crucial role: it is underlying the use of positivity of two-point functions in \cite{Zamolodchikov:1986gt}, of the optical theorem in \cite{Komargodski:2011vj}, and of the strong subadditive inequality in \cite{Casini:2012ei,Casini:2004bw,Casini:2017vbe}.
At present, it remains unclear to what extent unitarity is a necessary ingredient and if the assumptions of these theorems could be relaxed to include at least some class of non-unitary models.
A non-physical counterexample to the necessity of unitarity is provided by the generalized $F$-theorem tests in non-integer dimensions \cite{Giombi:2014xxa,Fei:2015oha,Giombi:2015haa}, where it was shown to hold, despite the fact that CFTs in non-integer dimensions have been shown to be non-unitary \cite{Hogervorst:2015akt}, at least in the case of the Wilson-Fisher fixed point. 
On the other hand, as we will see below, a trivial counterexample to a generic $F$-theorem without unitarity is provided by a non-unitary generalized free field theory flow.\footnote{It should be remarked that generalized free fields, as the long-range model we will consider here, evade also another hypothesis of the standard proofs, that is, locality. Long-range models in particular do not have a local energy-momentum tensor, which plays a crucial role in the standard proofs of the $c$- and $a$-theorems. However, the embedding of such models in a larger space \cite{Paulos:2015jfa} could perhaps provide a workaround for such proofs. In fact, for the special case of an integer number of extra dimensions, boundary versions of the $F$-theorem have indeed been proposed \cite{Gaiotto:2014gha} (see also \cite{Kobayashi:2018lil} for more information on monotonicity theorems in boundary or defect CFTs).}

In most applications, one knows from the start whether the theory of interest is unitary or not, or at least one has a good degree of confidence in that, and therefore testing the $F$-theorem in a theory satisfying the hypotheses of \cite{Casini:2011kv} is at most an interesting exercise.
However, in some cases ascertaining the unitarity, or lack thereof, of a theory can be challenging; for instance, in the Wilson-Fisher fixed point at non-integer dimensions \cite{Hogervorst:2015akt}, or in the $O(N)$ model at non-integer $N$ \cite{Maldacena:2011jn,Binder:2019zqc}, non-unitarity is a non-trivial result, manifesting itself only in operators of large dimension.
The main subject of this paper will be another non-trivial example, going in the opposite direction: a manifestly non-unitary model, which however in the large-$N$ limit has so far passed all the unitarity tests.

\subsection{Outline of the paper}

In this paper, we test the F-theorem in the long-range $O(N)^3$ bosonic tensor model introduced in \cite{Benedetti:2019eyl}. 
This model is one of many examples of field theories with a melonic large-$N$ limit \cite{Klebanov:2016xxf,Gubser:2017qed,Giombi:2017dtl,Bulycheva:2017ilt,Prakash:2017hwq,Giombi:2018qgp,Benedetti:2019rja,Chang:2018sve,Popov:2019nja,Lettera:2020uay,Benedetti:2020iku} (see also \cite{Klebanov:2018fzb,Benedetti:2020seh} for reviews).
The model of  \cite{Benedetti:2019eyl}, which we consider here, is one of the most extensively studied, and it has numerous interesting features: 
\begin{itemize}
\item the long-range kinetic term is such that the three quartic interactions are perturbatively marginal. One of them is actually exactly marginal in the large-$N$ limit, while the other two acquire an anomalous dimension. The renormalization group flow of the couplings admits four interacting fixed points, parametrically dependent on the marginal coupling.
 \item the two-point function can be obtained exactly in the large-$N$ limit, where it is \textit{melonic}. The non-perturbative resummation of two-point melonic diagrams is not very consequential: it amounts to a finite rescaling. However, the theory is defined only in a certain range of the marginal coupling, for which this resummation is convergent. 
 \item the four-point function is obtained as geometric series in a Bethe-Salpeter kernel, which at large $N$ is exactly a one-rung ladder kernel. Contrary to the $O(N)$ vector model, for which the Bethe-Salpeter kernel is local, in the $O(N)^3$ model this kernel is bilocal. This leads to important consequences.
 
 \item although it does not posses a local stress energy tensor, the model has renormalization group fixed points that are conformally invariant, and not only scale invariant \cite{Benedetti:2020yvb}.
\item despite having one purely imaginary coupling (the marginal one), all indications so far are that the large-$N$ CFTs at the fixed point are unitary \cite{Benedetti:2019ikb,Benedetti:2020yvb}.
\end{itemize}

The last point is somewhat delicate. Due to the imaginary coupling, the finite-$N$ model is non-unitary; however, the large-$N$ results show that imaginary parts of critical exponents are suppressed in $1/N$ \cite{Benedetti:2020sye}. At leading order of the large-$N$ expansion, the fixed points of the non-marginal couplings are real, with real exponents, parametrically depending only on the square of the marginal imaginary coupling \cite{Benedetti:2019eyl}. The spectrum and OPE coefficients of bilinear operators at large $N$ has also been computed \cite{Benedetti:2019ikb} and found to be real; similarly for some quartic operators the large-$N$ results are consistent with unitarity \cite{Benedetti:2020yvb}. Of course this evidence is far from being a proof, as we cannot exclude that non-unitarity might manifest itself in some operators of large dimension.
Therefore, this model does not fall straightforwardly in the domain of applicability of the $F$-theorem, as proved so far, and testing it is non-trivial. 

Our main result is the confirmation that the F-theorem holds for the long-range $O(N)^3$ model. We outline here the content of the paper and the main steps leading to this conclusion. 

\paragraph{Flow between Gaussian CFTs.} As a warm up, in section~\ref{sec:gaussian_CFT}, we consider two Gaussian CFTs with action 
$\frac{1}{2} \int \dd x\phi(x)(-\partial^2)^{\zeta} \phi(x)$, one being the standard short-range action $\zeta =1$ and the second one having $\zeta \neq 1$ (also known as generalized free field theory). We examine the flow between two such CFTs and find that, on the one hand, the RG always flows in the infrared towards lower $\zeta$, while on the other the sphere free energy is concave with a maximum at $\zeta=1$. An RG trajectory flowing from a short-range model $\zeta=1$ to a long-range one $0<\zeta<1$ thus satisfies the F-theorem. However a trajectory starting in the ``strong short-range'' regime $\zeta >1$ is such that the free energy increases in the IR. This gives a trivial counter-example of the F-theorem for non-unitary (and non-local) theories. 


\paragraph{Revisiting the $O(N)$ model: conformal partial wave expansion.} Next, in section~\ref{sec:ON_model}, we revisit the vector $O(N)$ model and rederive its sphere free energy, previously obtained in  \cite{Klebanov:2011gs}. While the result of this section is not new, we use it as an opportunity to introduce the set of techniques relevant for the rest of the paper. 

First we briefly recall the formalism of the two-particle irreducible (2PI) effective action, which is particularly natural for discussing the free energy of large-$N$ models.
At large $N$, the free energy of the interacting fixed point is the same as in the free theory, hence to test the $F$-theorem one needs to go beyond the leading order.
The diagrams contributing to the free energy at next-to-leading (NLO) order in $1/N$ are resummed to $F_{\rm NLO} = \frac{1}{2} \Tr\ln( 1-K )$, where $K$ is the Bethe-Salpeter kernel, which at the relevant order is a local operator.
We note that the component of the four-point function which is one-particle irreducible in the $s$-channel, ${\cal F}_s$,
writes in terms of  $K$ and the two-point function $G$ as ${\cal F}_s = (1-K)^{-1} GG$.

Second, we analyze $ {\cal F}_s $ and $F_{\rm NLO}$ using the conformal partial waves (CPW) expansion, which we review in appendix~\ref{app:CPW}. The use of this technique for the sphere free energy is new, and it has the advantage that it can be generalized to models with different kernels $K$. 
In the $O(N)$ model, the CPW expansion of the bilocal to bilocal identity operator includes the principal series and an additional non-normalizable state, as the dimension of the field is smaller than $d/4$. When evaluating the four-point function ${\cal F}_s$ in the critical model, one finds that the Bethe-Salpeter kernel is zero on the principal series, and infinite on the non-normalizable state, thus 
${\cal F}_s$ reduces to the free contribution ${\cal F}^{\rm free}_s$ restricted to the principal series. This implies the well-known result that the spectrum of the critical $O(N)$ model at large $N$ is the same as in the free theory, except that the $\phi^2$ operator is replaced by its shadow \cite{Gubser:2002vv}.

Using the same CPW expansion of the identity for $F_{\rm NLO}$, we obtain
a zero contribution from the principal series and a non-trivial one from the isolated non-normalizable state. This isolated contribution gives the only non-trivial part and reproduces by itself the value of $F_{\rm NLO}$ computed in \cite{Klebanov:2011gs}.

\paragraph{The long-range $O(N)^3$ model on the sphere.}
 In section~\ref{sec:ON3-model}, we study the long-range $O(N)^3$ model on the sphere.
The situation is similar to the one in the $O(N)$ model: 
 at the first non-trivial order in $1/N$, i.e.\ at next-to-next-to-leading (NNLO) order, the diagrams contributing  to the free energy  (up to an exceptional diagram which is finite and equal at all the fixed points) are resummed to $F_{\rm NNLO} = \frac{1}{2} \Tr\ln( 1-K )$. However, in stark contrast to the $O(N)$ model, $K$ is now a bilocal operator, and the CPW expansion is the only available non-perturbative method of evaluation.

Two different regimes are encountered. In the infrared (IR) CFT, $F_{\rm NNLO}$ has a CPW expansion restricted to the principal series only. Such expansion of the free energy can be evaluated non-perturbatively (numerically) showing in particular how the CPW expansions in CFT can be used to resum infinite classes of vacuum diagrams.

In contrast, for the ultraviolet (UV) CFT one needs to add a non-normalizable state besides the principal series, because the dimension of one of the primary operators in the $\phi \phi\sim \sum c_{\phi\phi O} \; O$ operator product expansion (OPE) descends below $d/2$. Although the inclusion of a non-normalizable state is reminiscent of the $O(N)$ model, the situation is conceptually and practically different, as will be explained. 

Due to the inclusion of this non-normalizable state, the free energy decreases between the ultraviolet and the infrared CFT, as in the $F$-theorem. In perturbation theory, this jump can be seen as the reversal of the sign of an infinite class of diagrams due to the flow of a coupling constant.

\paragraph{Comments on future directions}
Independently of the $F$-theorem, the free energy is a central quantity in statistical field theory, and computing it to higher orders of the $1/N$ expansion can be of interest for many statistical or condensed matter models. The closest example to our setting is certainly the Sachdev-Ye-Kitaev model \cite{Sachdev:1992fk,Kitaev}, which is a model of $N$ strongly interacting Majorana fermions in dimension $1$ with a melonic large-$N$ limit. The appropriate generalization of the CPW expansion, taking into account the Grassmann nature of the fields and the presence of an extra discrete series in the basis of bilocal functions, has already been considered in \cite{Maldacena:2016hyu}, where among other things the part of the free energy which is singular in the limit of zero temperature was computed. One could imagine using a similar computation as ours in order to compute also the finite part of the free energy.
In fact, as our main calculation concerns a long-range model, it would more naturally be compared to the computation of the free energy of the Gross-Rosenhaus model \cite{Gross:2017vhb}. Indeed, in this model, the UV part of the action is replaced by a quadratic bilocal term which leads to a line of fixed point in the IR, similarly to what happens for our model. A $O(N)^3$ tensor version of the Gross-Rosenhaus model (i.e.\ a long-range version of the Klebanov-Tarnopolsky model \cite{Klebanov:2016xxf}) would of course also be a natural candidate for such a computation. We expect that the main difference between the Gross-Rosenhaus model and a long-range Klebanov-Tarnopolsky model would be  the graph of Fig.~\ref{fig:monstru}, which  arises at next-to-next-to-leading order only for the latter. This generalization is beyond the scope of the present paper but would be interesting to study it in the near future.

\paragraph{Appendices.} 
We give abundant details on notations and computations in several appendices. Appendices \ref{app:useful} and \ref{app:sphereCFT} give formulas and definitions for CFTs on the sphere. In appendices \ref{app:F_GFFT} and \ref{app:C_dim_reg} we give details for the computations of the free energy for generalized free field theories (GFFT) with short and long-range covariances. In appendix~\ref{app:CPW}, we review the basics of conformal partial wave expansion.
In appendix~\ref{app:monster} we prove that the exceptional diagram of the $O(N)^3$ model at NNLO has a non-vanishing finite part (i.e.\ it contributes to $F$), although it does not contribute in the assessment of the $F$-theorem.
Lastly, in appendices~\ref{app:I_eps} and \ref{app:NumericsLargeJ}, we give details on intermediate results in the computation of the sphere free energy for the $O(N)^3$ model.

\section{Flow between Gaussian CFTs}
\label{sec:gaussian_CFT}

As a warm-up, and for later reference, let us consider the following quadratic action:
\be \label{eq:GaussianS}
S_{\rm Gauss}[\phi] = \f12 \int \dd x  \, \phi(x) (-\p^2)^{\z} \phi(x) + \f{\l}{2} \int \dd x  \, \phi(x) (-\p^2) \phi(x) \,,
\ee
with $0<\z<1$. The non-integer power of the Laplacian is defined in momentum space simply as $p^{2\z}$, or in position space by a convolution with a non-local kernel, see \eqref{eq:fracLapl_flat_phys}-\eqref{eq:flatCinv}.
The second term in \eqref{eq:GaussianS} is the standard short-range free action of scalar fields, while the first is a generalized free field theory (GFFT), constituting the free part of interacting long-range scalar models.
The models we will consider in the next two sections are short-range and long-range, respectively, hence this simple example will also allow us to introduce some useful results for later on.

The coupling $\l$ has mass dimension $2\z-2<0$, hence it is an irrelevant coupling for the GFFT. Since the theory is Gaussian, the Renormalization Group (RG) flow is rather trivial: the two-point function in momentum space is $(p^{2\z}+\l p^2)^{-1}$ and goes to the GFFT propagator $1/p^{2\z}$ for $p\to 0$, while for $p\to\infty$ it goes to the canonical free theory propagator $1/p^2$ (up to normalization).
Therefore, we have a flow between two Gaussian CFTs.

The flow is rather standard from the GFFT side, as the operator $ \phi \p^2 \phi$ is a primary in the OPE spectrum of $\phi\times \phi$ in the GFFT (e.g.\ \cite{Benedetti:2019ikb}), and it has scaling dimension greater than $d$.
On the other hand, the flow is somewhat unusual from the canonical free CFT side, as the non-local operator $ \phi \p^{2\z} \phi$ is not in the CFT spectrum.
One possible way to write the perturbation in the UV in the framework of conformal perturbation theory is to introduce an additional field, following the idea proposed in \cite{Behan:2017emf} for the short-range/long-range Ising transition. We thus rewrite the action with a second field $\chi$:
\be \label{eq:GaussianS-chi}
S_{\rm Gauss-2}[\phi,\chi] = \f12 \int \dd x  \, \phi(x) (-\p^2) \phi(x)  + \f12 \int \dd x  \, \chi(x) (-\p^2)^{-\z} \chi(x) + \f{\im}{\sqrt{\l} } \int \dd x  \, \chi(x) \phi(x) \,.
\ee
Integrating out the field $\chi$ we recover \eqref{eq:GaussianS}, up to a rescaling $\phi\to\phi/\sqrt{\l}$.\footnote{The two formulations of the theory are only equivalent if we restrict to the set of correlators of operators built only with fields $\phi$, plus the mixing operator $\chi(x) \phi(x)$. In particular it would make no sense to integrate out $\phi$ in \eqref{eq:GaussianS-chi}.}
This formulation is non-standard in other ways, namely the need of an imaginary coupling (that could however be absorbed into the field $\chi$), and the negative power of the Laplacian, the latter leading to unusual features about the thermodynamic limit, such as inequivalence of statistical ensembles \cite{Campa:2009rev}.
An important difference between \eqref{eq:GaussianS} and \eqref{eq:GaussianS-chi} is that, in the second case, the UV theory has an additional, albeit decoupled, degree of freedom, the field $\chi$. 

We want to test the $F$-theorem on this flow between Gaussian CFTs. To that end, we place the fixed-point  theories on a spherical background, which is done by the standard procedure recalled in appendix~\ref{app:sphereCFT}. 
In practice, the local Laplacian is replaced by the Weyl covariant version (i.e.\ the operator in \eqref{eq:prop-def} with the choice \eqref{eq:b_W}), while the non-local one by the more complicated operator \eqref{eq:D_z} (with kernel \eqref{eq:sphereCinv}).

As we are interested in the difference between the free energies at the two limiting theories, the overall normalization of the functional integral over $\phi$ is not important and will be omitted. However, in the formulation with the action \eqref{eq:GaussianS-chi} the auxiliary field $\chi$ should better have a unit normalized Gaussian functional measure
\be
\int [{\rm d} \chi] \, e^{-S_{\rm Gauss-2}[0,\chi]} = 1 \,,
\ee
so that integrating it out leads to to the functional integral for \eqref{eq:GaussianS}, with the same normalization. Equivalently, this is demanded by imposing that the UV theories obtained from \eqref{eq:GaussianS-chi} or from \eqref{eq:GaussianS} have the same free energy.

For the weak form of the $F$-theorem we only need to compare the fixed points theories. These are GFFTs with different values of $\z$, hence it is straightforward to compute $F$.
The free energy parametrized by $\z$ (including the standard $\z=1$ case) is given by
\be \label{eq:F_GFFT}
F = \f12 \Tr[\ln C^{-1}] = \f12 \sum_{n\geq 0} D_n \, \ln\big(\om^{(\z)}_n \big) \,,
\ee
where $D_n$ is the multiplicity of the eigenvalues, see \eqref{eq:multi}.
The sum is clearly divergent, hence we need a regularization.

\begin{figure}[htb!]
	\centering
	\includegraphics[width=0.5\linewidth]{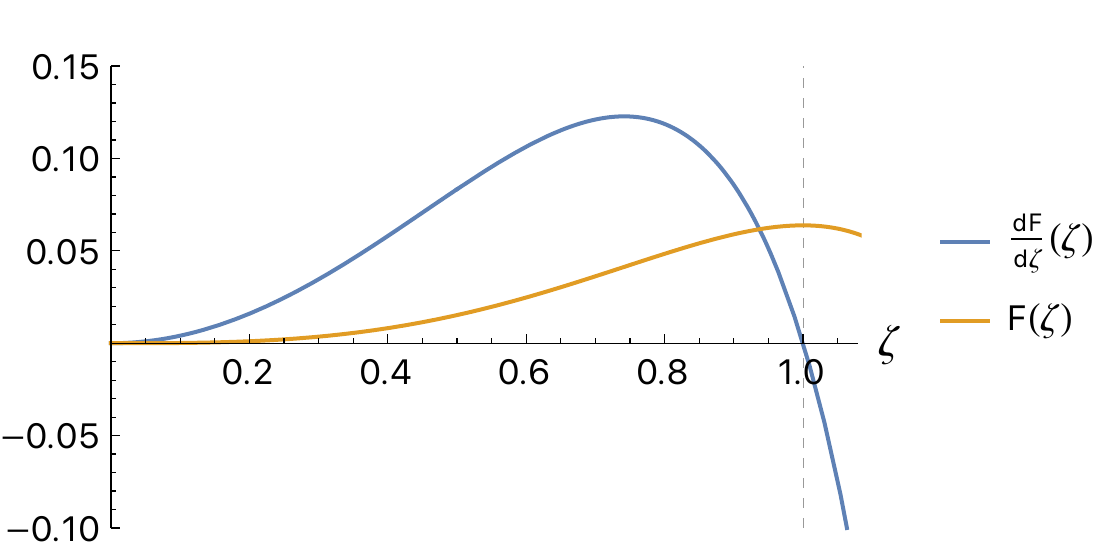}
	\caption{The free energy is $0$ when $\zeta=0$, grows with $\zeta$, and reaches its maximum at $\zeta=1$. We plot the $d=3$ case. The blue curve is the derivative of the free energy with respect to $\zeta$ and the orange curve is the free energy itself.}
	\label{fig:freeenergyfreetheory}
\end{figure}

The same kind of sum was encountered in \cite{Klebanov:2011gs} as the IR limit of a CFT perturbed by a double-trace operator.
Using dimensional regularization as in \cite{Diaz:2007an}, we find:
\begin{equation}
\frac{dF}{d\zeta}=-\zeta\frac{\sin(\pi \zeta)}{\sin(\pi d/2)}\frac{\Gamma(d/2-\zeta)\Gamma(d/2+\zeta)}{\Gamma(1+d)} \label{dF/dz} \,,
\end{equation}
which only has poles for $d$ even. See appendix~\ref{app:F_GFFT} for the detailed computation. When $d=3$, \eqref{dF/dz} simplifies to

\begin{equation}
\frac{d F}{d \zeta}=\frac{1}{24} \pi  \zeta  \left(1-4 \zeta ^2\right) \tan (\pi  \zeta ) \,,
\end{equation}
that is positive for $0 \le \zeta \le1 $. An immediate consequence is that the free energy $F$ grows with $\zeta$ (see Fig.~\ref{fig:freeenergyfreetheory}), showing that an RG trajectory flowing from a short-range free Gaussian model ($\z=1$) to a long-range Gaussian model ($\z<1$) satisfies the F-theorem.

It is interesting to consider the model \eqref{eq:GaussianS} with $\z>1$, for which the GFFT is non-unitary.\footnote{As can be seen from the K\"all\'en-Lehmann spectral representation of the propagator \cite{Benedetti:2020seh}, or from the fact that the unitarity bound $\D>d/2-1$ is violated.}
We also restrict to $\z<d/2$, to keep $\D>0$ and avoid the logarithmic two-point function at $\D=0$.
In this case, the role of UV and IR limiting theory is exchanged, with the $\z>1$ GFFT flowing in the IR to the standard free theory with $\z=1$. And since $\z=1$ is a maximum for  $F$, we find that the free energy increases in the IR. We thus have a trivial counterexample to the $F$-theorem for non-unitary theories.

\section{The $O(N)$ model revisited}
\label{sec:ON_model}

In this section we look at the three-dimensional $O(N)$ model in the short-range case, i.e.\ $\zeta=1$ and $d=3$,
\be	\label{eq:ON-action}
\begin{split}
S[\phi] &=  \frac{1}{2} \int \dd x \, \phi_{a} (  - \partial^2)\phi_{a} + 
	\frac{ m^{2}}{2} \int \dd x \, \phi_{a}  \phi_{a} + \frac{\l}{4 N} \int \dd x \,  (\phi_{a} \phi_{a})^2 \, , 
\end{split}
\ee
where repeated indices are summed over the range $a = 1, \cdots, N$.
Although this case has been studied before in \cite{Klebanov:2011gs} and we only reproduce here the known result, we will do this by a different method. This helps us prepare the ground for the next chapter. The new elements of our analysis are the following.
First, we will frame the discussion within the 2PI effective action formalism (for which we follow \cite{Benedetti:2018goh,Berges:2004yj}). 
Second, we will show how the result of \cite{Klebanov:2011gs} is reproduced by means of a conformal partial wave expansion.

\subsection{The sphere free energy at leading order in the large-$N$ expansion}

At large $N$, the leading-order (LO) 2PI effective action is of order $N$, and reads
\be \label{eq:ON_Gamma_2PI}
\begin{split}
\mathbf{\G}_{\rm LO}[G] &=  N\left(\f12 \int_{x,y} C_1^{-1}(x,y) G(y,x)  + \f12 \int_{x,y} \ln (G^{-1})(x,y) + \f{m^{2}}{2} \int_x G(x,x) + \f{\l}{4} \int_x G(x,x)^2 \right) \\
& =  \f{N}{2} \Tr\left[ C_1^{-1} G +  \ln (G^{-1}) + m^{2} G + \f{\l}{2}  \cB \right] \,,
\end{split}
\ee
where $\int_x = \int \dd x \sqrt{g(x)}$, and $C_1^{-1}=-\partial^2$ is written as the inverse of the free propagator, which is defined in \eqref{eq:C_1}.
In the second line we introduced a compact trace notation, as well as the double-propagator operator
\be \label{eq:bubble-kernel}
\cB(x,y) = G(x,y)^2\,.
\ee
The logarithmic term $\ln(G^{-1})$ should be understood in terms of its eigenvalues (in particular when $G$ coincides or is proportional to the free propagator $C_1$), or by a formal expansion around the free propagator.

The first two terms in \eqref{eq:ON_Gamma_2PI} are very generic, it is the one-loop part of the effective action. The rest should in general be given by a sum over all the vacuum 2PI diagrams, built from the vertices of the theory, but with a generic propagator $G(x,y)$, to be determined self-consistently at a second stage.
The sum of diagrams becomes manageable in the large-$N$ expansion, and at the leading order written in \eqref{eq:ON_Gamma_2PI} only two diagrams survive, the mass tadpole, and the interaction double-tadpole, or figure eight.
\begin{figure}[htb]
\begin{center}
\begin{minipage}{0.3\textwidth}
\includegraphics[scale=0.75]{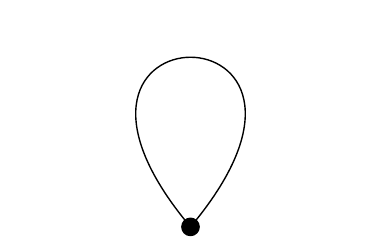}
\end{minipage}
\begin{minipage}{0.3\textwidth}
\includegraphics[scale=0.75]{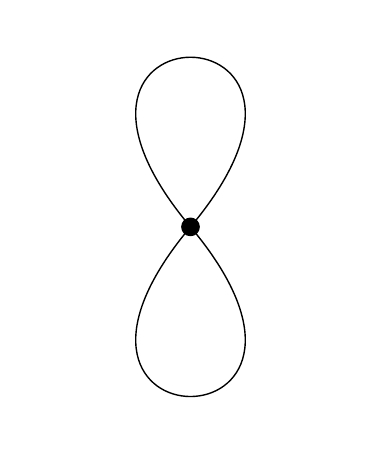}
\end{minipage}
 \caption{The only two vacuum 2PI diagrams occurring in the $O(N)$ model at large $N$. The tadpole on the left has a two-valent mass vertex, while the figure-eight on the right has a $\l$ vertex.} 
 \label{fig:vacuum-2PI-ON}
 \end{center}
\end{figure}

The true full two-point function of the model is found by solving the Schwinger-Dyson (SD) equations, which are obtained as the field equations of the 2PI effective action:
\be
\f{\d \G}{\d G} = 0 \,.
\ee
From \eqref{eq:ON_Gamma_2PI} we find the following form of the SD equations:
\be
G^{-1}(x,y) = C_1^{-1}(x,y) + \left(m^2+ \l G(x,x) \right) \f{\d(x-y)}{\sqrt{g(x)}} \,,
\ee
where we used
\be
\f{ \d G(u,w)}{\d G(x,y)} = \f12 \f{\d(x-u)\d(y-w)+\d(x-w)\d(y-u)}{\sqrt{g(u)}\sqrt{g(w)}} \,.
\ee
Clearly, it is enough to tune the bare mass:
\be \label{eq:mass-tadpole}
m^2 = -\l C_1(x,x) \,,
\ee
in order to cancel the on-shell tadpole and obtain trivially the solution $G=C_1$.

The free energy $F$ is obtained by evaluating the 2PI effective action on shell, i.e.\ by substituting the solution of the SD equations into \eqref{eq:ON_Gamma_2PI}.
Evaluating it on the sphere resolves the IR problem arising from the fact that $F$ is proportional to the volume of the $d$-dimensional background space. However, there are still UV divergences originating from the functional traces, and from the evaluation of the propagator at coincident points.
As we will now review, after appropriate regularization, the result is that the tadpole terms will drop out and thus the leading-order free energy is the same as that of the free theory.

When replacing $G=C_1$ in \eqref{eq:ON_Gamma_2PI}, the first two terms should reproduce ($N$ times) the free energy of the GFFT \eqref{eq:F_GFFT} at $\z=1$.
We have already discussed the regularization and evaluation of $\Tr[\ln C_1^{-1}]$ in the previous section, but what about the first term in \eqref{eq:ON_Gamma_2PI}? Clearly, it corresponds to a divergent contribution $\Tr[\mathbf{1}]\sim \d(0)$. A similar term (with the opposite sign) is however discarded in the typical derivation of the 2PI effective action \cite{Benedetti:2018goh}, for which the first term would otherwise read $\Tr[ (C_1^{-1}-G^{-1}) G]$, and therefore the two cancel out. In any case, when regularized by analytical continuation in the dimension, such terms vanish identically, as shown in \eqref{eq:sum_multiplicity}.

The bare mass was chosen in \eqref{eq:mass-tadpole} to cancel the tadpole in the SD equation, but the cancellation does not occur in the free energy.
However, as we show in appendix~\ref{app:tadpole-SR}, analytic continuation in the dimension gives $C_1(x,x)=0$, hence the contribution of the tadpole terms  to the free energy vanishes.
Therefore, at leading order the free energy of the $O(N)$ critical theory is the same as for the free theory, which in $d=3$ reads: 
\be
F_{\rm LO} = \mathbf{\G}_{\rm LO}[C_1] =  \f12 \Tr[\ln C_1^{-1}] = \f{N}{16} \left( \ln 4 - \f{3\,\z(3)}{\pi^2} \right)\,.
\ee
In order to find a non-trivial result one thus need to consider the subleading corrections to $F$.

\subsection{The next-to-leading order of the large-$N$ expansion}
\label{sec:ON-model-NLO}

\begin{figure}[htbp]
\centering
\includegraphics[width=0.5\textwidth]{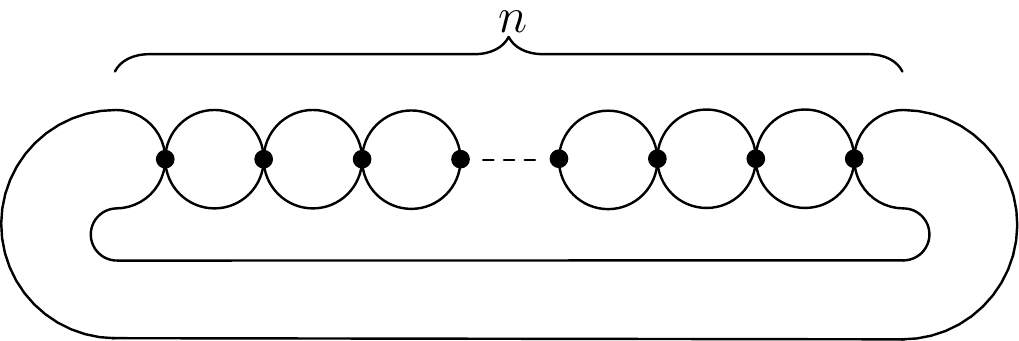}
\caption{Chain of bubbles with $n\geq 1$ vertices. The two-point kernel $\cB$ corresponds to one of the bubbles in the chain.}
\label{fig:bubble_chain}
\end{figure}

The next-to-leading order (NLO) contribution to the 2PI effective action is of order $N^0$, and reads \cite{Benedetti:2018goh}
\be \label{eq:ON_Gamma_2PI_NLO}
\mathbf{\G}_{\rm NLO}[G] =  \f12 \Tr[\ln(\mathbf{1}+ \l \cB)] \,,
\ee
where $\cB$ is the two-point kernel \eqref{eq:bubble-kernel}, corresponding to a single bubble in the chain of bubbles represented in Fig.~\ref{fig:bubble_chain}.
This is the same contribution found in  \cite{Klebanov:2011gs} by introducing an intermediate field, and their result will thus be reproduced: on shell we replace $G(x,y)\to G_{\rm LO}(x,y)=C_1(x,y)$,\footnote{Notice that the NLO part of the on-shell $G(x,y)$ only contributes at NNLO and beyond. This is obvious by expanding $\mathbf{\G}_{\rm NLO}[G_{\rm LO}+G_{\rm NLO}/N]$, and it is true also for $\mathbf{\G}_{\rm LO}[G_{\rm LO}+G_{\rm NLO}/N]$ because by construction $\f{\d \mathbf{\G}_{\rm LO}}{\d G}[G_{\rm LO}]=0$.} and thus the two-point kernel $\cB(x,y)$  is a two-point function of an operator of dimension $\D=d-2$, which on the sphere is diagonalized by the spherical harmonics.
After going to eigenvalues, and taking the bare coupling to infinity in order to tune to the fixed point,\footnote{\label{foot:FPcoupling}The bare coupling in \eqref{eq:ON_Gamma_2PI_NLO} should be expressed as a function of the renormalized coupling, at the leading order of the large-$N$ expansions, as the NLO part would only contribute to NNLO in the effective action. At LO, we have $\l=g/(1-b(d) g/\mu^{4-d})$, where $b(d)$ is a positive constant with a pole at $d=4$, $\mu$ is the renormalization scale, and $g$ is the renormalized coupling. The fixed point of the beta function for the dimensionless coupling $\tilde{g}=g/\mu^{4-d}$ is at $\tilde{g}_\star=1/b(d)$, hence we have $\l\to\infty$.} one is left with the same type of computation as in \eqref{eq:F_GFFT}. In $d=3$, the result is 
\be
F_{\rm NLO} = \mathbf{\G}_{\rm NLO}[C_1] = - \f{\z(3)}{8\pi^2} \,,
\ee
and thus the free energy of the IR fixed point is smaller than the one of the UV free theory.

However, the fact that the NLO contribution to the 2PI effective action could be expressed in terms of a two-point kernel is a very peculiar feature of the $O(N)$ model at large $N$.
In particular, it does not generalize to other models, such as the melonic ones that we will discuss in the following,
for which the NNLO (the NLO is vanishing) contribution is expressed in terms of a four-point kernel.
Therefore, as a warm-up to our next computation, we would like to recover the $O(N)$ model result of   \cite{Klebanov:2011gs} by expressing the NLO part of the effective action as
\be \label{eq:ON_Gamma_2PI_NLO_2}
\mathbf{\G}_{\rm NLO}[G] =  \f12 \Tr[\ln(\mathbb{I}-K)] \,,
\ee
where we introduced the Bethe-Salpeter four-point kernel
\be \label{eq:K-doubletrace}
K(x_1,x_2,x_3,x_4) = - \l G(x_1,x_3) G(x_2,x_4) \d(x_3-x_4) \,,
\ee
and by a slight abuse of notation we use the same symbol for the trace, which in this case refers to a trace of a bilocal to bilocal operator, that is a four-point kernel that acts by integration on two arguments, e.g.\ $\Tr[K]= \int_{x_1,x_2} K(x_1,x_2,x_1,x_2)$.

Under quite general assumptions the Bethe-Salpeter kernel of a CFT is diagonalized by the basis of three-point functions (see appendix~\ref{app:CPW}), hence once we go on shell ($G(x,y)\to C_1(x,y)$) and tune to the fixed point, we should obtain a representation of the kernel (or functions of it) by applying it on the resolution of the identity \eqref{eq:res-id-symm-2}. 
This is not so straightforward for the kernel \eqref{eq:K-doubletrace}. As explained in footnote~\ref{foot:FPcoupling}, the fixed point is at $\l\to\infty$, which requires us to work at finite $\l$ and then take the limit. However, for finite $\l$, and this being a dimensionful coupling for $d<4$, the kernel \eqref{eq:K-doubletrace} does not have the right conformal properties to ensure that its convolution with a three-point function \eqref{eq:3pt} transforms as the three-point function itself.  Thus the  kernel \eqref{eq:K-doubletrace} applied on a three-point function cannot be proportional to it. Unless of course the proportionality constant is zero or infinite. 

On the principal series, we have ${\rm Re}(2\D-h)<0$, hence when the delta function in \eqref{eq:K-doubletrace} acts on \eqref{eq:3pt} the result is zero: the three-point functions with $h\in\cP_+$ are indeed eigenfunctions of \eqref{eq:K-doubletrace} with \emph{vanishing eigenvalue}.
On the other hand, for the isolated contribution in \eqref{eq:res-id-symm-2} we have $2\D-h=0$, hence the delta function gives a finite result, which is not\footnote{At $h=2\D$ and $J=0$, the three-point function \eqref{eq:3pt} is proportional to $C(x_3,x_0)C(x_4,x_0)$, and the action of $K$ on it generates a bubble. The result can be evaluated most easily in Fourier space, and it is found to be proportional to $\l b(d) p_1^{-2} p_2^{-2}(p_1+p_2)^{d-4}$, with the same constant $b(d)$ that appeared in footnote~\ref{foot:FPcoupling}. Due to the factor $(p_1+p_2)^{d-4}$, the result is not proportional to the Fourier transform of $C(x_3,x_0)C(x_4,x_0)$.} proportional to $ \la \phi(x_1) \phi(x_2) \cO_{2\D}(x_0) \ra_{\rm cs}$. 
However, the conformal theory is reached at $\l\to\infty$, hence the three-point function with $h=2\D$ and $J=0$  is formally an eigenfunction with \emph{infinite eigenvalue}.

This has the interesting consequence that for the right-amputated four-point function, which is a geometric series in $K$ (see \eqref{eq:F_s-K}) and hence the coupling appears in the denominator of the resummed series, we obtain  (see appendix~\ref{app:CPW} for the notation):
\be
\begin{split}
\cF_s(x_1,x_2,x_3,x_4)  &= \int \dd y_1 \dd y_2 (\mathbb{I}-K)^{-1}(x_1,x_2,y_1,y_2) C_1(y_1,x_3) C_1(y_2,x_4) \\
& =   \sum_{J\in \mathbb{N}_0^{\text{even}}}  \int_{\f{d}{2}}^{\f{d}{2}+\im\infty}  \f{{\rm d}h}{2\pi\im} \r(h,J) \, \cN^{\D}_{h,J}  \cN^{\D}_{\htilde,J} \, \Psi_{h,J}^{\D,\D,\D,\D}(x_1,x_2,x_3,x_4)  \, .
\end{split}
\ee
Therefore, the isolated contribution that is present in the free theory (see \eqref{eq:cF-GFFT-extra} with $\D=d/2-1$) is suppressed in the critical theory, and the conformal partial wave expansion is the same as in \eqref{eq:cF-GFFT}, even tough $\D=d/2-1<d/4$.
We thus recover the well-known result that the spectrum of the critical $O(N)$ model (i.e.\ the poles of the integrand) at large $N$ is the same as in the free theory for $J>0$, while for $J=0$ it has the $\phi^2$ operator replaced by its shadow.

Applying the above formalism to the free energy on the sphere is however trickier than for the four-point function.
Because of the conformal nature of the basis of three-point functions, we expect the eigenbasis of $K$ on the sphere to be obtained by Weyl mapping:\footnote{Applying the same mapping in \eqref{eq:CPW}, the $\Om(z)$ factors cancel, and we obtain an overall factor $\Om(x_3)^{-d}\Om(x_4)^{-d}$, which reconstructs the correct factors of $1/\sqrt{g}$ for the delta functions in \eqref{eq:res-id-symm-2}.}
\be
\la \phi_{\D}(x_3)  \phi_{\D}(x_4) \cO_h^{\m_1 \cdots \m_J}(x_0) \ra_{\rm cs} \to \Om(x_3)^{-\D}\,\Om(x_4)^{-\D}\,\Om(x_0)^{-h}\,\la \phi_{\D}(x_3) \phi_{\D}(x_4) \cO_h^{\m_1 \cdots \m_J}(x_0) \ra_{\rm cs} \,.
\ee
In fact, for a Bethe-Salpeter kernel with the good conformal transformation properties, the Weyl mapping would give
\be
K(x_1,x_2,x_3,x_4)  \to \Om(x_1)^{-\D}\,\Om(x_2)^{-\D}\, \Om(x_3)^{\D-d}\,\Om(x_4)^{\D-d}\, K(x_1,x_2,x_3,x_4) \,,
\ee
and thus all the $\Om(x_3)$and $\Om(x_4)$ factors would drop out in the convolution, leading to a consistent eigenvalue equation.
As explained above, the kernel in \eqref{eq:K-doubletrace} does not transform in such a nice way at finite $\l$, and we have instead\footnote{The two transformations agree for $\D=d/4$, i.e.\ when $\l$ is dimensionless, but for the short-range $O(N)$ model this only happens at $d=4$, where there is no interacting fixed point.}
\be
K(x_1,x_2,x_3,x_4)  \to \Om(x_1)^{-\D}\,\Om(x_2)^{-\D}\, \Om(x_3)^{-\D-d/2}\,\Om(x_4)^{-\D-d/2}\, K(x_1,x_2,x_3,x_4) \,.
\ee
However, the three-point functions with operator in the principal series are still zero modes of $K$.
Therefore, inserting the resolution of the identity inside the trace in \eqref{eq:ON_Gamma_2PI_NLO_2}, we see that a non-vanishing contribution can only come from the isolated term at $h=2\D$.\footnote{One might worry that although the three-point functions with operator in the principal series have a vanishing eigenvalue, the trace of the conformal partial wave is divergent, and thus the product is undefined. However, as we will see in the following section, there is a clean way to regularize the trace of the conformal partial wave, which does not affect the eigenvalues, and hence the product is indeed zero in the present case.}
It turns out that at finite $\l$ this reproduces the same series as \eqref{eq:ON_Gamma_2PI_NLO}, and hence we can from here on follow again the same steps as in  \cite{Klebanov:2011gs} and obtain the same result. In fact, it is easy to check that the convolution of $K$ with \eqref{eq:extra} equals $K$.
While this is tautological (as we have inserted the identity and the contribution of the principal series is zero, the isolated contribution must reproduce the full identity), one can check directly that indeed the isolated contribution in \eqref{eq:cF-GFFT-extra} acts alone as the identity operator on $K$, and we are back to having to evaluate the chains of bubbles, for which we can follow the steps of \cite{Klebanov:2011gs}. 
Although from a practical point of view we have not gained anything by applying the conformal partial wave expansion to this problem, nevertheless this small detour taught us about the importance of the isolated contributions in such formalism, and it serves as a test of the method, in view of the application in the next section.


\section{The long-range $O(N)^3$ model}
\label{sec:ON3-model}

In this section, we study the long-range $O(N)^3$ tensor model with $0<\zeta<1$ and $d<4$ on the sphere. We first discuss the Schwinger-Dyson equation and then the free energy at next-to-next-to-leading order in this context.

The fundamental field is a real rank-$3$ tensor field, $\phi_{a^1a^2 a^3}(x)$,
transforming under $O(N)^3$ with indices distinguished by the position (typically labelled by a color).  Denoting $\mba = (a^1,a^2,a^3)$, the action of the model on flat space is:
\be	\label{eq:ON3-action}
\begin{split}
S[\phi] &=  \frac{1}{2} \int \dd x \, \phi_{\mba} (  - \partial^2)^{\zeta}\phi_{\mba} + 
	\frac{ m^{2\zeta}}{2} \int \dd x \, \phi_{\mba}  \phi_{\mba} \\
	&\qquad + \frac{1}{4} \int \dd x \, \left[ \im \lambda \hat{\delta}^t_{\mba \mbb\mbc\mbd} + \lambda_1 \hat{P}^{(1)}_{\mba\mbb; \mbc\mbd}
	+ \lambda_2 \hat{P}^{(2)}_{\mba\mbb; \mbc\mbd } \right] \phi_{\mba} \phi_{\mbb} \phi_{\mbc} \phi_{\mbd} \, , 
\end{split}
\ee
where repeated tensor indices are summed over $a^i = 1, \cdots, N$ and we introduced the projectors:
	\begin{equation}
		\hat{P}^{(1)}_{\mba\mbb; \mbc\mbd} \, = \, 3 (\hat{\delta}^p_{\mba\mbb;\mbc\mbd} - \hat{\delta}^d_{\mba\mbb;\mbc\mbd}) \, , \qquad 
		\hat{P}^{(2)}_{\mba\mbb; \mbc\mbd} \, = \, \hat{\delta}^d_{\mba\mbb;\mbc\mbd} \, .
	\end{equation}
and the rescaled operators:
\be \label{eq:deltas}
\hat{\delta}^t_{\mba\mbb\mbc\mbd}=\frac{1}{N^{3/2}} \, \delta^t_{\mba\mbb\mbc\mbd} \,,
\quad \hat{\delta}^p_{\mba\mbb;\mbc\mbd}=\frac{1}{N^{2}} \, \delta^p_{\mba\mbb;\mbc\mbd}\,,
\quad \hat{\delta}^d_{\mba\mbb;\mbc\mbd}=\frac{1}{N^{3}} \, \delta^d_{\mba\mbb;\mbc\mbd}\, ,
\ee
with
\be
\begin{split} \label{eq:deltas-nohat}
&\delta^t_{\mba \mbb\mbc\mbd}  = \delta_{a^1 b^1}  \delta_{c^1 d^1} \delta_{a^2 c^2}  \delta_{b^2 d^2 } \delta_{a^3 d^3}   \delta_{b^3 c^3} \, , \\
	\delta^p_{\mba\mbb; \mbc\mbd } &= \frac{1}{3} \sum_{i=1}^3  \delta_{a^ic^i} \delta_{b^id^i} \prod_{j\neq i}  \delta_{a^jb^j}  \delta_{c^jd^j} \,,
	\qquad  \delta^d_{\mba\mbb; \mbc\mbd }  = \delta_{\mba \mbb}  \delta_{\mbc \mbd} \,.
\end{split}
\ee
Here $t$ stands for \emph{tetrahedron},  $p$ for \emph{pillow}, and $d$ for \emph{double-trace}. Such names refer to the graphical representation of the respective pattern of contraction of indices \cite{Benedetti:2019eyl}.
The faithfully acting symmetry group of the model is $S_3 \times O(N)^3/\mathbb{Z}_2^2$, where the quotient by $\mathbb{Z}_2^2$ is to eliminate the redundancy of having a $\mathbb{Z}_2$ as subgroup in each $O(N)$, and $S_3$ is the permutation group acting on the indices. Symmetry under the latter is often called ``color symmetry", due to using color as a label distinguishing the indices. For the sake of simplicity we just refer to the model as the $O(N)^3$ model.

The power of the Laplacian is taken to be  $\z=d/4$, so that  $\D =d/4$ and the quartic couplings are marginal by power counting. Moreover, one assumes $d<4$ so that the kinetic term is non-local, the propagator is unitary, and local derivative interactions are irrelevant.
It should be noted that the usual dimensional regularization setting $d=4-\epsilon$ is not suitable for this model, as at $d=4$ the model becomes short-range and its renormalization properties differ drastically. The appropriate $\epsilon$ regularization for the long-range model is to set $\zeta=(d+\epsilon)/4$, corresponding to $\Delta=(d-\epsilon)/4$, at fixed $d<4$.

Lastly, the imaginary unit multiplying the tetrahedron interaction is introduced so that in the present notation the large-$N$ OPE spectrum is real for real $\l$ \cite{Benedetti:2019eyl,Benedetti:2019ikb}. This is to be contrasted to the short-range  $O(N)^3$ model of \cite{Klebanov:2016xxf,Giombi:2017dtl}, for which the tetrahedral term is necessarily real due the fixed point condition, and as a consequence the model contains operators of complex dimension. 

The main feature of the $O(N)^3$ model is that in the large-$N$ limit its perturbative expansion is dominated by melonic diagrams built on $\l$ vertices, mixed with cactus diagrams built on $\l_1$ and $\l_2$ vertices \cite{Benedetti:2019eyl}. The melonic dominance leads to new non-trivial CFTs at the interacting fixed points of the renormalization group.

As we review below, the two-point function can be kept conformal, and proportional to the free two-point function, by simply tuning the bare mass to cancel power divergences in the Schwinger-Dyson equation. In order to claim that we have a CFT of course we also need to find a fixed point for the couplings, and eventually show that this has not only scale invariance but full conformal invariance. This task has been carried out (at large $N$) in a series of papers \cite{Benedetti:2019eyl,Benedetti:2019ikb,Benedetti:2020yvb}, and we will not review it here.
We just recall the main results:
\begin{itemize}
\item The tetrahedral coupling $\l$ is exactly marginal.
\item All the  insertions of melonic two-point functions in the diagrams can be eliminated by rescaling each coupling by the square of an appropriate function $\cZ(\l)$, which we will define below. In particular, we define $g= \l \cZ(\l)^2$.
\item In the space of the other two couplings (whose renormalized and rescaled version we denote $g_1$ and $g_2$) we have four interacting fixed points parametrized by $g$:
\be \label{eq:FP-ON3}
g_{1\pm} = \pm \sqrt{g^2} + \cO(g^2) \,, \qquad g_{2\pm} = \pm \sqrt{3 g^2} + \cO(g^2) \,,
\ee
out of which one is IR stable, one is UV stable, and two are saddles (see Fig.~\ref{fig:trajectory}). All four collapse to the trivial fixed point for $g\to 0$.
\begin{figure}[htbp]
\centering
\includegraphics[scale=1]{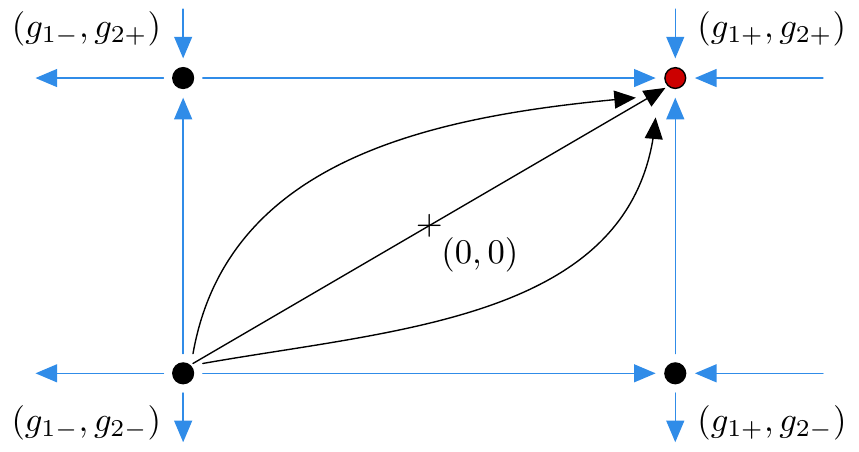}
\caption{Trajectory between the fixed points in the space $(g_1,g_2)$. The red dot is the IR stable fixed point.}
\label{fig:trajectory}
\end{figure}

\item $n$-point functions of the fundamental fields are proved to be conformal.
\item The OPE spectrum in the singlet sector $\phi_{\mba} \times\phi_{\mba}$ has been computed analytically for small $\l$, and numerically at a non-perturbative level: it consists of bilinear operators with real OPE coefficients and with real dimension of the form
\be
h_{n,J} = 2\D+2n+J+\eta_{n,J}(g^2)\,,
\ee
with $n,J/2\in \mathbb{N}_0$ and $\eta_{n,J}(0)=0$. These are operators that in the free limit schematically take the  form $\phi_{\mba} \p^{\m_1} \cdots \p^{\m_J} (\p^2)^n \phi_{\mba}$, and for $n=0$ are also known as ``double twist" operators.
\end{itemize}

It is important to notice that while for $(n,J)\neq (0,0)$ the anomalous dimensions $\eta_{n,J}(g^2)$ are analytic at $g^2=0$ and identical at all four fixed points, the anomalous dimension of $\phi_{\mba} \phi_{\mba}$ has a branch cut: 
\be
\eta_{0,0}(g^2) =   \pm \frac{2\sqrt{3 g^2}}{\Gamma(d/2)(4\pi)^{d/2}} \left( 1 + O(g^2)\right) \,,
\ee
with the plus sign (resp. the minus sign) corresponding to the value at the IR (resp. UV) fixed point of $g_2$.
This is the reason why the coefficient of the tetrahedron interaction in \eqref{eq:ON3-action} needs to be purely imaginary in order to avoid complex scaling dimensions.

We should also point out that the OPE spectrum in the non-singlet sector $\phi_{\mba} \times\phi_{\mbb}$ has not been computed in generality, but the case $\hat{P}^{(1)}_{\mba\mbb; \mbc\mbd}(\phi_{\mbc} \times\phi_{\mbd} ) $ is a straightforward generalization of the singlet case, obtained by the replacement $g^2\to g^2/3$, as will become clear later.

\subsection{Schwinger-Dyson equation for the two-point function}
\label{sec:SDE_tensor}

We start here with a brief review of the solution of the Schwinger-Dyson equation, with the slight modification of working on the spherical background.

\begin{figure}[htb]
\begin{center}
\begin{minipage}{0.3\textwidth}
\includegraphics[scale=0.75]{vacuum_mass.pdf}
\end{minipage}
\begin{minipage}{0.3\textwidth}
\includegraphics[scale=0.75]{vacuum-2PI-8.pdf}
\end{minipage}
\begin{minipage}{0.3\textwidth}
\includegraphics[scale=0.75]{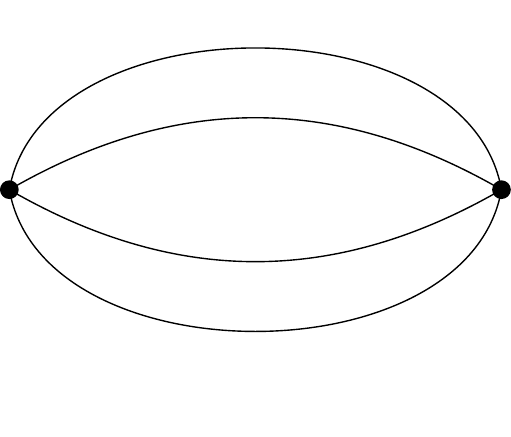}
\end{minipage}
 \caption{The only three types of vacuum 2PI diagrams occurring at large $N$. The tadpole on the left has a two-valent mass vertex. The figure-eight in the middle has a $\l_2$ vertex, and it is of the same type as in the $O(N)$ model. The melon diagram on the right has two $\l$ vertices, and it is characteristic of tensor models.} \label{fig:vacuum-2PI}
 \end{center}
\end{figure}

For large $N$, the dominant 2PI diagrams are depicted in Fig.~\ref{fig:vacuum-2PI}.
The resulting leading-order 2PI effective action is of order $N^3$, and reads
\be \label{eq:ON3_Gamma_2PI}
\begin{split}
\mathbf{\G}[G]  &= N^3\left(\f12 \int_{x,y} C^{-1}(x,y) G(y,x)  + \f12 \int_{x,y} \ln (G^{-1})(x,y) + \f{m^{2\zeta}}{2} \int_x G(x,x) \right.\\
&\qquad\quad \left.+ \f{\l_2}{4} \int_x G(x,x)^2 +\f{\l^2}{8}\int_{x,y} G(x,y)^4\right)\\
& = \f{N^3}{2} \Tr\left[  C^{-1} G +  \ln (G^{-1}) + m^{2\z} G + \f{\l_2}{2}  \cB +\f{\l^2}{4} \cB^2\right] \,,
\end{split}
\ee
where we used the notation introduced in \eqref{eq:ON_Gamma_2PI}, and we used the symmetry of $G(x,y)$ to write the melon integral as a trace of a convolution of two $\cB(x,y)$. The latter is of course an arbitrary choice, and we could as well write it as the trace of the convolution of $G(x,y)$ with $G(y,x)^3$, which will be a useful point of view later on.

The SD equations obtained from \eqref{eq:ON3_Gamma_2PI}:
\be \label{eq:SDeq}
G^{-1}(x,y) =  C^{-1}(x,y)+ \left(m^{d/2} +\l_2  G(x,x) \right) \f{\d(x-y)}{\sqrt{g(x)}} + \l^2     G(x,y)^3 \, ,
\ee
should be understood in the sense of distributions, that is, denoting by $\cS(x,y)$ the right-hand side of \eqref{eq:SDeq}, we should demand that in the limit in which any regularization is removed we obtain:
\be
\int \dd y \sqrt{g(y)} \,  \cS(x,y) \int \dd z \sqrt{g(z)} \, G(y,z) \phi(z) = \phi(x) \,.
\ee

Tuning the bare mass to cancel the tadpole and the divergent part of the melonic integral,
and taking the ansatz  
\begin{equation}
G_{\star}(x,y)=\cZ C(x,y) = \cZ \frac{c(\Delta)}{s(x,y)^{2\Delta}} \,,
\label{eq:ansatz}
\end{equation}
where the coefficient $c(\D)$ is defined in \eqref{eq:freeC-flat}, we obtain
\begin{align}  \label{eq:SDeq2}
\l^2 \cZ^4 \int \dd z \sqrt{g(z)}\, C(x,z)\left( C(z,y)^3  -\f{\delta(z-y)}{\sqrt{g(z)}} B \right) =(1-\cZ)\f{\delta(x-y)}{ \sqrt{g(x)}} \,.
\end{align}
The constant $B$ comes from the mass counterterm
$m^{d/2} = - \l_2  C(x,x) - B$
and should be chosen so as to cancel the divergence of the convolution of $C$ with $C^3$ and yield a delta function as a result.
Since for $\D=d/4$ we have $C(x,y)^3 = c(\D)^3/s(x,y)^{2(d-\D)}$, by comparison to footnote~\ref{foot:Dz-subtr} we have: 
\be
B = c(\D)^3  \int_{s(u,y)>r} \dd u \sqrt{g(u)}\,\f{1}{s(u,y)^{2(d-\Delta)}} - \f{c(\D)^3}{c(d-\D)}\f{\G(d-\D)}{\G(\D)} \,.
\ee
In the spirit of defining (power-) divergent quantities by analytic continuation, we might instead simply set $B=0$.
Either way we obtain for $\cZ$ the equation:
\be
\l^2 \cZ^4 \f{c(\D)^3}{c(d-\D)} = 1-\cZ \,,
\ee
or equivalently:
\be \label{eq:Catalan-Z}
\cZ=1+\l^2 \cZ^4\frac{4\Gamma(1-d/4)}{d(4\pi)^d\Gamma(3d/4)} \,,
\ee
which is the same equation as in flat space. The solution of this equation is the generating function of 4-Catalan (or Fuss-Catalan) numbers:
\be \label{eq:4Catalan}
\cZ(\l) = \sum_{n=0}^{+\infty} \f{1}{4n+1} \binom{4n+1}{n} \left( - \l^2 \f{c(\D)^3}{c(d-\D)}\right)^n \, ,
\ee
and can also be written in an explicit closed form (see\ \cite{Bonzom:2011zz}). Here we just point out that it has a square root singularity at
\be \label{eq:lambda_c}
\l^2_c =  - \f{3^3}{2^8}  \f{c(d-\D)}{c(\D)^3} =    \f{3^3}{2^8} \frac{d(4\pi)^d\Gamma(3d/4)}{4\Gamma(1-d/4)}  \,,
\ee
at which $\cZ(\l_c)=4/3$. 
The series \eqref{eq:4Catalan} resums all the melonic insertions in the two-point function and the critical point corresponds to the radius of convergence of this series, thus determining the maximal value of the coupling for which the model is defined.

In conclusion, at large $N$ the two-point function of the long-range $O(N)^3$ model is proportional to the free propagator, with proportionality constant satisfying \eqref{eq:Catalan-Z}.
This holds both on flat and spherical background, and the two-point function on $S^d$ is obtained from the two-point function in $\mathbb{R}^d$ by simply replacing the distance in $\mathbb{R}^d$ by the chordal distance in $S^d$, as expected.

\subsection{The sphere free energy at leading order in the large-$N$ expansion}
\label{sec:ON3-model-LO}

We now want to evaluate the on-shell 2PI effective action, i.e.\ the free energy. We then start from \eqref{eq:ON3_Gamma_2PI}, and replace $G$ by the solution \eqref{eq:ansatz}: 
\be \label{eq:Gamma_2PI_on_shell}
\begin{split}
\mathbf{\G}[G] = & N^3\left(\f12 \cZ \Tr[C^{-1} C]  + \f12 \Tr[\ln (\cZ^{-1} C^{-1})] + \f{m^{2\zeta}}{2} \cZ \int_x C(x,x) \right.\\
&\qquad\quad \left.+ \f{\l_2 \cZ^2}{4} \int_x C(x,x)^2 +\f{\l^2 \cZ^4}{8}\int_{x,y} C(x,y)^4\right)\,.
\end{split}
\ee
We have five terms to evaluate, all of which are UV divergent. The first four are similar to those in the $O(N)$ model, except for some $\cZ $ coefficients and for the long-range exponent $\z=d/4<1$.
The latter plays no role in the first term, which is proportional to $\Tr[\mathbf{1}]$, and it is set to zero by analytic continuation in $d$, as before. Similarly, the $\ln(\cZ^{-1})\Tr[\mathbf{1}]$ coming from the second term can be dropped. Therefore, the first two terms reproduce \eqref{eq:F_GFFT}.

In order to compute the next two terms, we need first to compute $C(x,x)$ for long-range scaling:
\begin{equation}
C(x,x)=\frac{(d-1)!}{a^{d}\Gamma(d/2)(4\pi)^{d/2}}\sum_{n=0}^{\infty}\frac{D_n}{\omega_n^{(\zeta)}} \,.
\end{equation}
Using dimensional regularization, we find again $C(x,x)=0$, see appendix~\ref{app:tadpole-LR}. 

Finally, we need to evaluate the melon integral, which at leading order is the only qualitative difference with respect to the $O(N)$ model. We call it $M$:
\begin{equation} \label{eq:melon-int}
M= \int_{x,y}  \, C(x,y)^4 \,.
\end{equation}

In order to regulate the UV divergences we set $\Delta=\frac{d-\epsilon}{4}$, and obtain the $\epsilon$ regularized melon integral:
\begin{equation} \label{eq:Mepsilon}
M_\eps 
=c(\Delta)^4 \int_{x,y}  \,  \frac{1}{s(x,y)^{2(d-\epsilon)}} = c(\Delta)^4 (2 a)^{2\eps}\int \dd  x \int \dd  y \, \f{1}{(1+x^2)^\eps (1+y^2)^\eps |x-y|^{2(d-\eps)}}\,.
\end{equation}
Alternatively, as we already mentioned, $M$ can be thought of as the trace of the convolution $G(x,y)$ with $G(y,x)^3$. On shell and for $\Delta = d/4$ we have $G_{\star}(y,x)^3 \propto C(x,y)^3 \propto 1/s(x,y)^{3d/2} = 1/s(x,y)^{2(d-\D)}$ which is also the two-point function of the shadow field. The regularization can then be viewed as a shift by $\epsilon$ of the dimension of the shadow, that is the replacement $\wtD=d-\D\to\wtD-\eps$ at fixed $\D=d/4$:
\be  \label{eq:Mepsilon-2}
M_\eps = c(\Delta)^4 \int_{x,y}  \,  \la \phi_\D(x) \phi_\D(y) \ra_{\rm cs} \la \phi_{\wtD-\eps}(x) \phi_{\wtD-\eps}(y) \ra_{\rm cs} \,.
\ee
This point of view will be useful in the next subsection.

The  regularized melon integral \eqref{eq:Mepsilon} can be reduced to an integral over a single point by exploiting the homogeneity of the sphere and factoring out a volume of the d-sphere  $V_d=\int \dd z\, \Omega(z)^d$, given in \eqref{eq:Vol-Sd}.
The remaining integral, which has appeared for example in \cite{Cardy:1988cwa,Klebanov:2011gs},  can be computed for $\eps>d/2>0$ and then analytically continued to small $\eps$, leading to 
\begin{equation}\label{eq:vanishmelo}
M_\eps=\frac{a^{2\epsilon}\, \Gamma(\frac{d+\epsilon}{4})^4 \, \Gamma(-\frac{d}{2}+\epsilon)}{2^{3d-1}\,\pi^{d-1/2}\,\Gamma(\frac{d-\epsilon}{4})^4\,\Gamma(\frac{d+1}{2})\, \Gamma(\epsilon)} \,,
\end{equation}
which vanishes in the limit $\eps\to 0$, whenever $d$ is not an even number.
Given that $C(x,y)^3$ is proportional to $C^{-1}(x,y)$ (see \eqref{eq:SDeq2}), we can interpret this result as another instance of $\Tr[\mathbf{1}]$ being set to zero by analytic regularization.

In conclusion, at leading order the sphere free energy of the interacting long-range $O(N)^3$ model reduces to that of the free model, i.e.\ $N^3$ times the GFFT free energy \eqref{eq:F_GFFT}.
This is an interesting feature, shared with the $O(N)$ model. However, we do not know if there is any deeper reason behind it, or if it is just an accident of these specific large-$N$ limits dominated by tadpole or melonic diagrams.

\subsection{The next-to-next-to-leading order of the large-$N$ expansion}
\label{sec:ON3-model-NLO}

The free energy of the $O(N)^3$ model has a series expansion in $1/\sqrt{N}$ \cite{Carrozza:2015adg}.
At NLO the only 2PI diagram is a figure eight with one tetrahedron vertex \cite{Benedetti:2018goh}, and hence its contribution vanishes like the similar LO contributions from the $\l_2$ coupling.

\begin{figure}[htbp]
\centering
\includegraphics[width=0.4\textwidth]{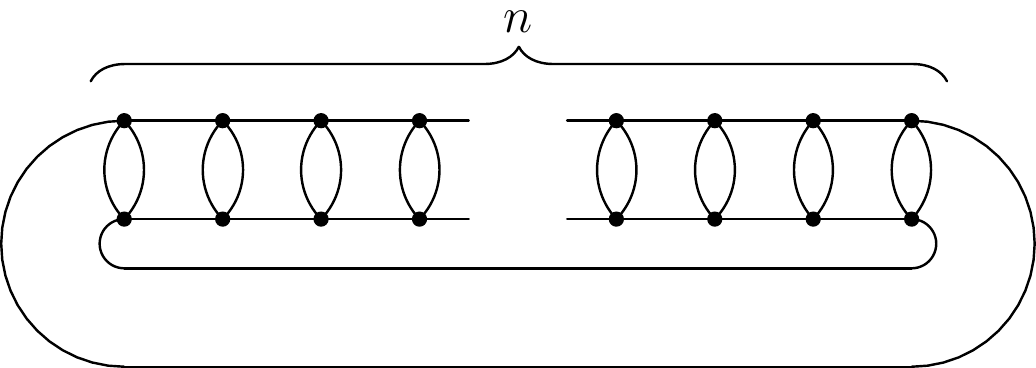}
\caption{A generic NNLO vacuum 2PI diagram having the form of a closed ladder with $n\geq 2$ rungs, and vertices corresponding to the tetrahedron interaction. Similar diagrams but  with a twist in the rails, thus forming a M\"obius strip, appear only at lower order in $N$.
}
\label{fig:ladders}
\end{figure}

\begin{figure}[htbp]
	\centering
	\includegraphics[width=0.20\linewidth]{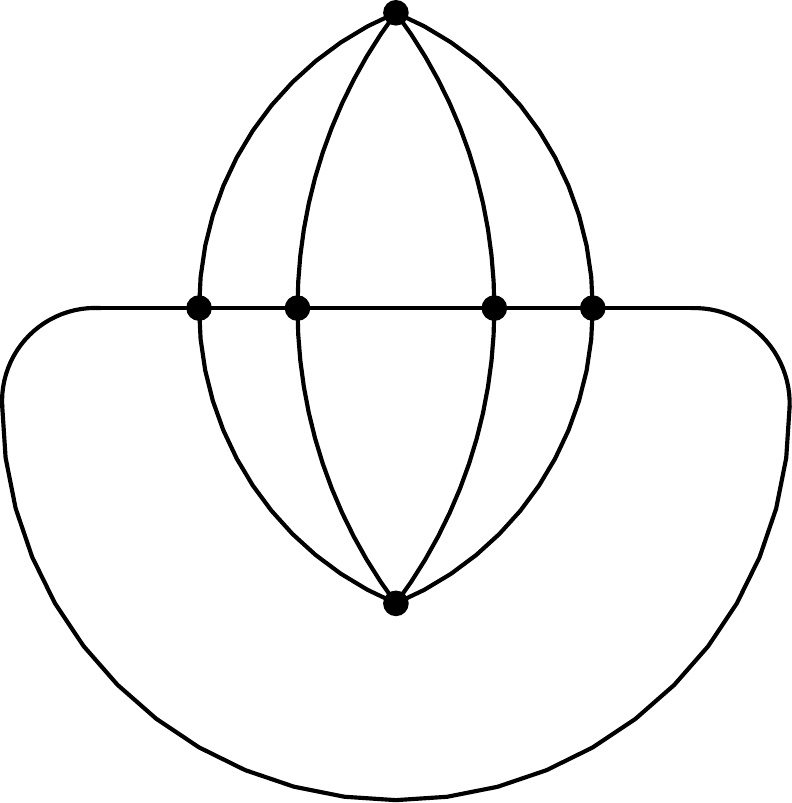}
	\caption{The unique NNLO vacuum 2PI diagram besides the ladders. All six vertices are tetrahedral.}
	\label{fig:monstru}
\end{figure}

At NNLO the combinatorics of the $O(N)^3$ model with only the tetrahedron interaction has been carefully studied in  \cite{Bonzom:2019yik}.
Restricting to 2PI diagrams, it turns out that there is an infinite family of ladder-like diagrams, closed in a planar way as shown in Fig.~\ref{fig:ladders}, plus one special diagram, shown in Fig.~\ref{fig:monstru}. It is straightforward to complement the analysis of  \cite{Bonzom:2019yik} by adding the effect of the pillow and double-trace interactions.
It turns out that we only need to add to those, diagrams obtained from the ladder diagrams by replacing one or more rungs (each made by two $\l$ vertices) with one or more local $\l_1$ vertices, as in the chain diagrams of the $O(N)$ model of Fig.~\ref{fig:bubble_chain}.\footnote{Similar diagrams with $\l_2$ instead of $\l_1$, appear only farther in the $1/\sqrt{N}$ expansion. In fact, the $\l_2$ coupling is associated to a double-trace interaction, which is the same as the $O(N)$ model interaction, with the replacement $N\to N^3$. Therefore, chains of bubbles with $\l_2$ vertices will only contribute at order $N^0$, as in the $O(N)$ model. }


As we will show in appendix~\ref{app:monster}, the graph of Fig.~\ref{fig:monstru} gives a finite contribution to the sphere free energy. However, since it only depends on the tetrahedron coupling, it takes the same value at all the fixed points, and thus it does not play a role in checking the F-theorem for this model. We will omit it in the rest of this section.

\begin{figure}[htbp]
\begin{center}
$$K_1 = -\l^2\, \vcenter{\hbox{\includegraphics[width=1.8cm]{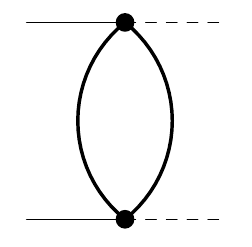}}} - \l_1 \, \vcenter{\hbox{\includegraphics[width=1.8cm]{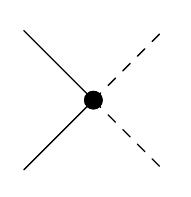}}}$$
 \caption{Graphical representation of the kernel $K_1$ \eqref{eq:ON3-K}. Solid lines represent full two-point functions, while dashed lines represent amputated external legs. For obvious reasons we call the first term ``ladder kernel" and the second ``local kernel".} 
 \label{fig:kernel}
 \end{center}
\end{figure}

We thus have an infinite series of diagrams that are a mixture of ladders and chains. Formally, they can be easily resummed in terms of a kernel that is the sum of a ladder kernel and a local kernel:\footnote{We have subtracted a $\Tr[K_1]$ from the expansion of the logarithm because its ladder contribution does not correspond to a 2PI diagram. As for its local contribution, it is another figure eight diagram, which evaluates to zero, and hence we can add or subtract it at will.}
%
\be \label{eq:ON3_Gamma_2PI_NNLO}
\mathbf{\G}_{\rm NNLO}[G] =  \f{N^2}{2} \left( \Tr[\ln(\mathbb{I}-K_1)] +  \Tr[K_1] \right)\,,
\ee
where $K_1$ is the familiar Bethe-Salpeter kernel of melonic theories, namely
\be \label{eq:ON3-K}
K_1(x_1,x_2,x_3,x_4) = - G(x_1,x_3) G(x_2,x_4) \Big(\l^2 \,  G(x_3,x_4)^2 +\lambda_1 \delta(x_3,x_4)\Big)     \,,
\ee
represented in Fig.~\ref{fig:kernel}.
%
It should be noted that this kernel displays two differences with respect to the usual ladder kernel of melonic theories found in the literature (e.g.\ \cite{Klebanov:2016xxf,Giombi:2017dtl}). First, we have a minus sign in the ladder kernel, due to the imaginary unit multiplying the tetrahedron interaction in \eqref{eq:ON3-action}. Second, a factor $3$ is missing from the ladder kernel. The reason for the latter can be understood from the full Bethe-Salpeter kernel obtained from the leading-order 2PI effective action, namely \cite{Benedetti:2019eyl}:
\be
\begin{split}
\label{eq:full-K}
\hat{K}_{\mba\mbb; \mbc\mbd}(x_1,x_2,x_3,x_4) & = -G(x_1,x_3) G(x_2,x_4) \Big(\l^2 \,  G(x_3,x_4)^2 +\lambda_1 \delta(x_3,x_4)\Big) \hat{P}^{(1)}_{\mba\mbb; \mbc\mbd} \\
& \quad - G(x_1,x_3) G(x_2,x_4)  \Big(3\l^2 \,  G(x_3,x_4)^2 +\lambda_2 \delta(x_3,x_4)\Big) \hat{P}^{(2)}_{\mba\mbb; \mbc\mbd} \\
& \equiv  K_1(x_1,x_2,x_3,x_4) \hat{P}^{(1)}_{\mba\mbb; \mbc\mbd} + K_2(x_1,x_2,x_3,x_4) \hat{P}^{(2)}_{\mba\mbb; \mbc\mbd}\,.
\end{split}
\ee
In most of the previous literature on these models the focus was on the spectrum of singlet operators, which can be obtained by the $s$-channel OPE of the four-point function $\la \phi_{\mba}(x_1)\phi_{\mba}(x_2) \phi_{\mbb}(x_3) \phi_{\mbb}(x_4)\ra$. The latter requires inverting $(\mathbb{I}-\hat{K})_{\mba\mbb; \mbc\mbd}$ and taking a trace on the tensor indices by contracting with $\delta_{\mba\mbb}\delta_{\mbc\mbd}$. In such case, the first term is zero because $\hat{P}^{(1)}$ is traceless in this channel, and thus we have a factor $3\l^2$ in the ladder part of the surviving kernel contribution. However, the index trace leading to \eqref{eq:ON3_Gamma_2PI_NNLO} is taken by contracting with $\delta_{\mba\mbc}\delta_{\mbb\mbd}$.  In this case, the $\hat{P}^{(1)}$ term gives the leading contribution of order $N^2$, while the $\hat{P}^{(2)}$ term only contributes at order $N^0$, thus explaining the absence of the factor $3$ in \eqref{eq:ON3-K}. 
Notice that the same part of the kernel is the relevant one for the four-point function $\hat{P}^{(1)}_{\mba\mbb; \mba'\mbb'} \hat{P}^{(1)}_{\mbc\mbd; \mbc'\mbd'} \la \phi_{\mba'}(x_1)\phi_{\mbb'}(x_2) \phi_{\mbc'}(x_3) \phi_{\mbd'}(x_4)\ra$, whose  $s$-channel OPE provides the spectrum of bilinear operators that are in a symmetric-traceless matrix representation of one of the $O(N)$'s, and the singlet one of the other two.

As before,  in order to evaluate the NNLO free energy we substitute \eqref{eq:ansatz} into \eqref{eq:ON3_Gamma_2PI_NNLO}, and we use the conformal partial wave formalism reviewed in appendix~\ref{app:CPW}.
Inserting the resolution of the identity \eqref{eq:res-id-nonsymm} inside the trace in \eqref{eq:ON3_Gamma_2PI_NNLO}, we find the following formal expression:
\be 
\begin{split}
F_{\rm NNLO} &= \f{N^2}{2} \sum_{J\in \mathbb{N}_0}  \int_{\f{d}{2}}^{\f{d}{2}+\im\infty}  \f{{\rm d}h}{2\pi\im} \r(h,J) \,\big(\ln(1-k(h,J))+k(h,J)\big)\, \cN^{\D}_{h,J}  \cN^{\wtD}_{\htilde,J} \, \Tr[ \Psi_{h,J}^{\D,\D,\wtD,\wtD} ]  \,,
\end{split}
\label{eq:F_NNLO}
\ee
with notation defined in appendix~\ref{app:CPW}.
In addition, we here have $\D=d/4$, and the kernel eigenvalue 
\begin{equation}
k(h,J)=-\f{g^2}{(4\pi)^d}
 \frac{\Gamma(-\frac{d}{4}+\frac{h+J}{2})\Gamma(\frac{d}{4}-\frac{h-J}{2})}{\Gamma(\frac{3d}{4}-\frac{h-J}{2})\Gamma(\frac{d}{4}+\frac{h+J}{2})} \,,
\label{eq:ON3-k}
\end{equation}
where $g$ is the effective tetrahedral coupling $g= \l\, \cZ(\l)^2 $, which resums all the two-point melonic insertions. The latter are absent by construction in the 2PI effective action, but reappear when going on shell, i.e.\ when replacing the generic $G$ by the solution of the SD equations $G_\star(x,y) = {\cal Z}(\l) C(x,y)$. By writing all quantities in terms of $g$ we can keep ignoring the melonic insertions and use the free propagator, but we should restrict its range to $|g|<g_c\equiv \l_c \, \cZ(\l_c)^2 $, because of the square root singularity at the critical coupling \eqref{eq:lambda_c}.

It will actually be convenient to consider the derivative of the free energy in order to get rid of the logarithm:
\be \label{eq:F-derviative}
- g\f{\p}{\p g}F_{\rm NNLO}  = N^2 \sum_{J\in \mathbb{N}_0}  \int_{\f{d}{2}}^{\f{d}{2}+\im\infty}  \f{{\rm d}h}{2\pi\im} \r(h,J) \,\f{k(h,J)^2}{1-k(h,J)}\, \cN^{\D}_{h,J}  \cN^{\wtD}_{\htilde,J} \, \Tr[ \Psi_{h,J}^{\D,\D,\wtD,\wtD} ]   \,.
\ee

A striking feature of \eqref{eq:F_NNLO}, or \eqref{eq:F-derviative}, is that the kernel eigenvalue is only sensitive to the ladder part of the kernel, because the local part has vanishing eigenvalue on the principal series.\footnote{This is straightforward for ${\rm Re}(h)>d/2$, and it is extended by analytic continuation to the principal series and beyond.} 
This fact can be puzzling, as the diagrams having $\l_1$ vertices are necessary at the perturbative level: expressing $\l_1$ as a series in the renormalized coupling $g_1$ and in $g$, they have to cancel the UV divergences of the ladder diagrams  \cite{Benedetti:2019eyl}.
Nevertheless, the result of the resummed series of diagrams, evaluated at the fixed point, where $g_1$ takes a specific $g$-dependent value, turns out to be expressible in terms of only ladder diagrams.
 This is a familiar situation in the four-point function of these models \cite{Benedetti:2019eyl,Benedetti:2019ikb},\footnote{As well as in the fishnet model \cite{Grabner:2017pgm,Kazakov:2018qbr}.} and it is due to the fact that in the conformal limit, the local kernel has zero eigenvalues.  
The resummed series captures the contribution of the chain diagrams in a subtle manner. When evaluating \eqref{eq:F-derviative} using conformal partial waves, only the ladder kernel contributes, and thus one needs to integrate over the principal series an analytic function of $g^2$. However, the result of the integration (for $J=0$) is a non-analytic function with a $\sqrt{g^2}$ branch cut. In the perturbative expansion such a branch cut can only come from the $\lambda_1$ diagrams, due to the branch cut in the $g_{1\pm}$ fixed points \eqref{eq:FP-ON3}. 
Therefore, the non-perturbative resummation of the ladder diagrams automatically includes the contribution of the chain diagrams as well, which is a very non-trivial fact.
We will provide a cross check of this statement below.

The problem with the expression \eqref{eq:F-derviative} is that the trace of the conformal partial wave is divergent. From \eqref{eq:CPW} we have:
\be \label{eq:TrCPW}
\Tr[ \Psi_{h,J}^{\D,\D,\wtD,\wtD} ] = \int_{x_1,x_2,z}  \,  \la \phi_{\D}(x_1) \phi_{\D}(x_2) \cO_h^{\m_1\cdots \m_J}(z) \ra_{\rm cs}\la \phi_{\wtD}(x_1) \phi_{\wtD}(x_2) {\cO}_{\htilde}^{\m_1\cdots \m_J}(z)\ra_{\rm cs} \,.
\ee
Formally this integral is conformally invariant, but as a consequence it is also divergent because of the infinite volume of the conformal group. Notice that the same type of integral appears as a natural pairing (or inner product) of $n$-point functions \cite{Karateev:2018oml}; however, in that case one divides by the volume of $SO(d+1,1)$ (or in other words, one considers a gauge-fixed version of the integral) in order to define a finite pairing. 
In our case, we do not have the freedom to divide the free energy by a diverging quantity: the idea of the $F$-theorem is  that instead we should look at the finite part of the free energy. This might be hiding behind some divergence, which we have to regulate and subtract.
The melon integral in \eqref{eq:melon-int} is an example of the same kind: for $\D=d/4$ it is proportional to a pairing of two-point structures $\int \dd x_1 \dd x_2   \, \la \phi_{\D}(x_1) \phi_{\D}(x_2)  \ra_{\rm cs}\la  \phi_{\wtD}(x_1) \phi_{\wtD}(x_2)\ra_{\rm cs}$, and it is divergent for the same reason as above. We have regularized the melon integral by analytic continuation in \eqref{eq:Mepsilon-2}, which is equivalent to subtracting the divergent part, and we have found a vanishing finite part.
We are now going to employ a similar analytic continuation in order to extract the finite part of \eqref{eq:TrCPW}.

In the case of \eqref{eq:TrCPW}, setting $\Delta=\frac{d-\epsilon}{4}$ everywhere would not help, as the dependence of the integrand on $\D$ drops out. It is thus useful to take the second point of view we presented on the regularization of the melon integral and shift only the dimensions of the shadow operators. That is, we define:
\be \label{Iepsilon}
\cI_\eps(J) =  \int_{x_1,x_2,z}  \,  \la \phi_\D(x_1) \phi_\D(x_2) \cO_h^{\m_1\cdots \m_J}(z) \ra_{\rm cs}\la \phi_{\wtD-\eps}(x_1)\phi_{\wtD-\eps}(x_2) {\cO}_{\htilde-\eps}^{\m_1\cdots \m_J}(z)\ra_{\rm cs} \,.
\ee
This analytic regularization breaks the conformal invariance of the integral, but not its translation invariance. Therefore, on flat space there is still a space-time volume divergence, which is instead regularized on the sphere. UV divergences (at coincident points) are still there, but will be cured in an appropriate range of $\eps$, and then we will use analytic continuation to take the limit $\eps\to 0$.

It can be shown (see appendix~\ref{app:I_eps}) that all the $J$-dependence in \eqref{Iepsilon} can be factored out of the integral
\begin{equation}
\cI_\eps (J)= \frac{\Gamma(d-2+J)\Gamma(\tfrac{d-2}{2})}{2^J\Gamma(d-2)\Gamma(\tfrac{d-2}{2}+J)} \cI_\eps (0)\,,
\end{equation}
with 
\begin{equation}\label{eq:I0e}
\cI_\eps (0)=\int \dd x_1 \dd x_2   \dd  z 
\frac{\left( \Omega(x_1)\Omega(x_2) \Omega(z)\right)^d}{s(x_1,x_2)^{d-\eps}s(x_1,z)^{d-\eps} s(x_2,z)^{d-\eps}}\,.
\end{equation}
Because of the homogeneity of the sphere we are free to set $z=0$, and factor out  the volume of the $d$-sphere  $V_d=\int \dd z\, \Omega(z)^d$, given in \eqref{eq:Vol-Sd}. The integral $\frac{\cI_\eps(0)}{V_d}$ has already been computed  in \cite{Cardy:1988cwa} and the results is\footnote{In appendix~\ref{app:I_eps} we provide a detailed derivation.}
\begin{equation} \label{eq:I_eps}
	\frac{\cI_\eps(0)}{V_d}=(2a)^{3\epsilon-d}\frac{\pi ^d \Gamma \left(\frac{\epsilon }{2}\right)^3 \Gamma \left(\frac{3 \epsilon }{2}-\frac{d}{2}\right)}{\Gamma \left(\frac{d}{2}\right) \Gamma \left(\epsilon\right)^3} \,,
\end{equation}
which has a finite limit for $\epsilon \rightarrow 0$ as long as $d$ is not an even number:
\begin{equation}
\lim_{\eps \to 0}	\frac{\cI_\eps(0)}{V_d}=\frac{8 \left(\pi ^d \Gamma \left(-\frac{d}{2}\right)\right)}{(2a)^d\Gamma \left(\frac{d}{2}\right)} \,.
\end{equation}
Reinserting the spin and volume factors we have:
\begin{equation}
\cI_\eps (J)= (2a)^{3\epsilon}\frac{\pi^{3d/2}\Gamma(\frac{\eps}{2})^3\Gamma(\frac{3\eps}{2}-\tfrac{d}{2})\Gamma(d-2+J)\Gamma(\tfrac{d-2}{2})}{2^J\Gamma(\eps)^3\Gamma(d-2)\Gamma(\tfrac{d-2}{2}+J)\Gamma(d)}\,.
\end{equation}
and after removing the regulator $\epsilon$ we get:
\begin{equation}
\cI_0 (J)= \frac{8\pi^{3d/2}\Gamma(-\tfrac{d}{2})\Gamma(d-2+J)\Gamma(\tfrac{d-2}{2})}{2^J\Gamma(d-2)\Gamma(\tfrac{d-2}{2}+J)\Gamma(d)} \,.
\end{equation}

The $\epsilon$ regularization thus provides a finite result for the trace of the conformal partial wave.
However, it turns out that it is important to consistently shift by $\epsilon$ also the normalization factor of the three-point function of shadow operators, as otherwise the resulting series in $J$ would diverge.
The product of normalization factors \eqref{eq:prod-cN} is then replaced, at large $J$, by:
\be
\cN^{\D}_{h,J}  \cN^{\wtD-\eps}_{\htilde-\eps,J} \sim \f{2^{3(d+\eps)/2+J}}{(2\pi)^{d}} \, J^{-3\eps} \left( 1 + \mathcal{O}(1/J) \right) \,.
\ee
This $J^{-3\eps} $ factor renders the series over $J$, whose coefficients otherwise behaves asymptotically as $1/J$, convergent.

We can now perform the integral on $h$ and sum over $J$ in \eqref{eq:F-derviative}.
As explained in appendix~\ref{app:NumericsLargeJ}, at large $J$, the integral behaves as $\frac{f(\epsilon)}{J^{1+3\epsilon}}$ with $f(\epsilon)$ an analytic function at $\epsilon=0$. We then write:
\be\label{eq:ladderssum}
\begin{split}
- g\f{\p}{\p g}F_{\rm NNLO}^{\eps}  =  &\, N^2 \int_{\f{d}{2}}^{\f{d}{2}+\im\infty}  \f{{\rm d}h}{2\pi\im} \r(h,0) \,\f{k(h,0)^2}{1-k(h,0)}\, \cN^{\D}_{h,0}  \cN^{\wtD-\eps}_{\htilde-\eps,0} \, \cI_\eps (0) \\
&+ N^2 \Bigg[ \sum_{J\in \mathbb{N}_+} \Bigg( \int_{\f{d}{2}}^{\f{d}{2}+\im\infty}  \f{{\rm d}h}{2\pi\im} \r(h,J) \,\f{k(h,J)^2}{1-k(h,J)}\, \cN^{\D}_{h,J}  \cN^{\wtD-\eps}_{\htilde-\eps,J} \, \cI_\eps (J) -\frac{f(\epsilon)}{J^{1+3\epsilon}}\Bigg) \\
 & \quad + \sum_{J\in \mathbb{N}_+} \frac{f(\epsilon)}{J^{1+3\epsilon}} \Bigg] \,.
\end{split}
\ee
The first sum is now convergent for $\epsilon=0$, and thus can be computed numerically, while the second sum gives and explicit pole in $\eps$.
We thus define the renormalized sphere free energy, or rather its derivative, as:
\be
- g\f{\p}{\p g}F_{\rm NNLO} = \lim_{\eps\to 0} \left(- g\f{\p}{\p g}F_{\rm NNLO}^{\eps} - \frac{N^2 f(0)}{3\epsilon} \right) \,,
\ee
which for example at $d=3$, $g=1$ and $a=1$ gives:
\be
- g\f{\p}{\p g}F_{\rm NNLO}   = 7.57 \times 10^{-4} \, N^2 \, .
\ee

\begin{figure}[htbp]
\centering
\includegraphics[scale=0.5]{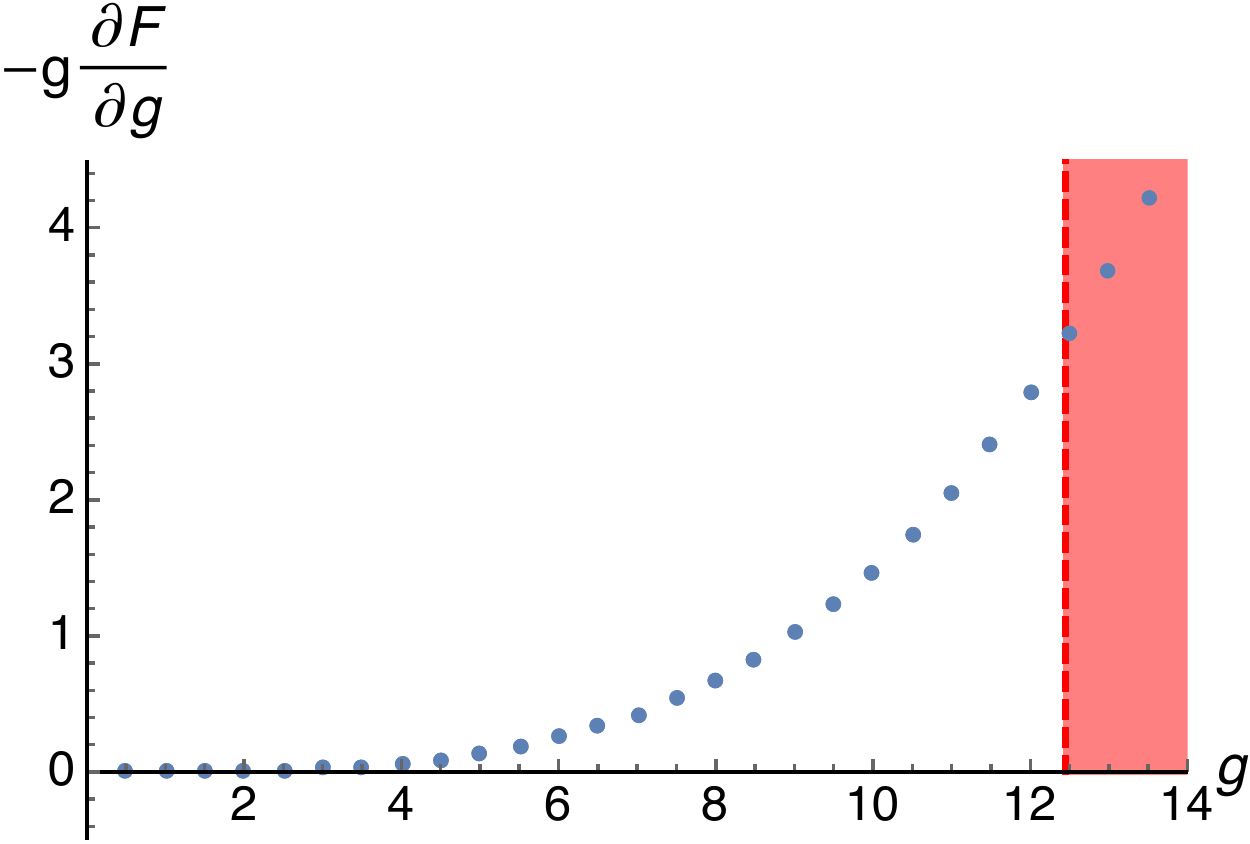}
\caption{Derivative of the free energy at $d=3$ and $a=1$. The red area corresponds to $g>g_c$, where nothing seems to happen, but in fact there is no $\l$ giving such values of $g$.}
\label{fig:plotDF}
\end{figure}

We computed this value at $d=3$, for $a=1$ and different values of $g$ up to $g_c\equiv \l_c \, \cZ(\l_c)^2$ (with $\l_c$ given in \eqref{eq:lambda_c}). The result is a positive convex function, vanishing at the origin, as shown in Fig.~\ref{fig:plotDF}.

\paragraph{The non-normalizable contribution.}
In the context of the F-theorem we are interested in the difference between the free energy at the UV fixed point and at the IR fixed point. The value of $g$ being the same at the two fixed points, and the above result being seemingly independent of $g_1$ or $g_2$, it would naively seem that the free energy is the same at the two fixed points. 

Things are however more subtle than this. As explained in appendix~\ref{app:CPW}, the resolution of the identity \eqref{eq:res-id-nonsymm} is valid in a functional space with appropriate integrability conditions, and the latter are violated by four-point functions whose $s$-channel OPE contains operators of dimension smaller than $d/2$.
It turns out that this is precisely what happens in the ultraviolet CFT, due to a primary operator in the OPE of $\hat{P}^{(1)}_{\mba\mbb; \mbc\mbd}(\phi_{\mbc} \times\phi_{\mbd} )$ whose dimension descends below $d/2$. 
In fact, at $J=0$, the equation $k(h,0)$ has two solutions $h_{\pm}$ lying respectively on the right and on the left of the integration contour $\cP_+$:
\begin{equation}  \label{eq:h_pm}
h_{\pm}=\frac{d}{2} \pm \frac{2\sqrt{g^2}}{\Gamma(d/2)(4\pi)^{d/2}} + \mathcal{O}(|g|^3) \,,
\end{equation} 
and in the UV, the physical dimension is actually the one on the left of the contour. 
Therefore, in evaluating the free energy of the UV theory by the CPW method we need to subtract these contributions from the operator being traced before applying it on the resolution of identity \eqref{eq:res-id-nonsymm}, and then add them back.
This amounts to including, besides  the principal series integral, an isolated non-normalizable contribution, as in \eqref{eq:cF-extra}.
That is, in the UV version of \eqref{eq:F-derviative} we have to add minus the residue of the integrand at $h=h_-$.

\begin{figure}[htbp]
\centering
\includegraphics[scale=0.5]{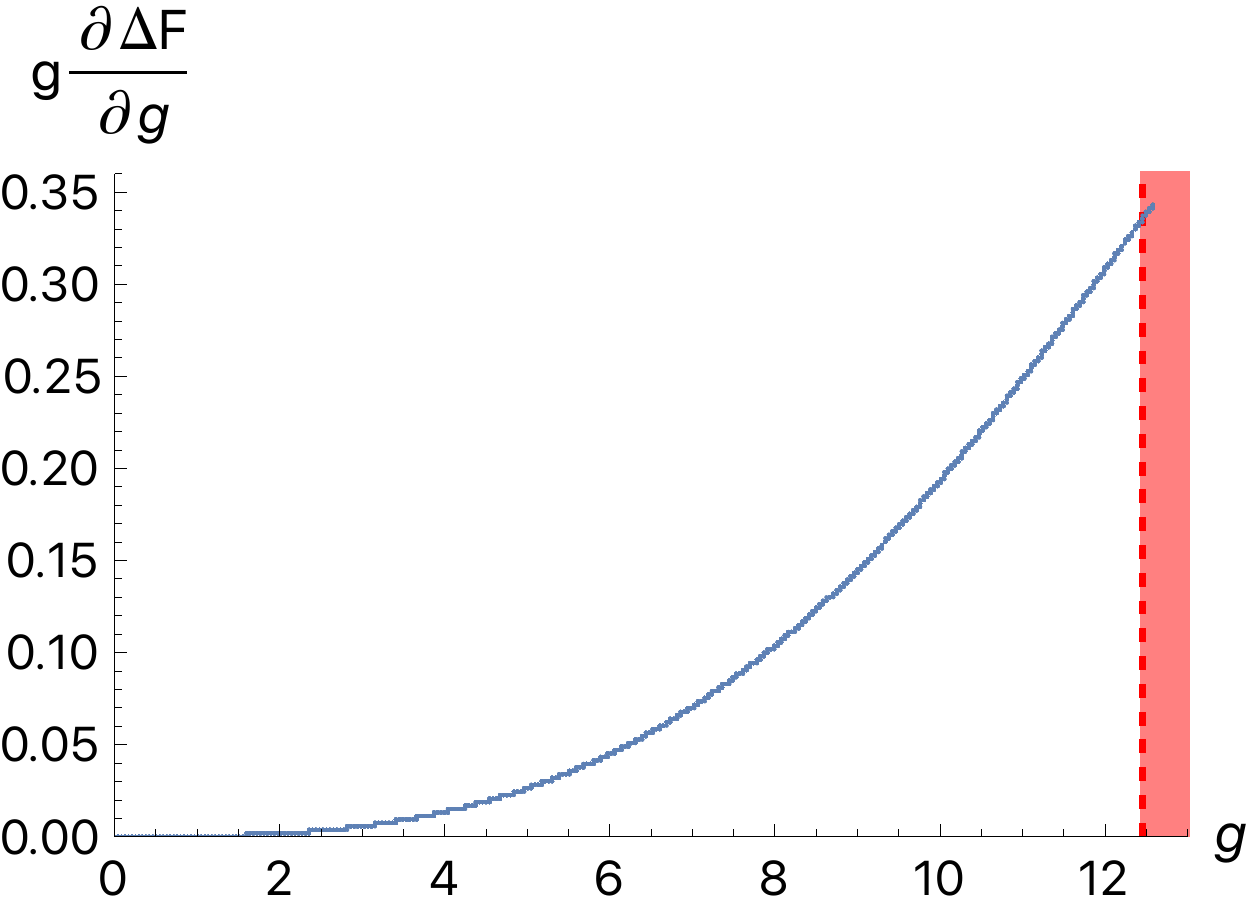}
\caption{Difference between the free energy in the UV and the free energy in the IR at $d=3$.  The red area corresponds to $g>g_c$.}
\label{fig:residue}
\end{figure}

With this in mind, it is clear that the difference between the free energy of the UV theory and the one of the IR theory is given precisely by the isolated non-normalizable contribution of the former.
Going again through the same regularization procedure as in the IR case, we thus find:
\begin{equation} \label{eq:dF_tm}
\begin{split}
g\f{\p}{\p g}\left(F_{\rm NNLO}^{UV}-F_{\rm NNLO}^{IR}\right)
&=  N^2\, {\rm Res}\left[ \r(h,0) \,\f{k(h,0)^2}{1-k(h,0)}\, \cN^{\D}_{h,0}  \cN^{\wtD}_{\htilde,0} \, \cI_0 (0) \right]_{h= h_{-}} \\
& =16\frac{\Gamma(-d/2) |g|^3}{2^{3d}\pi^{3d/2}\Gamma(d)} N^2 + \mathcal{O}(|g|^5)\,,
\end{split}
\end{equation}
which is positive for $2<d<4$. By numerical evaluation at finite $g$, it can be checked that the positivity remains valid also at all values of $g$, within the radius of convergence of the melonic series (see Fig.~\ref{fig:residue}).

\begin{figure}[htbp]
\centering
\includegraphics[scale=1]{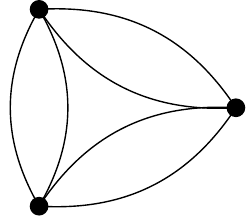}
\caption{Feynman graph contributing to the free energy at order $3$ in the coupling constant. The vertices are either two tetrahedron and one $g_1$ or three $g_1$.}
\label{fig:triangle}
\end{figure}

This result can also be checked perturbatively. At the fixed points the critical coupling $g_1$ is \cite{Benedetti:2019eyl,Benedetti:2020sye}:
\begin{equation}
g_{1\pm}=\pm \sqrt{g^2}\big( 1+ \mathcal{O}(g^2)\big) + g^2 \big(\psi(1)+\psi(d/2)-2\psi(d/4)  + \mathcal{O}(g^2)\big)   \,,
\end{equation}
where  $\psi(z)$ is the digamma function. When flowing from the UV to the IR, the fixed point value goes from $g_{1-}\simeq -\sqrt{g^2}$ to $g_{1+}\simeq \sqrt{g^2}$  (except for the vertical trajectories in  Fig.~\ref{fig:trajectory}, which we will discuss below): therefore, at leading order in $g$, graphs with an even number of vertices have the same amplitude in the UV as in the IR, while graphs with an odd number of vertices have opposite signs. The difference between the free energy in the UV and the free energy in the IR is thus expanded in odd powers of $|g|$. Up to order $|g|^3$, only the graph of Fig.~\ref{fig:triangle} contributes, where the vertices are either two tetrahedron and one $g_1$ or three $g_1$.

Using the expression of the kernel given in \eqref{eq:full-K} to fix the combinatorial factors, we have:
\begin{equation}
 g\f{\p}{\p g}(F_{\rm NNLO}^{UV}- F_{\rm NNLO}^{IR}) =2 |g|^3 N^2 \mathcal{A}  + \cO(|g|^5)\;,
\end{equation}
where we denoted $\mathcal{A}$ the amplitude of the graph in Fig.~\ref{fig:triangle}:
\begin{equation}
\mathcal{A}=c(\tfrac{d}{4})^6 \int d^dx d^dy d^dz \frac{\left( \Omega(x)\Omega(y) \Omega(z)\right)^d}{s(x,y)^{4\Delta}s(x,z)^{4\Delta} s(y,z)^{4\Delta}} \;.
\end{equation}

Using again dimensional regularization with $\Delta=\frac{d-\epsilon}{4}$, we notice that this amplitude is proportional to $\cI_\eps (0)$ in \eqref{eq:I0e}. We then obtain:
\begin{equation}
\mathcal{A}=8\frac{\Gamma(-d/2) }{2^{3d}\pi^{3d/2}\Gamma(d)} \quad \Rightarrow \quad
g\f{\p}{\p g}\left(F_{\rm NNLO}^{UV}-F_{\rm NNLO}^{IR}\right)=16\frac{\Gamma(-d/2) |g|^3}{2^{3d}\pi^{3d/2}\Gamma(d)} N^2+ \cO(|g|^5)\;,
\end{equation} 
in agreement with \eqref{eq:dF_tm}.

In conclusion, the difference between the sphere free energy at the fixed points $(g_{1-},g_{2-})$ or $(g_{1-},g_{2+})$ and the one at the fixed points $(g_{1+},g_{2-})$ or $(g_{1+},g_{2+})$ (see Fig.~\ref{fig:trajectory})\footnote{Notice that $(g_{1+},g_{2-})$ can be reached only through one trajectory emanating from $(g_{1-},g_{2-})$, while $(g_{1-},g_{2+})$ has only one trajectory to flow to an IR fixed point, i.e.\ $(g_{1+},g_{2+})$. Between $(g_{1-},g_{2-})$ and $(g_{1+},g_{2+})$ there is instead an infinite set of trajectories.} grows with growing $|g|$. Since the difference vanishes at $g=0$, we conclude that for the fixed points at fixed $g\neq 0$ the sphere free energy satisfies $F_{\rm NNLO}^{UV}>F_{\rm NNLO}^{IR}$, in accordance with the $F$-theorem.

\paragraph{Trajectories at fixed $g_1$.}
The reader will note in Fig.~\ref{fig:trajectory} the presence of two vertical lines: these are two trajectories at fixed $g_1$ connecting two different pairs of fixed points. As neither the tetrahedral coupling nor $g_1$ change along these trajectories, the above computation implies that at NNLO the free energy at the two ends of such trajectories is the same. We expect that in order to see a change of the free energy along these trajectories one would need to push the evaluation to lower orders in $1/N$.

The first non-trivial contribution involving the double trace coupling $g_2$, which does vary along the vertical trajectories, comes from ladder graphs generated by the double trace part $K_2 \hat P^{(2)}$ of the Bethe-Salpeter kernel \eqref{eq:full-K}. They essentially behave the same as the ladders generated by $K_1 \hat P^{(1)}$, up to replacing $g^2$ by $3g^2$, but the isolated contribution to the conformal partial wave expansion in the UV would in this case depend on the fixed-point value of $g_2$. 

However, it should be noted that, as such ladders appear only at order $N^0$, one should include the effect of the $1/N$ corrections to the on-shell two-point function as well as to the fixed-point values of the couplings. We know from \cite{Benedetti:2020sye} that such corrections have a drastic effect on the fixed-point theory: in order to find a non-trivial precursor of the large-$N$ theory, one needs to keep a finite $\eps$, with $\eps N\ll 1$; then, the lines of fixed points collapse to isolated points and the scaling dimensions acquire an imaginary part.
The whole analysis becomes much more involved, and since in this case the theory is manifestly non-unitary we do not expect the $F$-theorem to necessarily hold.


\section*{Acknowledgements}

We thank Igor Klebanov for useful comments.
R.G., S.H. and D.L. are supported by the European Research Council (ERC) under the European Union's Horizon 2020 research and innovation program (grant agreement No818066) and by Deutsche Forschungsgemeinschaft (DFG, German Research Foundation) under Germany's Excellence Strategy  EXC-2181/1 - 390900948 (the Heidelberg STRUCTURES Cluster of Excellence).

\appendix

\section{Useful formulas on $S^d$}
\label{app:useful}

%
%

We collect here some useful formulas about $S^d$ and the spectrum of the Laplace-Beltrami operator on it.

The $d$-dimensional round sphere $S^d$ can be defined by the equation
\be
\sum_{\bar{\m}=1}^{d+1} (\bar{X}^{\bar{\m}})^2 \equiv \sum_{\m=1}^{d} (X^\m)^2 + Z^2= a^2\, ,
\ee
where $\bar{X}^{\bar{\m}}=\{X^\m,Z\}$ are the Cartesian coordinates in the embedding space $\mathbb{R}^{d+1}$, and $a$ is the radius of the sphere.
In the northern and southern hemispheres, the equation can be solved as $Z_\pm(X)= \pm \sqrt{a^2-X^2}$, respectively, with $X^2=\sum_{\m=1}^{d} (X^\m)^2$.

It is convenient to describe the metric on $S^d$ through the (equatorial) stereographic projection to the $d$-dimensional flat space $\mathbb{R}^d$.
The stereographic projection from the North pole to the equatorial plane is obtained by the change of variables
\be \label{eq:x-stereo}
x^{\m}_\pm(X) = \f{X^\m}{1- Z_\pm(X)/a} \,.
\ee
Introducing polar coordinates on the equatorial plane, $\{r=\sqrt{x^2},\theta_1,\ldots,\theta_{d-1}\}$, and denoting $\theta_d$ the additional angular coordinate on $S^d$ (i.e.\ the geodesic distance from the North pole), the stereographic mapping reads simply
\be
r(\theta_d) = a \cot(\theta_d/2)\,.
\ee

In stereographic coordinates, the line element takes the following form:
\begin{equation}
ds^2=\frac{4a^2}{\left(1+x^2\right)^2}\sum_{i=1}^d \, dx_i{}^2\; , \qquad x^2\equiv \sum_{i=1}^d x_i{}^2 \,.
\end{equation}
The metric is conformally flat, i.e.
\be \label{eq:Weyl_g}
g_{\m\n}(x) = \Om(x)^2 \d_{\m\n} \,,
\ee
with
\be \label{eq:Omega}
\Om(x) = \frac{2a}{(1+x^2)} \,.
\ee
The square root of the determinant of the metric is then given $\sqrt{g}=\Om(x)^d$, 
and the Ricci scalar is $R=d(d-1)/a^2$.
The volume of the $d$-sphere is
\be \label{eq:Vol-Sd}
V_d = \int_{S^d} \dd x\, \sqrt{g(x)} =  \f{ 2 \pi^{\f{d+1}{2}}}{\G\left(\f{d+1}{2}\right)} a^d\, .
\ee

\

The eigenmodes of the scalar Laplacian on the sphere are the spherical harmonics (e.g. \cite{Rubin:1984,Samko-book})
\be \label{eq:spherHarmonics}
\Psi_{n,j} (\bar{X}) = \r^{-n}\, T^{(j)}_{\bar{\m}_1 \cdots \bar{\m}_n} \bar{X}^{\bar{\m}_1} \cdots \bar{X}^{\bar{\m}_n} \,,
\ee
where $n=0,1,2,...+\infty$, $\r=(\bar{X}^{\bar{\m}}\bar{X}_{\bar{\m}})^{1/2}$ and $T^{(j)}_{\bar{\m}_1 \cdots \bar{\m}_n}$ form a basis of constant traceless-symmetric tensors on $\mathbb{R}^{d+1}$, each basis element being labeled by a different value of $j$. Therefore, we take $j=1,2,...D_n$, with 
\be \label{eq:multi}
D_n=\frac{(n+d-2)!\, (2n+d-1)}{n!(d-1)!}\,.
\ee
The corresponding eigenvalues are
\be
\om_n=n(n+d-1)/a^2 \,.
\ee
They are independent of $j$, hence they have multiplicity $D_n$.

The addition theorem of spherical harmonics states that
\be \label{eq:additionTh}
\sum_{j=1}^{D_n}  \Psi_{n,j}(x) \Psi^*_{n,j}(y) = \f{D_n}{V_d}    P_n(\bar{X}\cdot \bar{Y}) \,,
\ee
where
\be
P_n(z) = \begin{cases} \f{n! (d-2)!}{(n+d-2)!} C_n^{(d-1)/2}(z)\,, &\; \text{if } d> 2\,,\\
T_n(z)\,, &\; \text{if } d= 2\,, \end{cases}
\ee
and $C_n^{\a}(z)$ and $T_n(z)$ are the Gegenbauer and Chebyshev polynomials, respectively.

\section{CFTs on $S^d$}
\label{app:sphereCFT}

Given a CFT on $\mathbb{R}^d$, and assuming that conformal invariance can be promoted to Weyl invariance\footnote{See \cite{Farnsworth:2017tbz} and references therein for the relation between Weyl and conformal invariance.} (possibly up to an anomaly), we can then define a corresponding CFT on $S^d$ by performing the Weyl transformation \eqref{eq:Weyl_g}, 
together with the transformation of primary fields
\be \label{eq:Weyl_Op}
\cO(x) \to \Om(x)^{-\D_{\cO}} \cO(x) \,,
\ee
such that $n$-point functions are obtained as
\be
\la \cO_1(x_1) \cdots \cO_n(x_n) \ra_{S^d} = \Om(x)^{-\D_1} \cdots \Om(x)^{-\D_n} \la \cO_1(x_1) \cdots \cO_n(x_n) \ra_{\mathbb{R}^d} \,.
\ee
In practice, the latter amounts to replacing the flat-space distances $|x-y|$ appearing in the conformal correlators with the chordal distance
\begin{equation}
s(x,y)=2a \frac{|x-y|}{(1+x^2)^{1/2}(1+y^2)^{1/2}} = |x-y| \Om(x)^{1/2}  \Om(y)^{1/2} \,.
\end{equation}
Notice that this is not the geodesic distance $\s(x,y)$ on $S^d$, but the Euclidean distance in the embedding space $\mathbb{R}^{d+1}$, i.e.\ $s(x(p),x(p'))=|X(p)-X(p')|$ where $X(p)$ is the embedding map $X: S^d\to \mathbb{R}^{d+1}$ while $x(p)$ is the stereographic map $x: S^d\to \mathbb{R}^{d}$. Of course, the two are trivially related by a trigonometric relation: $s(x,y)=2 a \sin(\s(x,y)/2a)$.

We can check that for a usual free scalar, the propagator on $S^d$ matches the flat one with the flat distance replaced by the chordal distance, if we  appropriately tune the non-minimal coupling with the curved background of the sphere.
The covariance, or propagator, $C_1(x,y;b)$ is the solution of the equation
\be \label{eq:prop-def}
(-\nabla_x^2 + b) C_1(x,y;b) = \f{\d(x-y)}{\sqrt{g}} \, ,
\ee
where $b$ is a constant (of squared-mass dimension),  $\nabla^2= \nabla^\m \nabla_\m$ is the covariant Laplacian on the $d$-sphere, and we specified by a subscript the coordinate on which derivatives as in this case there could be an ambiguity.
Due to homogeneity of space the propagator depends only on the geodesic distance $\s(x,y)$, hence we will also write $C_1(\s;b)$ for the propagator.
The reason for the subscript 1 in the latter is that this corresponds to the case $\z=1$ of the more general propagator we will consider in the following subsection.

The propagator on $S^d$ has been computed in \cite{Allen:1985wd} directly solving \eqref{eq:prop-def}, or from an explicit mode sum in \cite{Dowker:1975tf}. Defining
\be
\g_\pm = \f{d-1}{2} \pm \sqrt{\f{(d-1)^2}{4}-a^2 b}\, ,
\ee
the propagator is given by
\be
C_1(\s;b) = a^{2-d} \f{ \G(\g_+) \G(\g_-)}{\G(d/2)\, 2^d\, \pi^{d/2}}\, {}_{2}F_1(\g_+,\g_-;d/2; \cos(\s/2a)^2) \, ,
\ee
where  ${}_{2}F_1(\a,\b;\g;z)$ is the hypergeometric function.

The case of a Weyl invariant free scalar field is obtained with the choice 
\be \label{eq:b_W}
b= b_W\equiv  \f{d-2}{4(d-1)} R = \f{d(d-2)}{4a^2}\,,
\ee 
for which $\g_+=d/2$ and $\g_-=(d-2)/2$. In this case, the hypergeometric function reduces to a simple power and we find:
\be \label{eq:C_1}
C_1(x,y;b_W) = a^{2-d} \f{  \G(d/2-1)}{ 2^d\, \pi^{d/2}}\, (\sin(\s(x,y)/2a) )^{2-d} =  \f{  \G(d/2-1)}{ 4\, \pi^{d/2}}\, s(x,y)^{2-d}\, ,
\ee
which is precisely the massless free scalar propagator of flat space with the replacement $|x-y|\to s(x,y)$.

\subsection{Generalized free field theory}
\label{app:GFFT}

We now consider the case of a conformal generalized free field  theory (GFFT), i.e. a long-range massless Gaussian theory, sometimes called a mean field theory. It is worth discussing it in some detail because it is the simplest case of (non-local) CFT, and because typical long-range models can be defined as perturbations of a GFFT.
By definition this is a CFT whose only non-vanishing connected $n$-point function is the two-point function, which however has a scaling exponent $\D\neq d/2-1$, and which moreover we take to be in the range $\D\in(0,d/2)$.
On flat space, the two-point function is 
\begin{equation} \label{eq:freeC-flat}
C_{\rm flat}(x,y)=\frac{c(\Delta)}{|x-y|^{2\Delta}} \,, \qquad c(\Delta)=\frac{\Gamma(\Delta)}{2^{d-2\Delta}\pi^{d/2}\Gamma(\frac{d}{2}-\Delta)} \,.
\end{equation}
Writing $\D=d/2-\z$, such GFFT can be obtained from a functional integral with the action\footnote{We typically assume $0<\z<1$. The restriction to $\z<1$ is imposed to preserve reflection positivity (unitarity in Lorentzian signature), but also because $\z>1$ would correspond to a strong short-range rather than long-range action, and moreover the operator with $\z=1$ would in that case be a relevant perturbation. The restriction to $\z>0$ is instead chosen to avoid a strong long-range action, with its associated unusual thermodynamic features  \cite{Campa:2009rev}.}
\be \label{eq:flatGFFT}
S_{\rm GFFT}[\phi] = \f12 \int \dd x  \, \phi(x) (-\p^2)^{\z} \phi(x) \,,
\ee
where the fractional power of the Laplacian can be defined in many equivalent ways \cite{Kwasnicki:2017}, among which in particular as the inverse of the ``Riesz potential'' \eqref{eq:freeC-flat}.
The easiest definition is of course in Fourier space, where $(-\p^2)^{\z}$ is defined as the multiplication operator $p^{2\z}$, which is the inverse of the Fourier transform of \eqref{eq:freeC-flat}. 
Going to position space one finds instead a representation as a hypersingular integral operator:\footnote{This can be derived by first writing
\be \nn
p^{2\z} = \f{1}{\G(-\z)} \int_0^{+\infty} {\rm d} t \f{e^{-t p^2}-1}{t^{1+\z}} \,,
\ee
whose validity is trivially checked by rescaling $t\to t/p^2$ and recognizing that the integral reduces to $p^{2\z}$ times the Cauchy-Saalsch\"utz representation of $\G(-\z)$ for $0<\z<1$.
The singular integral representation is then found by going back to position space and exchanging the order of integration  \cite{Stinga:2009}.
}
\be \label{eq:fracLapl_flat}
(-\p^2)^{\z} \phi(x) = \lim_{r\to 0} \,  c(d-\Delta) \int_{|x-y|>r} \dd y \,\f{\phi(y)-\phi(x)}{|x-y|^{2(d-\Delta)}} 
\,.
\ee
In the physics literature such representation is often expressed as
\be  \label{eq:fracLapl_flat_phys}
(-\p^2)^{\z} \phi(x) = \int \dd y \, C^{-1}_{\rm flat}(x,y) \phi(y) \,,
\ee
with  convolution kernel
\be \label{eq:flatCinv}
C^{-1}_{\rm flat}(x,y) = \frac{c(d-\Delta)}{|x-y|^{2(d-\Delta)}} \,,
\ee
without any subtraction term. For $\z>0$ the convolution is a formal divergent expression (a ``hypersingular'' integral), which is to be interpreted through analytic continuation from $\z<0$. For simplicity we will stick to this point of view.

In order to place the GFFT on the $d$-sphere, we can apply again the Weyl mapping to \eqref{eq:freeC-flat}, and thus write
\begin{equation} \label{eq:freeC}
C(x,y)= \Om(x)^{-\D}\Om(y)^{-\D} C_{\rm flat}(x,y) =\frac{c(\Delta)}{s(x,y)^{2\Delta}} \,. 
\end{equation}
Constructing an action associated to such propagator requires as usual identifying the inverse propagator, and from this the type of non-minimal coupling to the background geometry that is needed in order to obtain a conformal theory.

The covariance $C(x,y)$ in \eqref{eq:freeC} is also known as the Riesz potential, and it can be written
\be \label{C-harmonics}
C(x,y) = \sum_{n\geq 0} \sum_{j=1}^{D_n} \f{1}{\om^{(\z)}_n}  \Psi_{n,j}(x) \Psi^*_{n,j}(y) = \f{1}{V_d}  \sum_{n\geq 0} \f{D_n}{\om^{(\z)}_n} P_n(\bar{X}\cdot \bar{Y})\,,
\ee
where we used the addition theorem \eqref{eq:additionTh}.

The inverse of \eqref{eq:freeC} is defined by the equation
\be
\int \dd z \sqrt{g(z)} \, C^{-1}(x,z) \int \dd y \sqrt{g(y)} \, C(z,y) \phi(y) = \phi(x)\,,
\ee
or
\be \label{eq:defCinv}
\int \dd z \sqrt{g(z)} \, C^{-1}(x,z) C(z,y) = \f{1}{\sqrt{g}} \d(x-y) \, .
\ee
Given that on flat space \eqref{eq:flatCinv} is the inverse of \eqref{eq:freeC-flat}, it is easily seen that the above equations are solved by
\be \label{eq:sphereCinv}
\begin{split}
C^{-1}(x,y) &=  \Om(x)^{\D-d} \, \Om(y)^{\D-d} \, C_{\rm flat}^{-1}(x,y) 
= \frac{c(d-\Delta)}{s(x,y)^{2(d-\Delta)}}\,,
\end{split}
\ee
whose convolution should again be interpreted by analytic continuation.
This is of the expected form we would obtain by the Weyl mapping applied to $C_{\rm flat}^{-1}(x,y)$, formally viewed as  the two-point function of the shadow operators \cite{Ferrara:1972uq} of dimension $\widetilde{\D}=d-\D$.
It also means that defining, for  $\z=d/2-\D$, the operator whose kernel is \eqref{eq:sphereCinv} as\footnote{\label{foot:Dz-subtr}As a subtracted hypersingular integral, a rigorous covariant expression is given by (see \cite{Samko:2003,Samko-book}):
\be \nn
\cD_\z \phi (x) = \lim_{r\to 0} c(d-\Delta) \int_{s(x,y)>r} \dd y \sqrt{g(y)}\,\f{\phi(y)-\phi(x)}{s(x,y)^{2(d-\Delta)}} +\f{\G(d-\D)}{\G(\D)} \phi(x) \,.
\ee
}
\be
\cD_\z \phi (x) =  \int \dd y \, \sqrt{g(y)}\, C^{-1}(x,y) \phi(y) \,,
\ee
we find that 
\be
\cD_\z \phi (x) = \Om(x)^{\D-d} (-\p^2)^{\z} \left( \Om(x)^{\D} \phi(x) \right) \,.
\ee 
Given the Weyl transformations \eqref{eq:Weyl_g}, \eqref{eq:Weyl_Op} relating the flat space to the sphere, one recognizes in such a relation the definition of conformally covariant operator of order $\z=d/2-\D$, or conformal biweight $(\D,d-\D)$ \cite{Branson:1997,gonzalez2016recent}.

Therefore, the action replacing \eqref{eq:flatGFFT} on the sphere is
\be
S_{\rm GFFT}[\phi] = \f12 \int \dd x  \, \sqrt{g(x)}\\, \phi(x) \cD_\z \phi(x) \,.
\ee
However, calling $\cD_\z$ a conformal ``fractional Laplacian'' would be deceiving, as it turns out that the operator $\cD_\z$ is not of the form $(-\nabla^2 +b)^{\z}$:
the conformal Laplacian of biweight $(\D,d-\D)$ on the $d$-sphere can be related to the Laplace-Beltrami operator by the expression  \cite{Branson:1995}
\be \label{eq:D_z}
\cD_\z = a^{-2\z} \,\f{ \G(a \cD_{1/2} +\f12 +\z) }{ \G(a \cD_{1/2} +\f12 -\z) } \,, \qquad \cD_{1/2} 
= a \, \sqrt{ -\nabla^2 + \left( \f{d-1}{2a}\right)^2 } \,,
\ee
which should of course be interpreted in terms of the eigenvalues
\be
\begin{split}
\om^{(\z)}_n = a^{-2\z} \,\f{ \G(n +\f{d}{2} +\z) }{ \G(n +\f{d}{2}-\z) } =  a^{-2\z} \,\f{ \G(a\, \om^{(1/2)}_n +\f12 +\z) }{ \G(a\, \om^{(1/2)}_n +\f12 -\z) } \,,\\
 \om^{(1/2)}_n = a^{-1}\, \left(n+\f{d-1}{2}\right) =   \sqrt{ \om_n + \left( \f{d-1}{2a}\right)^2 }
=  \sqrt{ \om^{(1)}_n +  \f{1}{4a^2} } \,.
\end{split}
\ee
In the last equality we introduced $\om_n = n(n+d-1)/a^2$, the eigenvalues of Laplace-Beltrami operator on the sphere. 
Notice that for $\z=1$ we have $\cD_1=-\nabla^2+b_W$, as expected, and that for $n\to+\infty$ the eigenvalues of $\cD_\z$ do asymptotically approach $n^{2\z}$, as for a Laplacian to  the power $\z$.
The eigenvalues $\om^{(\z)}_n$ are of course the inverse of the eigenvalues of the Riesz potential \eqref{eq:freeC}, which were known since long to mathematicians (e.g. \cite{PavSam84}), and have been later rederived also in the physics literature \cite{Gubser:2002vv}.

\

Notice that, denoting $c_n\equiv n +\f{d}{2}-\z$, we can write
\be
\f{1}{a^{2\z} \om^{(\z)}_n} = \f{\G(c_n)}{\G(c_n+2\z)} = \f{1}{\G(2\z)} B(c_n,2\z) \,,
\ee
where $B(x,y)$ is the Euler beta function. Therefore, we have various useful representations, among which in particular the following integral representation:
\be \label{eq:EulerBeta}
\f{1}{a^{2\z} \om^{(\z)}_n} = \f{1}{\G(2\z)} \int_0^1 dt\, t^{c_n-1} (1-t)^{2\z-1} = \f{1}{\G(2\z)} \int_0^{+\infty} ds\, e^{-s\, c_n}  (1-e^{-s})^{2\z-1}\,.
\ee
One way to introduce a UV cutoff in the theory is then to replace the beta function with the incomplete beta function, i.e.\ truncating the upper end of $t$-integration at $1-e^{-s_0}$, or the lower end of the $s$-integration at $s_0>0$. From the latter one can see that such cutoff is roughly proportional to an exponential $e^{-s_0 c_n}$.
This should be compared to the flat space representation 
\be \label{eq:EulerGamma}
\f{1}{p^{2\z}} = \f{1}{\G(2\z)} \int_0^{+\infty} ds\, e^{-s\, p}  s^{2\z-1} \,,
\ee
which again can be regularized by replacing the integral representation of the gamma function with that of the incomplete gamma function.
This is in the same spirit of what was done in \cite{Benedetti:2019eyl}, where however the representation of $\G(z)$ was used instead of $\G(2\z)$, i.e. the exponential cutoff was with respect to $p^2$ rather than $p$.
Notice that as expected the two $s$-integral representations in \eqref{eq:EulerBeta} and \eqref{eq:EulerGamma} coincide in the deep UV (small $s$) but differ in the IR (large $s$).

\section{Computation of the free energy for GFFT}
\label{app:F_GFFT}

In this appendix we present a detailed computation of the following sum:
\begin{equation}
F=\frac{1}{2}\sum_{n=0}^{\infty} D_n \ln \left(a^{-2\zeta}\frac{\Gamma(n+d/2+\zeta)}{\Gamma(n+d/2-\zeta)}\right)
\label{eq:free_GFFT}.
\end{equation}
This computation was done in \cite{Diaz:2007an}, but we reproduce it here with more details.

For $d>0$, this sum is divergent. We will compute it in the regime $2\zeta-2<d<0$ and then perform an analytic continuation to deduce the result for $d>0$. This computation can also be done for $\zeta>1$. For $k-1<\zeta <k \; ,  k\geq 2$, the sum \eqref{eq:free_GFFT} would have to be computed in the range $2\zeta-2k<d<0$. However, the computation is very similar than the one for $0<\zeta<1$ and leads to the same result so we will detail here only the computation for $0<\zeta<1$. 

Let us first show that the sum of multiplicity is zero in this regularization. 

\begin{equation} \label{eq:sum_multiplicity}
	\begin{split}
\sum_{n=0}^{\infty} D_n&= \sum_{n=0}^{\infty} \frac{(n+d-2)!(2n+d-1)}{n!(d-1)!}=\sum_{n=0}^{\infty} \frac{(n+d-1)!}{n!(d-1)!}\frac{2n+d-1}{n+d-1} \\
& =\sum_{n=0}^{\infty}  \frac{(n+d-1)!}{n!(d-1)!}\frac{n}{n+d-1} +  \sum_{n=0}^{\infty} \frac{(n+d-1)!}{n!(d-1)!} \\
& =\sum_{n=1}^{\infty}  \frac{(n+d-2)!}{(n-1)!(d-1)!} +(1-1)^{-d} = \sum_{n=0}^{\infty}  \frac{(n+d-1)!}{(n)!(d-1)!}+0 = (1-1)^{-d}=0 \; . 
 	\end{split}
\end{equation}

The term $\ln(a^{-2\zeta})$ can thus be dropped from the expression of $F$. Taking the derivative with respect to $\zeta$ of the remaining expression, we obtain:

\begin{equation}
\frac{dF}{d\zeta}=\frac{1}{2}\sum_{n=0}^{\infty} D_n\left( \psi(n+d/2+\zeta)+\psi(n+d/2-\zeta)\right) \, .
\end{equation}

We will now use the following integral representation of the digamma function:
\begin{equation}
\psi(z)=\int_0^{\infty} dt \left(\frac{e^{-t}}{t}-\frac{e^{-tz}}{1-e^{-t}}\right) \, ,
\end{equation}
which is valid for $z>0$. 

Leaving out the $n=0$ term \footnote{For the case $1<\zeta<2$, we would also need to leave out the term $n=1$.}, we then get:
\begin{equation}
	\begin{split}
\frac{dF}{d\zeta}&=\frac{1}{2}\left(\psi(d/2+\zeta)+\psi(d/2-\zeta)\right)+\frac{1}{2}\sum_{n=1}^{\infty} D_n\int_0^{\infty} dt \left(\frac{2e^{-t}}{t}-\frac{e^{-t(n+d/2)}}{1-e^{-t}}\left(e^{-t\zeta}+e^{t\zeta}\right)\right) \\
&=\frac{1}{2}\left(\psi(d/2+\zeta+1)+\psi(d/2-\zeta+1)-\frac{1}{d/2+\zeta}-\frac{1}{d/2-\zeta}\right)\\
& \qquad +\frac{1}{2}\sum_{n=1}^{\infty} D_n\int_0^{\infty} dt \left(\frac{2e^{-t}}{t}-\frac{e^{-t(n+d/2)}}{1-e^{-t}}\left(e^{-t\zeta}+e^{t\zeta}\right)\right) \, ,
	\end{split}
\end{equation}
where we have used $\psi(z)=\psi(1+z)-\frac{1}{z}$. 

We can now use the integral representation of the digamma function for $\psi(n+d/2 \pm \zeta +1)$. Rearranging the terms and exchanging sum and integral, we obtain:
\begin{equation}
\begin{split}
\frac{dF}{d\zeta}=& -\frac{1}{2}\left(\frac{1}{d/2+\zeta}+\frac{1}{d/2-\zeta}\right)+\int_0^{\infty} dt \frac{e^{-t}}{t}\sum_{n=0}^{\infty} D_n \crcr
& \qquad -\frac{1}{2} \int_0^{\infty} \frac{e^{-t\zeta}+e^{t\zeta}}{1-e^{-t}}e^{-td/2}\left(e^{-t}+\sum_{n=1}^{\infty}D_n e^{-tn}\right) \ .
\end{split}
\end{equation}
Again, as the sum of the multiplicities is zero, the second term vanishes. The remaining sum is:
\begin{align}
	\begin{split}
\sum_{n=1}^{\infty}D_n e^{-tn}&= \sum_{n=1}^{\infty} \frac{(n+d-2)!(2n+d-1)}{n!(d-1)!}e^{-tn} \\
&= 2 \sum_{n=1}^{\infty} n\frac{(n+d-2)!}{n!(d-1)!}e^{-tn} + \sum_{n=1}^{\infty} \frac{(n+d-2)!}{n!(d-2)!}e^{-tn} \\
&= 2e^{-t}  \sum_{n=0}^{\infty} \frac{(n+d-1)!}{n!(d-1)!}e^{-tn} +(1-e^{-t})^{-(d-1)}-1 \\
&= 2e^{-t} (1-e^{-t})^{-d}+(1-e^{-t})^{-(d-1)}-1  = (1-e^{-t})^{-d}(1+e^{-t})-1 \, . 
	\end{split}
\end{align}
Substituting this result into $\frac{dF}{d\zeta}$ and changing variables $u=e^{-t}$, we obtain:
\begin{equation}
\frac{dF}{d\zeta}=-\frac{1}{2}\left(\frac{1}{d/2+\zeta}+\frac{1}{d/2-\zeta}\right)-\frac{1}{2}\int_0^1 du\, u^{d/2-1}(u^{\zeta}+u^{-\zeta})\left((1-u)^{-d-1}(1+u)-1\right) \ .
\end{equation}

We can now compute the last integral using the regular as well as the subtracted integral representations of the beta function \footnote{In the case $1<\zeta<2$, we also need the following subtracted integral representation of the beta function:
\begin{equation*}
B(a,b)-B(a,1)+(b-1)B(a+1,1)=\int_0^1 dt \, t^{a-1}\left((1-t)^{b-1}+(b-1)t-1\right) \; , \; a>-2 \, , \; b>0 \, .
\end{equation*}}
\begin{equation}
	\begin{split}
B(a,b)&= \int_0^1 dt\, t^{a-1}(1-t)^{b-1} \; , \; a,b>0 \\
B(a,b)-B(a,c)&= \int_0^1 dt \, (1-t)^{a-1}\left(t^{b-1}-t^{c-1}\right) \; , \; a>-1 \, , \; b,c>0 \, .
	\end{split} 
\label{eq:Beta_int}
\end{equation}

Indeed, we have:
\begin{equation}
	\begin{split}
&\int_0^1 du\, u^{d/2+\zeta-1}\left((1-u)^{-d-1}(1+u)-1\right) \\
&=\int_0^1 du \left(u^{d/2+\zeta}(1-u)^{-d-1} +u^{d/2+\zeta-1}\left((1-u)^{-d-1}-(1-u)^{1-1}\right) \right) \\
&= B(d/2+\zeta+1,-d)+B(d/2+\zeta,-d)-B(d/2+\zeta,1) = 2\zeta\frac{\Gamma(d/2+\zeta)\Gamma(-d)}{\Gamma(\zeta-d/2+1)}-\frac{1}{\zeta+d/2} \,,
	\end{split}
\end{equation}
and similarly:
\begin{align}
\int_0^1 du\, u^{d/2-\zeta-1}\left((1-u)^{-d-1}(1+u)-1\right)&= -2\zeta\frac{\Gamma(d/2-\zeta)\Gamma(-d)}{\Gamma(-\zeta-d/2+1)}-\frac{1}{-\zeta+d/2} \, .
\end{align}

Thus we obtain:
\begin{equation}
\frac{dF}{d\zeta}=\zeta \Gamma(-d)\left(\frac{\Gamma(d/2-\zeta)}{\Gamma(1-\zeta-d/2)}-\frac{\Gamma(d/2+\zeta)}{\Gamma(1+\zeta-d/2)}\right) =-\zeta\frac{\sin(\pi \zeta)}{\sin(\pi d/2)}\frac{\Gamma(d/2-\zeta)\Gamma(d/2+\zeta)}{\Gamma(1+d)}\, , 
\end{equation}
which can be analytically continued to $d>0$ not even, and is valid for any value of $\zeta$, except at the poles at $\z=d/2+k$.

\section{Computation of $C(x,x)$ in dimensional regularization}
\label{app:C_dim_reg}

From the expansion of $C(x,y)$ in spherical harmonics in \eqref{C-harmonics}, we find that at coinciding points we have
\be
C(x,x) = \f{1}{V_d} \, \sum_{n\geq 0} \f{D_n}{\om^{(\z)}_n}   \,,
\label{eq:covxx}
\ee
which of course is divergent and needs regularization. We employ here analytic continuation in the dimension $d$, treating separately the two cases $\z=1$ and $\z<1$.

\subsection{$\z=1$ case}
\label{app:tadpole-SR}

With $\zeta=1$, the expression of the covariance simplifies to:
\begin{equation}
C_1(x,x)=\frac{a^{2-d}}{\Gamma(d/2)(4\pi)^{d/2}}\sum_{n=0}^{\infty} \frac{4\,\Gamma(n+d-1)(2n+d-1)}{n!(2n+d)(2n+d-2)} \,.
\end{equation}

The Weyl invariant coupling  $b= b_W\equiv  \f{d(d-2)}{4a^2}$ in \eqref{eq:prop-def} provides an IR regularization by removing the zero mode, for $d\neq 2$.
However, the sum is divergent for $d>2$ and needs a regularization.
A convenient approach is to compute it for $0<d<2$, where it converges, and where we find $C(x,x)=0$, thanks to a cancellation between the $n=0$ contribution (negative because $b_W<0$ in this range of dimensions), and the rest of the series. 
We then analytically continue the result to $d>2$. 

Let us begin by rewriting the sum as
\begin{equation}
	\begin{split}
\sum_{n=0}^{\infty}&  \frac{\Gamma(n+d-1)(2n+d-1)}{n!(n+d/2)(n+d/2-1)}= \sum_{n=0}^{\infty} \frac{\Gamma(n+d-1)}{n!(n+d/2-1)}+\sum_{n=0}^{\infty}\frac{\Gamma(n+d-1)}{n!(n+d/2)} \\
&=\frac{\Gamma(d-2)}{d/2-1}\sum_{n=0}^{\infty}\frac{1}{n!}\frac{(d-1)_n(d/2-1)_n}{(d/2)_n}+\frac{2\Gamma(d-2)}{d}\sum_{n=0}^{\infty}\frac{1}{n!}\frac{(d-1)_n(d/2)_n}{(d/2+1)_n} \,,
	\end{split}
\label{eq:sumd2}
\end{equation}
where $(b)_n=b(b+1)\dots (b+n-1)=\frac{\Gamma(b+n)}{\Gamma(b)}$ and we have used $\frac{(b)_n}{(b+1)_n}=\frac{b}{b+n}$. 

We then recognize the hypergeometric function of argument $1$:
\begin{equation}
{}_2F_1(a,b,c,1)=\sum_{n=0}^1 \frac{(a)_n(b)_n}{n!(c)_n}=\frac{\Gamma(c)\Gamma(c-b-a)}{\Gamma(c-b)\Gamma(c-a)} \,,
\end{equation} 
which is valid for ${\rm Re}(b),{\rm Re}(c)>0$ and ${\rm Re}(c-a-b)>0$. 

In order to apply this formula, we thus need $d>0$ and $d<2$: in the first sum of \eqref{eq:sumd2} we have $a=d/2-1,b=d-1,c=d/2$ and in the second sum we have $a=d/2,b=d-1,c=d/2+1$. 

We thus get:
\begin{equation}
	\begin{split}
C_1(x,x)&\propto 2\Gamma(d-2)\left(\frac{\Gamma(d/2)\Gamma(2-d)}{(d-2)\Gamma(1-d/2)}+\frac{\Gamma(d/2+1)\Gamma(2-d)}{d\Gamma(2-d/2)}\right) \\
&=\frac{2\Gamma(d-2)\Gamma(d/2)\Gamma(2-d)}{\Gamma(1-d/2)}\left(\frac{1}{d-2}+\frac{d}{2d(1-d/2)}\right)=0 \; .
	\end{split}
\end{equation}

\subsection{$\z<1$ case}
\label{app:tadpole-LR}

We want to compute the following sum:
\begin{equation}
\frac{a^{2\zeta-d}(d-1)!}{\Gamma(d/2)(4\pi)^{d/2}}\sum_{n=0}^{\infty}\frac{D_n}{a^{2\zeta}\omega_n^{(\zeta)}}\,.
\end{equation}

We use the integral representation of \eqref{eq:EulerBeta}, which in the range $2\zeta-2<d<0$ is valid for $n>0$. Taking out the $n=0$ term, and exchanging the sum and the integral, we obtain:
\begin{equation}
C(x,x)=\frac{a^{2\zeta-d}(d-1)!}{\Gamma(d/2)(4\pi)^{d/2}}\Bigg[\frac{\Gamma(d/2-\zeta)}{\Gamma(d/2+\zeta)}+\frac{1}{\Gamma(2\zeta)}\int_0^{\infty}ds (1-e^{-s})^{2\zeta-1}e^{-s(\frac{d}{2}-\zeta)}\sum_{n=1}^{\infty}D_n e^{-s n}\Bigg]\,.
\end{equation}

The remaining sum was already computed in the previous appendix and doing the change of variable $u=e^{-s}$ we obtain:
\begin{equation}
\begin{split}
 C(x,x)&=\frac{a^{2\zeta-d}(d-1)!}{\Gamma(\frac{d}{2})(4\pi)^{d/2}}\Bigg[\frac{\Gamma(\frac{d}{2}-\zeta)}{\Gamma(\frac{d}{2} +\zeta)}+ \frac{1}{\Gamma(2\zeta)}\int_0^1 du\, u^{d/2-\zeta-1}(1-u)^{2\zeta-1}\left((1-u)^{-d}(1+u)-1\right)\Bigg] \\
&=\frac{a^{2\zeta-d}(d-1)!}{\Gamma(d/2)(4\pi)^{d/2}}\Bigg[\frac{\Gamma(d/2-\zeta)}{\Gamma(d/2+\zeta)}\\
& \qquad + \frac{1}{\Gamma(2\zeta)}\int_0^1 du\, \left(u^{d/2-\zeta}(1-u)^{2\zeta-d-1}+u^{d/2-\zeta-1}\left((1-u)^{2\zeta-d-1}-(1-u)^{2\zeta-1}\right)\right)\Bigg]\,.
\end{split}
\end{equation}

We perform the integral using the regular and subtracted representations of the Beta function \eqref{eq:Beta_int} and obtain:
\begin{equation}
	\begin{split}
C(x,x)&=\frac{a^{2\zeta-d}(d-1)!}{\Gamma(\frac{d}{2})(4\pi)^{d/2}}\Bigg[\frac{\Gamma(\frac{d}{2}-\zeta)}{\Gamma(\frac{d}{2}+\zeta)}+\frac{\Gamma(\frac{d}{2}-\zeta+1)\Gamma(2\zeta-d)}{\Gamma(-\frac{d}{2}+\zeta+1)\Gamma(2\zeta)}+\frac{\Gamma(\frac{d}{2}-\zeta)\Gamma(2\zeta-d)}{\Gamma(\zeta-\frac{d}{2})\Gamma(2\zeta)}-\frac{\Gamma(\frac{d}{2}-\zeta)}{\Gamma(\zeta+\frac{d}{2})}\Bigg] \\
&= \frac{a^{2\zeta-d}(d-1)!\Gamma(d/2-\zeta)\Gamma(2\zeta-d)}{\Gamma(d/2)(4\pi)^{d/2}\Gamma(2\zeta)\Gamma(\zeta-d/2)}\Bigg[\frac{d/2-\zeta}{\zeta-d/2}+1\Bigg] =0\,.
	\end{split}
\end{equation}

\section{Basics of conformal partial wave expansion}
\label{app:CPW}

We provide here some important formulas and background on the conformal partial wave expansion used in the main body of the paper. The main results of this appendix have been derived by Dobrev et al.\ in \cite{Dobrev:1976vr,Dobrev:1975ru,Dobrev:1977qv}, and largely revived in recent years \cite{Caron-Huot:2017vep,Simmons-Duffin:2017nub,Liu:2018jhs,Karateev:2018oml}.\footnote{These methods have been at the heart of a very active field in recent years, see for example their use with Mellin amplitudes  \cite{Mack:2009mi,Costa:2012cb}, their application to the Sachdev-Ye-Kitaev model \cite{Maldacena:2016hyu,Murugan:2017eto}, to the bootstrap crossing equations \cite{Gadde:2017sjg,Hogervorst:2017sfd,Sleight:2018ryu,Sleight:2018epi}, and to the construction of an AdS/CFT map \cite{deMelloKoch:2018ivk,Aharony:2020omh}.} 
Here we mostly follow the notation of \cite{Benedetti:2021qyk}, where a more detailed review can be found.

We work on flat space and comment at the end on the straightforward extension to the sphere. In the conformal limit, the Bethe-Salpeter kernel $K(x_1,x_2,x_3,x_4)$ of scalar fields of dimension $\D$ is diagonalized by functions with the structure of a conformal  three-point function of the type
\be \label{eq:3pt}
\la \phi_{\D}(x_3) \phi_{\D}(x_4) \cO_h^{\m_1 \cdots \m_J}(x_0) \ra_{\rm cs} = 
\f{ Z^{\m_1} \cdots Z^{\m_J} - \text{``traces"} }{ |x_{34}|^{2\D-h} |x_{30}|^{h} |x_{40}|^{h} } \,,   \;\;\;\; Z^{\m} = \f{|x_{30}||x_{40}|}{|x_{34}|} \left(\f{x_{30}^{\m}}{|x_{30}|^2} - \f{x_{40}^{\m}}{|x_{40}|^2}\right) \,,
\ee
where $x_{ij}=x_i-x_j$, such that
\be \label{eq:eigK}
\int_{x_3x_4} \, K(x_1,x_2,x_3,x_4) \, \la \phi_{\D}(x_3) \phi_{\D}(x_4) \cO_h^{\m_1 \cdots \m_J}(x_0) \ra_{\rm cs}
= k(h,J) \, \la \phi_{\D}(x_1) \phi_{\D}(x_2) \cO_h^{\m_1 \cdots \m_J}(x_0) \ra_{\rm cs} \,.
\ee
The subscript  ``cs" stands for conformal structure, meaning that the three-point function is just a notation for the structure on right-hand side. In particular, there is no structure constant, and the operator $\cO_h^{\m_1 \cdots \m_J}(z)$, of conformal dimension $h$ and in the spin-$J$ symmetric-traceless representation of the rotation group, is in general not part of the spectrum of the CFT. We denote $\phi_{\D}(x)$ a generic scalar primary of dimension $\D$, without introducing any flavor/color index structure, which we assume to be already diagonalized, as for example in \eqref{eq:full-K}.

The precise form of the kernel eigenvalue depends on the specific model. In the case of the long-range $O(N)^3$ model studied in section~\ref{sec:ON3-model}, we have:
\begin{equation}
k(h,J)=-\f{g^2}{(4\pi)^d}
 \frac{\Gamma(-\frac{d}{4}+\frac{h+J}{2})\Gamma(\frac{d}{4}-\frac{h-J}{2})}{\Gamma(\frac{3d}{4}-\frac{h-J}{2})\Gamma(\frac{d}{4}+\frac{h+J}{2})} \,.
\end{equation}

Multiplied by the following normalization factor ($\htilde=d-h$ is the dimension of the shadow operator \cite{Ferrara:1972uq})
\be \label{eq:norm3pt}
\begin{split}
\cN^{\D}_{h,J} = & \f{2^{(2\D+h+J)/2}}{(2\pi)^{d/2}} \left(  \frac{\G(\tfrac{\htilde+J+2\D-d}{2}) \G(\tfrac{h+J+2\D-d}{2})  }{ \G(\tfrac{\htilde+J-2\D+d}{2}) \G(\tfrac{h+J-2\D+d}{2}) }   \right)^{1/2}\frac{\G(\tfrac{h+J }{2})}{ \G(\tfrac{\htilde+J}{2})  } \,,
\end{split}
\ee
 the three-point functions \eqref{eq:3pt}  with fixed ${\rm Re}(\D)\in (d/4,3d/4)$  form a complete and orthonormal basis in an appropriate space of bilocal functions \cite{Dobrev:1976vr,Dobrev:1977qv}, the basis elements being labeled by the spin $J\in \mathbb{N}_0$, the position $x_0\in \mathbb{R}^d$, and the scaling dimension $h\in \cP_+$,
where
\be \label{eq:principal-series}
\cP_+ = \left\{ h \Bigm| h=\f{d}{2}+\im r,\, r\in\mathbb{R}_+ \right\}\,,
\ee
labels the principal series representations of the conformal group. More precisely, the space of bilocal functions $\cV_\D$ can be defined as the space of smooth functions $f(x_1,x_2)$ that are square integrable with respect to the scalar product 
\be \label{eq:scalar-prod}
\begin{split}
(f_1 , f_2) =  \int_{x_1\ldots x_4 } \overline{f_1(x_1,x_2)}  C^{-1}(x_1,x_3) C^{-1}(x_2,x_4) f_2(x_3,x_4) \,,
\end{split}
\ee
i.e.\ $(f,f)<\infty$, and satisfy the asymptotic boundary condition $f(x_1,x_2) \sim |x_1|^{-2\D}$ for $|x_1|\to\infty$ and similar for $|x_2|\to\infty$. Here, we have assumed that the bilocal functions have no symmetry under permutation of their two arguments, and we denoted\footnote{In this appendix we use $C(x,y)$ to denote the full-two-point function $\la\phi(x)\phi(y)\ra$ of the CFT, for which we are free to choose the same normalization as the one we used for the GFFT, even if the theory we have in mind is in general interacting.}
 $C(x_1,x_3) =c(\D) / |x_1-x_3|^{2\D}$.
Similarly, we can introduce the shadow space $\cV_{\wtD}$ with its basis of three-point functions defined as above but with $\D$ replaced by its shadow $\wtD=d-\D$. Since the two-point function of $\phi_\D$ and that of $\phi_{\wtD}$ are the inverse of each other (see \eqref{eq:flatCinv}), we can write the analogue of the scalar product \eqref{eq:scalar-prod} for $\cV_{\wtD}$ by replacing $C^{-1}$ with $C$, the two-point function of  $\phi_\D$.

The relation between $\cV_{\D}$ and $\cV_{\wtD}$ can better be understood in terms of raising and lowering of indices by the metric associated to the scalar product on them.
Let us denote $f^{x_1x_2}$, with contravariant indices $x_1,x_2$, the elements of $\cV_\D $, signaling that $f$ has dimension $\Delta$ on each of its arguments.
The factor $g_{x_1x_2;x_3x_4 }=C^{-1}(x_1,x_3)C^{-1}(x_2,x_4)$ in the scalar product in \eqref{eq:scalar-prod}  is a metric on $\cV_\D $ with covariant indices, that is with dimension $\tilde \D = d-\D$ on each of its arguments. The inverse metric is $g^{x_1x_2;x_3x_4} =  C(x_1,x_3)C(x_2,x_4) $ and the contraction on an index (integral over the position) has dimension $-d$. The metric and its inverse allow one to lower respectively raise indices, i.e. map $\cV_{\D}$ to its dual $\cV_{\tilde \D}$. 
The mapping holds also for the basis elements:\footnote{The three-point functions are not in $\mathcal{V}$, as they are not integrable, but they form a basis in the continuous sense, just like the Fourier basis does for ${\rm L}^2(\mathbb{R}^d)$.}
\be \label{eq:shadowTrans}
\begin{split}
\int  & \dd x_3 \dd x_4 \, C^{-1}(x_1,x_3)C^{-1}(x_2,x_4) \, \la \phi_{\D}(x_3) \phi_{\D}(x_4) \cO_h^{\m_1 \cdots \m_J}(x_0) \ra_{\rm cs} \,  \cN^{\D}_{h,J} \\
&=  \la \phi_{\wtD}(x_1) \phi_{\wtD}(x_2) \cO_h^{\m_1 \cdots \m_J}(x_0) \ra_{\rm cs} \, \cN^{\wtD}_{h,J} \,.
\end{split}
\ee
The completeness relation, or resolution of the identity, reads
\be \label{eq:res-id-nonsymm}
\begin{split}
\mathbb{I}(x_1,x_2,x_3,x_4) &\equiv  \d(x_1-x_3)\d(x_2-x_4)  \\
&=  \sum_{J\in \mathbb{N}_0}  \int_{\f{d}{2}}^{\f{d}{2}+\im\infty}  \f{{\rm d}h}{2\pi\im} \r(h,J) \, \cN^{\D}_{h,J}  \cN^{\wtD}_{\htilde,J} \, \Psi_{h,J}^{\D,\D,\wtD,\wtD}(x_1,x_2,x_3,x_4)  \,,
\end{split}
\ee
where the equality holds in a distributional sense when acting to the left (integration over $x_3$ and $x_4$) on $\cV_{\D}$, or to the right (integration over $x_1$ and $x_2$) on $\cV_{\tilde \D}$.
We have introduced  the Plancherel weight 
\be
\begin{split}
\r(h,J) &= \f{\G(\tfrac{d}{2}+J)}{2 (2\pi)^{d/2} J!} \f{\G(\htilde-1)\G(h-1)}{\G(\f{d}{2}-h)\G(\f{d}{2}-\htilde)} (h+J-1) (\htilde+J-1)
\,,
\end{split}
\ee
and the conformal partial wave, defined as
\be \label{eq:CPW}
 \Psi_{h,J}^{\D,\D,\wtD,\wtD}(x_1,x_2,x_3,x_4) = \int \dd  z \, \la \phi_{\D}(x_1) \phi_{\D}(x_2) \cO_h^{\m_1\cdots \m_J}(z) \ra_{\rm cs}\la \phi_{\wtD}(x_3) \phi_{\wtD}(x_4) {\cO}_{\htilde}^{\m_1\cdots \m_J}(z)\ra_{\rm cs} \,.
\ee
We notice that the product of normalization factors of the basis simplifies to
\be \label{eq:prod-cN}
\cN^{\D}_{h,J}  \cN^{\wtD}_{\htilde,J} = \f{2^{3d/2+J}}{(2\pi)^{d}} \,.
\ee

Any endomorphism $\cE:\cV_\D \to \cV_\D$ associated to a conformal kernel can be diagonalized by convoluting the kernel with the appropriate resolution of the identity, e.g.
\be
\begin{split}
\cE(x_1,x_2,x_3,x_4) & = \int \dd y_1 \dd y_2 \, \cE(x_1,x_2,y_1,y_2) \, \mathbb{I}(y_1,y_2,x_3,x_4) \\
&=  \sum_{J\in \mathbb{N}_0}  \int_{\f{d}{2}}^{\f{d}{2}+\im\infty}  \f{{\rm d}h}{2\pi\im} \r(h,J) \, \L_{\cE}(h,J) \, \cN^{\D}_{h,J}  \cN^{\wtD}_{\htilde,J} \, \Psi_{h,J}^{\D,\D,\wtD,\wtD}(x_1,x_2,x_3,x_4) \,,
\end{split}
\ee
where $\L_{\cE}(h,J)$ is the eigenvalue of $\cE$, satisfying an equation similar to \eqref{eq:eigK}.
Using the following relation between the conformal partial waves and the conformal blocks $\cG_{h,J}$ \cite{Dolan:2011dv,Simmons-Duffin:2017nub},
\be \label{eq:CPW-block} 
\Psi_{h,J}^{\D,\D,\wtD,\wtD}(x_1,x_2,x_3,x_4) =  \left(-\tfrac{1}{2}\right)^J  \left(S_{\tilde{h},J}\, \cG_{h,J}^{\D,\D,\wtD,\wtD}(x_1,x_2,x_3,x_4)  +    S_{h,J}\, \cG_{\tilde{h},J}^{\D,\D,\wtD,\wtD}(x_1,x_2,x_3,x_4) \right) \,,
\ee
with
\be
S_{h,J} =\f{\pi^{d/2} \G(h-\f{d}{2}) \G(h+J-1) \G(\f{\tilde{h}+J}{2})^2}{\G(h-1) \G(d-h+J) \G(\f{h+J}{2})^2} \,,
\ee
one can then write
\be
\begin{split}
\cE(x_1,x_2,x_3,x_4) &=  
\sum_{J\in \mathbb{N}_0}  \left(-\tfrac{1}{2}\right)^J \int_{\f{d}{2}-\im\infty}^{\f{d}{2}+\im\infty}  \f{{\rm d}h}{2\pi\im} \r(h,J)\, \L_{\cE}(h,J) \, \cN^{\D}_{h,J}  \cN^{\wtD}_{\htilde,J} \, S_{\tilde{h},J}\, \cG_{h,J}^{\D,\D,\wtD,\wtD}(x_1,x_2,x_3,x_4) \,,
\end{split}
\ee
where we used the symmetry of the measure factor $\r(h,J) \, \cN^{\D}_{h,J}  \cN^{\wtD}_{\htilde,J} $ under shadow reflection $h\to\htilde$  to extend the integration to negative imaginary parts and keep only one conformal block term.

Acting by convolution on the last two arguments of $\cE(x_1,x_2,x_3,x_4)$ with the inverse metric, we obtain an operator mapping $\cV_{\wtD}$ to $\cV_{\D}$, with a similar conformal partial wave expansion, except that the $\wtD$ arguments are replaced by $\D$:
\be \label{eq:4ptCPW}
\begin{split}
\int \dd y_3 \dd y_4 \,& \cE(x_1,x_2,y_3,y_4) C(y_3,x_3) C(y_4,x_4)\\
&=  \sum_{J\in \mathbb{N}_0}  \int_{\f{d}{2}}^{\f{d}{2}+\im\infty}  \f{{\rm d}h}{2\pi\im} \r(h,J) \, \L_{\cE}(h,J) \, \cN^{\D}_{h,J}  \cN^{\D}_{\htilde,J} \, \Psi_{h,J}^{\D,\D,\D,\D}(x_1,x_2,x_3,x_4) \\
&=\sum_{J\in \mathbb{N}_0}  \left(-\tfrac{1}{2}\right)^J \int_{\f{d}{2}-\im\infty}^{\f{d}{2}+\im\infty}  \f{{\rm d}h}{2\pi\im} \r(h,J)\, \L_{\cE}(h,J) \, \cN^{\D}_{h,J}  \cN^{\D}_{\htilde,J} \, S_{\tilde{h},J}\, \cG_{h,J}^{\D,\D,\D,\D}(x_1,x_2,x_3,x_4)\,.
\end{split}
\ee
In this case, the product of normalization factors has a ratio of gamma functions, with its own poles:\footnote{These poles do not cross the principal series as long as $\D>d/4$. Similarly, the poles of $\cN^{\wtD}_{h,J}  \cN^{\wtD}_{\htilde,J}$ stay away from $h=d/2$ for $\D<3d/4$. Together, these two conditions explain the condition on $\D$ mentioned below \eqref{eq:norm3pt}.}
\be \label{eq:prod-cN-2}
\cN^{\D}_{h,J} \cN^{\D}_{\htilde,J}=  \f{2^{(2\D+d/2+J)}}{(2\pi)^{d}}  \frac{\G(\tfrac{\htilde+J+2\D-d}{2}) \G(\tfrac{h+J+2\D-d}{2})  }{ \G(\tfrac{\htilde+J-2\D+d}{2}) \G(\tfrac{h+J-2\D+d}{2}) }  \,.
\ee
For $\cE=\mathbb{I}$, \eqref{eq:4ptCPW}, with $\L_{\mathbb{I}}(h,J)=1$, gives a conformal partial wave expansion of the inverse metric.
A similar expansion is obtained for the metric itself, replacing $C$ with $C^{-1}$ in the first line, and $\D$ with $\wtD$ in the expansion.

In the case of symmetric bilocal functions,\footnote{The corresponding space $\cV_\D^{\rm symm}$  is defined as before, except that the metric needs also symmetrization: $g^{\rm symm}_{x_1x_2;x_3x_4 }=\f12 (C^{-1}(x_1,x_3)C^{-1}(x_2,x_4)+C^{-1}(x_1,x_4)C^{-1}(x_2,x_3))$.} the completeness relation is of the same type, but with contribution only from even spin:
\be \label{eq:res-id-symm}
\begin{split}
\mathbb{I}_{\rm symm}(x_1,x_2,x_3,x_4) &\equiv  \f12\left( \d(x_1-x_3)\d(x_2-x_4) + \d(x_1-x_4)\d(x_2-x_3) \right) \\
&=  \sum_{J\in \mathbb{N}_0^{\text{even}}}  \int_{\f{d}{2}}^{\f{d}{2}+\im\infty}  \f{{\rm d}h}{2\pi\im} \r(h,J) \, \cN^{\D}_{h,J}  \cN^{\wtD}_{\htilde,J} \, \Psi_{h,J}^{\D,\D,\wtD,\wtD}(x_1,x_2,x_3,x_4)  \,. 
\end{split}
\ee

\paragraph{Non-normalizable contributions.} 
In practical applications, such as those we encounter in the bulk of this paper, some of the hypotheses behind what we just reviewed can be violated.
Typically, we have two possible situations:
\begin{enumerate}
\item A four-point kernel $\cE(x_1,x_2,x_3,x_4)$ with right conformal transformation might nevertheless not be an endomorphism on $\cV_{\D}$ (or a map $\cV_{\wtD}\to \cV_{\D}$) because its action on an element $f\in\cV_{\D}$ (or  $\tilde{f}\in\cV_{\wtD}$) leads to a function not satisfying the integrability condition associated to the scalar product \eqref{eq:scalar-prod}.
\item The scalar field dimension might lie outside the range $(d/4,3d/4)$. This is in particular the case of the standard free theory (or the critical $O(N)$ model at large $N$) with $\D=d/2-1<d/4$, for $d<4$.
\end{enumerate}

In both cases, we can still use the conformal partial wave machinery, as long as we take care of deforming the contour of integration over $h$, or isolating the non-normalizable contributions from the four-point kernel \cite{Simmons-Duffin:2017nub}.

The typical example of a four-point kernel which is not an endomorphism is a physical four-point function of one scalar field $\phi$ whose
$s$-channel OPE contains operators of dimension smaller than $d/2$.
The identity operator is one such operator and it is always present, hence we always need to subtract the contribution that is disconnected in the  $s$-channel, $C(x_1,x_2)C(x_3,x_4)$, before applying the expansion \eqref{eq:4ptCPW}.\footnote{Convoluting $C(x_1,x_2)C(x_3,x_4)$ with $f(x_3,x_4)\in\cV_{\D}$, we obtain a new function proportional to $C(x_1,x_2)$. Regardless of whether the proportionality constant is finite or not, $C(x_1,x_2)$ is not square integrable with respect to the scalar product in \eqref{eq:scalar-prod}, and therefore it is not in $\cV_{\D}$.} Similarly, if the field $\phi$ has $\D<d/2$ and the three-point function with itself is non-vanishing, then we need to subtract the contribution that is one-particle reducible in the  $s$-channel (the $s$-channel skeleton tree diagram). It is then useful to define
\be \label{eq:F_s}
\begin{split}
\cF_s(x_1,x_2,x_3,x_4)   \equiv &\, \langle{\phi(x_1) \phi(x_2) \phi(x_3) \phi(x_4)} \rangle 
 -   C(x_1,x_2)  C(x_3,x_4) 
\\
 &-\int \dd y_1 \dd y_2 \langle \phi(x_1) \phi(x_2) \ph(y_1) \rangle C^{-1}(y_1,y_2) \langle \phi(y_2) \phi(x_3) \phi(x_4) \rangle \,,
\end{split}
\ee
which is obtained in terms of the Bethe-Salpeter kernel $K$ as (e.g.\ \cite{Benedetti:2021qyk})
\be \label{eq:F_s-K}
\cF_s(x_1,x_2,x_3,x_4)  = \int \dd y_1 \dd y_2 (\mathbb{I}-K)^{-1}(x_1,x_2,y_1,y_2) C(y_1,x_3) C(y_2,x_4)\,.
\ee
Applying to $\cF_s$ the expansion in the last line of \eqref{eq:4ptCPW}, and pushing the integration contour to the right, the integral is reduced to a sum over the residues at the poles of the integrand (the poles of $\L_{(\mathbb{I}-K)^{-1}}(h,J)$ being now the solutions of $k(h,J)=1$). Together with the sum over $J$, this reproduces the operator product expansion of the four-point function in the $s$ channel, if no other physical operators have dimension smaller than $d/2$. If instead other primaries have dimension smaller than $d/2$, then on the right of the principal series we pick their shadow pole; 
this must be corrected by deforming the contour in the conformal block representation to keep only the physical poles on the right, or equivalently (because of \eqref{eq:CPW-block}), by adding to the expansion the appropriate $\Psi$ contributions:
\be \label{eq:cF-extra}
\begin{split}
\cF_s(x_1,x_2,x_3,x_4)  
&=  \sum_{J\in \mathbb{N}_0}  \int_{\f{d}{2}}^{\f{d}{2}+\im\infty}  \f{{\rm d}h}{2\pi\im} \f{\r(h,J)}{1-k(h,J)}  \, \cN^{\D}_{h,J}  \cN^{\D}_{\htilde,J} \, \Psi_{h,J}^{\D,\D,\D,\D}(x_1,x_2,x_3,x_4) \\
&\quad - \sum_{i,J} {\rm Res}\left[  \f{\r(h,J) }{1-k(h,J)}  \, \cN^{\D}_{h,J}  \cN^{\D}_{\htilde,J} \, \Psi_{h,J}^{\D,\D,\D,\D}(x_1,x_2,x_3,x_4) \right]_{h=h_i(J)<d/2}\,,
\end{split}
\ee
where $h_i(J)$ are the physical solutions of $k(h,J)=1$ on the left of the principal series.
These isolated contributions are exactly analogue to the contributions we subtracted from the four-point function to define $\cF_s$. If such operators are present, we first subtract them from $\cF_s$, then we use the resolution of the identity to decompose the subtracted $\cF_s$, and finally we add them back as in \eqref{eq:cF-extra} to give the expansion of $\cF_s$ itself.

An example of the second situation in which \eqref{eq:4ptCPW} fails, namely when the dimension of the scalar field is outside the range $(d/4,3d/4)$, can be obtained from a generalized free field theory. Consider $\cF_s$ for a GFFT:
\be
\begin{split}
\cF_s^{GFFT}(x_1,x_2,x_3,x_4)  
&= C(x_1,x_3) C(x_2,x_4) + C(x_1,x_4) C(x_2,x_3) \,,
\end{split}
\ee
which is also twice the inverse metric of $\cV_{\D}^{\rm symm}$.
Using \eqref{eq:4ptCPW} with $\cE=\mathbb{I}$, but restricted to even spins to provide the symmetrization, we find
\begin{align}
\label{eq:cF-GFFT}
&\cF_s^{GFFT}(x_1,x_2,x_3,x_4)  
= \crcr
&2 \sum_{J\in \mathbb{N}_0^{\text{even}}}  \left(-\tfrac{1}{2}\right)^J \int_{\f{d}{2}-\im\infty}^{\f{d}{2}+\im\infty}  \f{{\rm d}h}{2\pi\im} \r(h,J)\, \cN^{\D}_{h,J}  \cN^{\D}_{\htilde,J} \, S_{\tilde{h},J}\, \cG_{h,J}^{\D,\D,\D,\D}(x_1,x_2,x_3,x_4)\,,
\end{align}
and from \eqref{eq:prod-cN-2} we know the normalization factors at  $J=0$ have a pole at $h=2\D$, that is the dimension of $\phi^2$. This pole crosses to the left of the principal series when $\D$ moves below $d/4$.
In that case, in order to recover the correct OPE, we must deform the integration contour in such a way that the pole at $h=2\D$ stays to its right, and the shadow pole at $h=d-2\D$ stays to its left.
Exploiting \eqref{eq:CPW-block}, it can be verified that the same result is obtained by writing, for $\D<d/4$:
\be \label{eq:cF-GFFT-extra}
\begin{split}
\f12 \cF_s^{GFFT}(x_1,x_2,x_3,x_4)  
&=  \sum_{J\in \mathbb{N}_0^{\text{even}}}  \int_{\f{d}{2}}^{\f{d}{2}+\im\infty}  \f{{\rm d}h}{2\pi\im} \r(h,J)  \, \cN^{\D}_{h,J}  \cN^{\D}_{\htilde,J} \, \Psi_{h,J}^{\D,\D,\D,\D}(x_1,x_2,x_3,x_4) \\
&\quad - {\rm Res}\left[  \r(h,0)  \, \cN^{\D}_{h,0}  \cN^{\D}_{\htilde,0} \, \Psi_{h,0}^{\D,\D,\D,\D}(x_1,x_2,x_3,x_4) \right]_{h=2\D}\,.
\end{split}
\ee

This has the exact same form as \eqref{eq:cF-extra}, but is quite different in nature. Contrary to the previous case, it is now the functional space itself that is deviating from the original definition. 
In particular, convoluting \eqref{eq:cF-GFFT-extra} with the  metric of $\cV_{\D}^{\rm symm}$, we obtain a modified resolution of the identity:
\be \label{eq:res-id-symm-2}
\begin{split}
\mathbb{I}_{\rm symm}(x_1,x_2,x_3,x_4) &\equiv  \f12\left( \d(x_1-x_3)\d(x_2-x_4) + \d(x_1-x_4)\d(x_2-x_3) \right) \\
&=  \sum_{J\in \mathbb{N}_0^{\text{even}}}  \int_{\f{d}{2}}^{\f{d}{2}+\im\infty}  \f{{\rm d}h}{2\pi\im} \r(h,J) \, \cN^{\D}_{h,J}  \cN^{\wtD}_{\htilde,J} \, \Psi_{h,J}^{\D,\D,\wtD,\wtD}(x_1,x_2,x_3,x_4)  \\
&\qquad -   {\rm Res}\left[ \r(h,0) \, \cN^{\D}_{h,0}  \cN^{\wtD}_{\htilde,0} \, \Psi_{h,0}^{\D,\D,\wtD,\wtD}(x_1,x_2,x_3,x_4) \right]_{h=2\D} \,.
\end{split}
\ee

Since \eqref{eq:prod-cN} has no poles, it would seem that the isolated contribution here is trivial. However, $\Psi_{2\D,0}^{\D,\D,\wtD,\wtD}(x_1,x_2,x_3,x_4)$ is a singular distribution proportional to $1/|x_{34}|^d\sim \Gamma(0) \d(x_{34})$; 
writing explicitly the limit involved in the definition of the residue, we find that  such singularity leads to a non-trivial isolated contribution to the resolution of the identity:
\be \label{eq:extra}
\begin{split}
-{\rm Res} &\left[  \r(h,0) \, \cN^{\D}_{h,0}  \cN^{\wtD}_{\htilde,0} \, \Psi_{h,0}^{\D,\D,\wtD,\wtD}(x_1,x_2,x_3,x_4) \right]_{h=2\D}\\
&= \lim_{\eps\to 0} \,  2\eps \,  \r(2\D-2\eps,0) \,  \f{2^{d/2}}{\pi^{d}} \, \Psi_{2\D-2\eps,0}^{\D,\D,\wtD,\wtD}(x_1,x_2,x_3,x_4)\\
&= \lim_{\eps\to 0} \, c(d/2-\epsilon) \, c(2\D-2\eps) \, c (d-2\D+2\eps) \, \Psi_{2\D-2\eps,0}^{\D,\D,\wtD,\wtD}(x_1,x_2,x_3,x_4)\\
&= c(2\D) \, c (d-2\D) \, \int \dd x_0 \f{1}{|x_{10}|^{2\D} |x_{20}|^{2\D}}  \f{1}{|x_{30}|^{d-2\D} |x_{40}|^{d-2\D}}  \d(x_{34}) \,,
\end{split}
\ee
where we used the distributional identity $\lim_{\epsilon\to 0} c(d/2-\epsilon) / |x|^{d-2\epsilon} = \delta(x)$.

\paragraph{Conformal partial waves on the sphere.}
Everything we discussed in this appendix transcribes mutatis mutandis for CFTs on the sphere via the Weyl mapping (i.e. replacing all the distances by chordal distance and adding the adequate volume factors). By conformality of all the integrals involved (e.g.\ in \eqref{eq:shadowTrans}, \eqref{eq:CPW}, and so on), the analytic properties of the conformal partial wave expansion are unchanged, and the effect of being on the sphere is only visible in the external points being multiplied by the appropriate $\Om(x)$ factors.

However,  in the evaluation of the sphere free energy we encounter traces of endomorphisms on $\cV_\D$,
\be
\Tr(\cE) =  
\int \dd x_1 \dd x_2 \, \cE( x_1,x_2 , x_1,x_2 ) \,,
\ee
which are divergent due to their conformal invariance. The regularization of such traces necessarily breaks conformal invariance, and as a result $\Om(x)$ factors survive, which play a crucial role in leading to finite results.
The precise choice of regularization and its effects are described in the bulk of the paper.

\section{The NNLO graph in Fig.~\ref{fig:monstru}}
\label{app:monster}

In this section we show that the contribution of the graph in
Fig.~\ref{fig:monstru} to the sphere free energy is finite. As it is not especially informative, we will not compute this contribution exactly.

The amplitude of any vacuum graph can be written as the convolution of a free covariance (representing any of its edges) and a two-point kernel (representing the amplitude of the remaining two-point amputated graph $G$):
\be
 I = \int_{x,y} C(x,y) A^G(x,y) \;.
\ee
In the case of the melon in \eqref{eq:melon-int} for instance $A^G(x,y) = C(x,y)^3$. Using dimensional regularization and $\Delta = \frac{d-\epsilon}{4}$, dimensional analysis leads to:
\be
 I = \int_{x,y} \frac{c(\Delta)}{s(x,y)^{ \frac{d- \epsilon}{2}} } \; \frac{A^G(\epsilon)}{s(x,y)^{\frac{3}{2}d - b \epsilon}} \;.
\ee
The precise scaling $b$ depends on the particular graph one considers (e.g. $b=3/2$ for the melon). Following the same steps leading to \eqref{eq:vanishmelo} we conclude that:
\be
 I \sim \frac{1}{\Gamma(\frac{1+2b}{4} \epsilon)} A^G(\epsilon) \;.
\ee

\begin{figure}[htbp]
	\centering
	\includegraphics[width=0.16\linewidth]{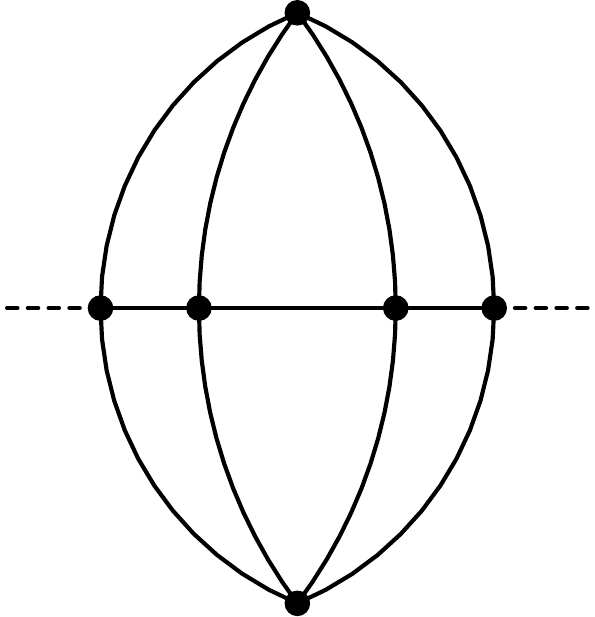}
	\caption{Amputated two-point diagram obtained by opening any edge in the graph of Fig.~\ref{fig:monstru}.}
	\label{fig:monstruopen}
\end{figure}

For the melon $A^G(\epsilon)\sim \epsilon^0$, that is the melon does not contribute to the sphere free energy in the $\epsilon\to 0$ limit. We show below that for the graph in Fig.~\ref{fig:monstru} opened on any of its edges, as depicted in Fig.~\ref{fig:monstruopen},
there exist two constants $A_1$ and $A_2$ such that: 
\be
\label{appeq:amplibound} 
\frac{A_1}{\epsilon} < A^G(\epsilon) < \frac{A_2}{\epsilon} \;.
\ee 
hence this graph brings a finite contribution to the sphere free energy. This contribution cancels between the different fixed points of interest as they all have the same tetrahedral coupling.

The remainder of this appendix is a rigorous proof of \eqref{appeq:amplibound}. It is quite lengthy and uses a set of techniques that, while standard, are quite apart from the ones used in the rest of the paper.

One can of course apply the techniques we discuss here to the ladder diagrams. Graph by graph the ladder diagrams are divergent, and present increasingly singular poles in $1/\epsilon$. Accounting for the diagrams with $\lambda_1$ vertices allows one to subtract the poles and one can in principle resum the finite parts to reproduce the finite part of \eqref{eq:ladderssum}. This graph by graph computation is exceedingly difficult. The main gain of the conformal partial waves techniques we introduced in the main body of this paper is to bypass this analysis entirely and directly give us the end result.

\paragraph{Going to flat space.}
We are interested in identifying the leading singular behavior in the $\epsilon\to 0$ limit of the amplitude $A^G(x,y)$ of amputated two-point graphs $G$. As already mentioned, dimensional analysis implies that, up to local terms:
$ A^G(x,y) =  A^G(\epsilon) / s(x,y)^{\frac{3}{2}d - b \epsilon}$
where $A^G(\epsilon)$ might display poles in $1/\epsilon$. As this is an ultraviolet divergence it is insensitive to the details of the infrared regularization hence the leading divergence is the same on the sphere and on flat space. From now on we work on flat space.

\paragraph{Why it is non-trivial.} In the long-range model two-point graphs are primitively power divergent. Once this local power divergence is dealt with (either set to zero in dimensional regularization or subtracted by a mass counterterm in other schemes) the only remaining primitively divergent graphs are the four-point ones. The latter bring $1/\epsilon$ poles that pile up if they come from subgraphs fully included in larger subgraphs (we review how this occurs below).

For two-point amputated graphs this naive power counting yields $A^G(\epsilon) \sim \epsilon^{-1}$ for the two-point melon (as it has four-point subgraphs) and $A^G(\epsilon)\sim \epsilon^{-2}$ for the two-point graph in Fig.~\ref{fig:monstruopen}, as it has a four-point subgraph that is a subgraph of another four-point subgraph. The non-trivial result of this section is that a detailed analysis of the sub divergences of these two graphs improves on the naive expectation, that is $A^G(\epsilon)\sim \epsilon^0$ for the melon and $A^G(\epsilon) \sim \epsilon^{-1}$ for the graph in Fig.~\ref{fig:monstruopen}

\paragraph{Structure of the amplitudes.}
On flat space the divergences and their subtractions are captured by the Bogoliubov--Parasiuk--Hepp--Zimmermann theorem \cite{Hepp:1966eg,Zimmermann:1969jj,rivasseau2014perturbative}.
For long-range models only two and four-point subgraphs
are primitively divergent and they are subtracted by applying local Taylor operators. 

The momentum and direct space formulation of the BPHZ theorem
are reviewed for instance in \cite{rivasseau2014perturbative} and
the parametric space version is discussed in 
\cite{Bergere:1974zh,Bergere:1980sm}. In flat space, using Schwinger parameters, the amplitude of an amputated graph in the long-range model with external momenta $p_i$, $V$ vertices and $E$ edges is $ \bar A_{\mu}^G(p_i)  = (2\pi)^d \delta(\sum_ip_i)    A_{\mu}^G(p_i)  $ with:
\be
\begin{split}
   A^G(p_i) & = 
   \frac{ 1 }{\Gamma(\zeta)^E (4\pi)^{\frac{d}{2}(E-V+1) }  }
 \int_0^{k^{-2} }  \left( \prod_{e\in G} d\alpha_e \; \alpha_e^{\zeta-1}  \right) \; 
 \frac{ e^{-\frac{ V(G)  } { U(G) } } } { U(G)^{d/2}  }   \;,\crcr
 U(G) & = \sum_{T \subset G } \prod_{e\notin T} \alpha_e  \;,\;\;
  V(G) =  \sum_{T_1,T_2 \subset G } (\sum_{i \in T_2} p_i)^2\;  \prod_{e\notin T_1\cup T_2} \alpha_e \; ,
\end{split}
\ee
where $T$ runs over the spanning trees in $G$, respectively $T_1,T_2$ run over the pairs of trees, and $i\in T_2$ runs over the external vertices that belong to the tree $T_2$. 
The integral is cut-offed by and infrared cutoff $k$ on the integration interval, but any other infrared cutoff will do.
The overall factor $ (2\pi)^d \delta(\sum_ip_i) $ reproduces the bare vertex. 
We work in dimensional regularization and we set $\zeta = \frac{d+\epsilon}{4}$. 
The amplitude at zero external momenta of $G$ is, up to prefactors:
\be
 A^G \sim  \int_0^{k^{-2}}  \left( \prod_{e\in G} d\alpha_e \; \alpha_e^{\zeta-1} \right) \; 
 \frac{ 1} { U(G)^{d/2}  }  = k^{-2\zeta E + d [E -V +1 ]}
 \int_0^{1}  \left( \prod_{e\in G} d\alpha_e \; \alpha_e^{\zeta-1} \right) \; 
 \frac{ 1} { U(G)^{d/2}  } \; .
\ee

If $G$ is a two-point graph its amplitude in momentum space writes:
\be\label{appeq:ampli2point}
 A^G(p) = p^{ \frac{d + \epsilon }{2} - V \epsilon} A^G(\epsilon) + A^G \; ,
\ee
where we denote somewhat abusively by $A^G(\epsilon)$ the coefficient of the scaling behavior in $p$. Note that for two-point graphs $A^G$ is zero in dimensional regularization, as it is power divergent. This is of course not the case for four-point graphs. In principle $A^G(\epsilon)$ is a function of $p/k$ and in the case of the melon it can be shown \cite{Benedetti:2019eyl} that it has a finite limit for $k\to 0$. 

\paragraph{Taylor operators and renormalized amplitudes.}
The Taylor operators acting on amplitudes are localization operators. For any $\gamma$ subgraph of $G$ we define:
\be
\tau_{\gamma} A^G(p_i) \equiv  A^{\gamma} \; A^{G/\gamma}(p_i) \;,
\ee
where $A^{\gamma}$ is the amplitude of $\gamma$ at zero momentum and $G/\gamma$ denotes the graph obtained from $G$ by contracting $\gamma$ to a point. As we deal with the long-range case the subtraction of local parts suffices in order to render the amplitudes ultraviolet finite.

We consider amputated subgraphs $\gamma$ of $G$. A subgraph $\gamma$ contains all the vertices hooked to its edges.
An inclusion forest \cite{Zimmermann:1969jj} of subgraphs of $G$ is a family ${\cal F}$ of subgraphs such that any two $\gamma_1 ,\gamma_2 \in {\cal F}$ are either totally disjoint (i.e. they do not share neither edges nor vertices) or one of them is fully contained in the other:
\be
 {\cal F} = \big\{\gamma \in G \big{|}\forall \gamma_1,\gamma_2  \text{ either } \gamma_1 \subset \gamma_2 \; \text{, or } \gamma_2 \subset \gamma_1 \; \text{, or }  \gamma_1 \cap \gamma_2 = \emptyset \big\} \;.
\ee

The renormalization operator  \cite{Hepp:1966eg,Zimmermann:1969jj,rivasseau2014perturbative} $R_G$ associated to the graph $G$ is a sum over the inclusion forests ${\cal F} \subset G$ 
of primitively divergent subgraphs (i.e. two and four-point subgraphs in our case) of a product of Taylor operators associated to the graphs in the forest:
\be
R_G =\sum_{\cal F \subset G } \prod_{\gamma \in \cal F} (-\tau_{\gamma} ) \;.
\ee
the forests include the empty forest, which contributes a $1$ to this formula. 

\begin{theorem}(BPHZ \cite{Bergere:1974zh,rivasseau2014perturbative}). 
The renormalized amplitude $R_GA^G(p_i)$ of any graph $G$ is ultraviolet convergent, that is $\lim_{\epsilon \to 0} R_GA^G(p_i)$ is finite.
\end{theorem}

\paragraph{How we will use the BPHZ theorem.}
We are interested in identifying the $1/\epsilon$ behavior of the bare amplitude of a two-point graph $G$. Separating the empty forest in the renormalization operator we have:
\be
R_G A^G(p)  = A^G(p) + \sum_{\cal F \subset G}^{ {\cal F} \neq \emptyset} \prod_{\gamma \in \cal F} (-\tau_{\gamma} ) A^G(p) \; ,
\ee
and the BPHZ theorem ensures that this expression is convergent in the $\epsilon\to 0$ limit.
It follows that the singular part of the coefficient $A^G(\epsilon)$ in \eqref{appeq:ampli2point} \emph{must be entirely canceled by the subtractions}, hence it equals the divergent part of the counterterms.

\paragraph{The melon.} Let us first consider the melon $G$. As primitively divergent subgraphs it has itself $\gamma = G$, and three subgraphs $\gamma_i,i=1,2,3$ made of two edges. It follows that:
\be\label{appeq:melofin}
 R_GA^G(p) = A^G(p) - \tau_\gamma A^G(p) - \sum_i \tau_{\gamma_i}A^G(p)  + \sum_i \tau_{\gamma}\tau_{\gamma_i}A^G(p) =
 A^G(p) - A^G  \; .
\ee
In this equation $ \tau_\gamma A^G(p)$ is just the local part $A^G$ of the melon (and in particular it is zero in dimensional regularization). Moreover, $ \tau_{\gamma}\tau_{\gamma_i}A^G(p) = \tau_{\gamma_i} A^G(p)$ because $G/\gamma_i$ is a tadpole graph, therefore the action of $\tau_{\gamma}$ is trivial.
It follows that the two terms summed over $i$ cancel exactly, which, together with \eqref{appeq:ampli2point}, yields:
\be
p^{ \frac{d + \epsilon }{2} - V \epsilon} A^G(\epsilon) = 
A(p) - A^G  =  R_GA^G(p) \; .
\ee
As $R_GA^G(p)$ has a finite limit for $\epsilon \to 0$ we conclude that $ A^G(\epsilon) $ has no poles in $1/\epsilon$.

\paragraph{The graph in Fig.~\ref{fig:monstruopen}.} This graph has many four-point subgraphs.
To identify them we label $x,y$ the two external vertices, $v_1,v_2$ the top and bottom vertices and $z_1,z_2$ the two vertices on the horizontal ($x$ and $z_1$ are connected by an edge). The list of two and four-point subgraphs of the graph in Fig.~\ref{fig:monstruopen} comprises: 
\begin{itemize}
 \item two kite graphs, see Fig.~\ref{fig:monstruopensubgraph2}, $\gamma_x$ (and $\gamma_y$) obtained by cutting the edges $(x,v_1)$, $(x,z_1)$ and $(x,v_2)$ (resp. 
  $(y,v_1)$, $(y,z_2)$ and $(y,v_2)$).
  \begin{figure}[htbp]
  	\centering
  	\includegraphics[width=0.18\linewidth]{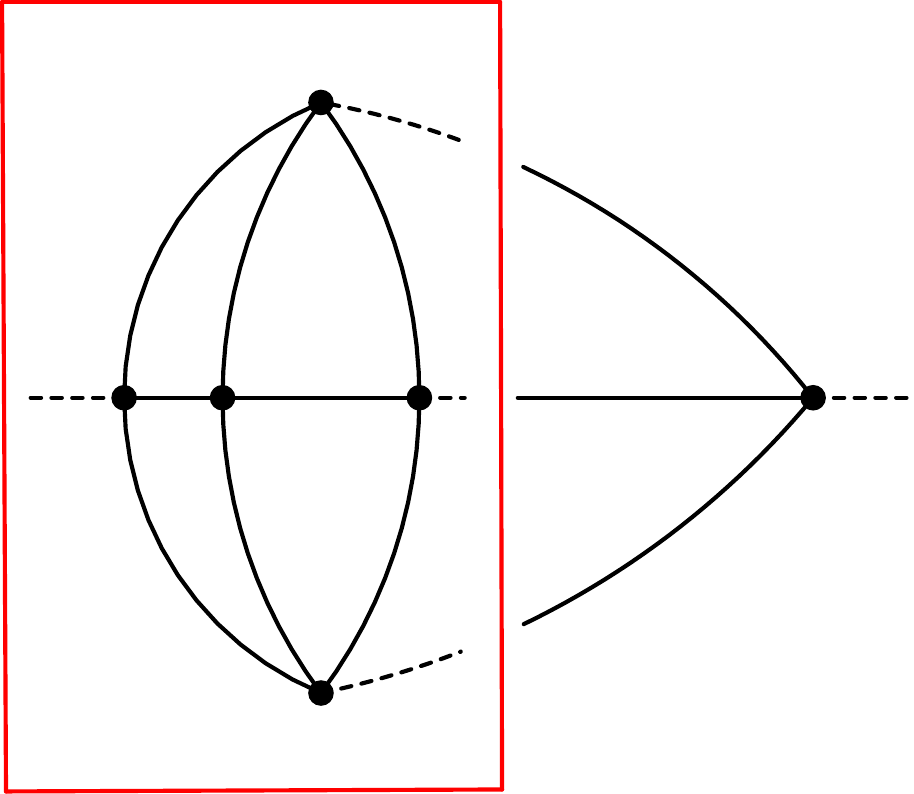}
  	\caption{Kite graph obtained by cutting $(y,v_1)$, $(y,z_2)$ and $(y,v_2)$.}
  	\label{fig:monstruopensubgraph2}
  \end{figure}
 \item 11 graphs $\gamma^{ab}$ (see Fig.~\ref{fig:monstruopensubgraph1}) obtained by cutting any of the internal edges $(a,b)$ in the graph.
 \begin{figure}[htbp]
 	\centering
 	\includegraphics[width=0.14\linewidth]{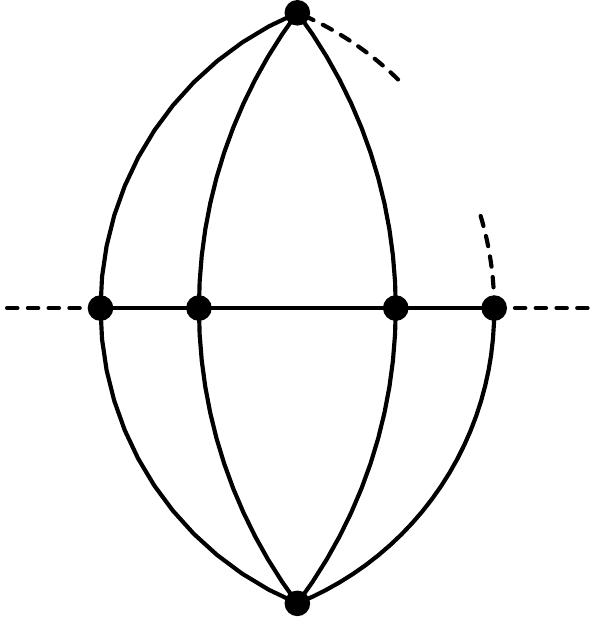}
 	\caption{Example of one $\gamma^{ab}$ graph, obtained by cutting one of the internal edges.}
 	\label{fig:monstruopensubgraph1}
 \end{figure}
 
 \item itself, that is $\gamma=G$
\end{itemize}

These graphs are organized in several inclusion forests:
\begin{itemize}
 \item the empty forest
 \item the one graph forests. They consist in either
    \begin{itemize}
     \item the graph $\gamma$
     \item one of the kite graphs $\gamma_x$ or $\gamma_y$
     \item any one of the remaining 11 graphs $\gamma^{ab}$
    \end{itemize}
  \item the two graphs forests. They are
    \begin{itemize}
    \item 11 of the form $\{\gamma, \gamma^{ab}\}$ for some  $ab$.
    \item three of the form $\{ \gamma^{xa},\gamma_x \} $ for the $a$s connected to $x$ by some edge
    respectively three of the form $\{\gamma^{ya} ,\gamma_y \} $ for the $a$s connected to $y$ by some edge in $G$
    \item two special ones  $\{ \gamma, \gamma_x\}$ and
    $\{ \gamma, \gamma_y\}$
    \end{itemize}

  \item the three graph forests. There are six of these ones
     \begin{itemize}
      \item three of the form $\{\gamma,\gamma^{xa}, \gamma_x\} $ for some $a$ and three of the form  $\{\gamma,\gamma^{ya}, \gamma_y\} $ for some $a$.
     \end{itemize}
\end{itemize}
Plenty of the contributions to the renormalized amplitude cancel
as for any $\gamma^{ab}$ , $G/\gamma^{ab}$ is contracted to a tadpole hence $\tau_{\gamma^{ab}}A^G = \tau_{\gamma}\tau_{\gamma^{ab}}A^G$ and $\tau_{\gamma^{ab}}\tau_{\gamma_i}A^G = \tau_{\gamma}\tau_{\gamma^{ab}}\tau_{\gamma_i}A^G$. The renormalized amplitude is then 
\be
 R_GA^G(p) = A^G(p) - A^G - 2 A^{\gamma_x}  \big[ A^{G/\gamma_x}(p) - A^{G/\gamma_x}  \big] \;.
\ee
Observe that $G/\gamma_x$ is the two-point melon graph, hence incidentally $A^{G/\gamma_x}(p) - A^{G/\gamma_x}$ is nothing but the subtracted melon $R_{G/\gamma_x} A^{G/\gamma_x}(p) $
Combining this with \eqref{appeq:ampli2point} we find that:
\be
p^{ \frac{d + \epsilon }{2} - V \epsilon} A^G(\epsilon) = 
A(p) - A^G  =  R_GA^G(p) + 2 A^{\gamma_x}  R_{G/\gamma_x} A^{G/\gamma_x}(p) \;,
\ee
and as the renormalized amplitudes on the right hand side above have finite limits at $\epsilon \to 0$, the leading divergence of $A^G(\epsilon)$ is the same as the local part of the kite graph $A^{\gamma_x}$.

\paragraph{The amplitude of the kite graph.} We will now conclude this section by proving that the amplitude of the kite graph has a pole of order $1$ in $\epsilon$. The amplitude of any four-point graph $G$  with no two-point subgraphs at zero external momenta:
\be
 A^{G} =  \int_0^{k^{-2}}  \left( \prod_{e\in G} d\alpha_e \; \alpha_e^{\zeta-1} \right) \; 
 \frac{ 1} { U(G)^{d/2}  }  = k^{-2\zeta E + d [E -V +1 ]}
 \int_0^{1}  \left( \prod_{e\in G} d\alpha_e \; \alpha_e^{\zeta-1} \right) \; 
 \frac{ 1} { U(G)^{d/2}  } \; ,
\ee
is a convergent integral over $\alpha$ for $\zeta = \frac{d+\epsilon}{4}$ but exhibits poles in $1/\epsilon$. 

For a graph with $E$ edges, we divide the integration interval into Hepp sectors, which we denote $\sigma$, that is total orderings of the $\alpha$ parameters $\alpha_{e_{\sigma(1)} } \le \alpha_{e_{\sigma(2) } } \le \dots  \le \alpha_{e_{\sigma(E)} }$. The edge $e_{\sigma(1)} $ is the most ultraviolet (lowest $\alpha$), $e_{\sigma(2)}$ is the next most ultraviolet and so on up to $e_{\sigma(E)}$, which is the most infrared. There are $E!$ sectors, as many as there are permutations. Up to the global scaling the amplitude is:
\be
A^G =  k^{ \epsilon- V \epsilon } \sum_{\sigma} A^G_{\sigma} \;,\quad
 A^G_{\sigma} =  \int_{0\le \alpha_{\sigma(1)} \le \dots \le \alpha_{\sigma(E)}  \le 1}  \left( \prod_{e\in G} d\alpha_e \; \alpha_e^{\zeta-1} \right) \;  \frac{ 1} { U(G)^{d/2}  }  \;.
\ee

The polynomial $U(G)$ is a sum of positive terms. It is thus bounded from below by any of its terms, and from above by the number of terms times the largest of them. The number of trees in a graph is bounded by the number of subsets of edges, that is $2^E$.

In each Hepp sector there is exactly one leading monomial corresponding to the tree $T_{\sigma}$ built by proceeding from $1$ to $E$ and at each step adding the edge $e_{\sigma(q)}$ if it does not form a loop. We thus ensure that the edges with the lowest possible $\alpha$ parameters are in the tree, hence the complement of $T_{\sigma}$ has the highest possible $\alpha$s, that is in the sector $\sigma$:
\be
0\le \alpha_{\sigma(1)} \le \dots \le \alpha_{\sigma(E)}  \le 1  \Rightarrow 2^E \prod_{e \notin T_{\sigma}} \alpha_e \ge  U(G) \ge
 \prod_{e \notin T_{\sigma}} \alpha_e \; .
\ee

Associated to a Hepp sector we have the set of ``high graphs'' $\gamma_q =\{e_{\sigma(1)} ,\dots e_{\sigma(q)}\}$ formed by the $q$ most ultraviolet edges. The subgraphs are considered amputated, that is they contain all the edges $\gamma_q =\{e_{\sigma(1)} ,\dots e_{\sigma(q)}\}$ and the vertices hooked to them, but not the external edges (which can be either genuine external edges of $G$, or some of the ``lower'' edges $e_{\sigma(q+1)}\dots e_{\sigma(E)}$). We remark that $\gamma_E = G$ and that $\gamma_q$ is not necessarily connected. By construction the leading tree $T_{\sigma}$ is a tree in every connected component of $\gamma_{q}$, and it is the unique tree with this property.

At fixed Hepp sector $\sigma$ we perform the diagonal change of variables:
\be
 \alpha_{\sigma(i)} = \prod_{j=i}^E t_j^2 \;,\qquad \alpha_{\sigma(E) } = t_E^2 \;, \; \; \alpha_{\sigma(E-1)} = t_{E-1}^2t_E^2  \;, \;\; \dots \;\; 
  \alpha_{ \sigma(1) } = t_1^2 t_2^2 \dots t_E^2 \;,
\ee
and we note that all the edges of $\gamma_q$ have a factor $t_q^2$. Now $T_{\sigma}$ is a tree in every connected component of $\gamma_q$, hence the number of edges of $\gamma_q$ not in $T_{\sigma}$ is $E(\gamma_q) - V(\gamma_q) + C(\gamma_q)$ with 
$ V(\gamma_q), E(\gamma_q)$ and $C(\gamma_q)$ the numbers of vertices, edges and connected components of $\gamma_q$.
Every other tree $T$ in $G$ will be a forest (a collection of trees) in some of the connected components of some $\gamma_q$s and the number of edges of such a $\gamma_q$ not belonging to $T$ is strictly larger that $E(\gamma_q) - V(\gamma_q) + C(\gamma_q)$, hence: 
\be
 U(G)\bigg{|}_{ \alpha_{\sigma(i)} = \prod_{j=i}^E t_j^2 } =  \prod_{i=1}^E  t_i^{2[E(\gamma_i) - V(\gamma_i)  + C(\gamma_i) ]  } \big[ 1 + O(t) \big] \;.
\ee
As the change of variables is diagonal we have:
\be
 A^G_{\sigma} = 2^{E} \int_0^1 \left(  \prod_{i=1}^E dt_i \right)
   \prod_{i=1}^E t_i^{ - 1 + 2 i \zeta - d [E(\gamma_i) - V(\gamma_i)  + C(\gamma_i) ] }   \;\frac{1}{ [1 + O(t)]^{d/2} } \;.
\ee
An upper/lower bound is obtained by using  $2^E > 1 + O(t) >1$. Denoting the convergence degree of $\gamma_i$ by 
$ \omega(\gamma_i) = 2 i\zeta - d [E(\gamma_i) - V(\gamma_i)  + C(\gamma_i) ]$, if all the convergence degrees are positive an upper/lower bound is:
\be
  \prod_{i=1}^E \frac{1}{\omega(\gamma_i)}  \le A^G_{\sigma} \le  2^E \; \prod_{i=1}^E \frac{1}{\omega(\gamma_i)}\;.
\ee

In order to conclude it is enough to examine the possible convergence degrees for all the subgraphs of $G$. The number of edges of $\gamma_i$ is exactly $i$ and as we deal with quartic vertices $2i = 4V(\gamma_i) - n(\gamma_i)$ with $n(\gamma_i)$ the number of external half edges of $\gamma_i$. 
Denoting $\gamma_i^{\rho}$ the connected components of $\gamma_i$, with $\rho = 1,\dots C(\gamma_i)$, and observing that the edges, vertices and external legs are distributed among the connected components, the convergence degree of $\gamma_i$ is:
\be
 \omega(\gamma_i) = 2 i\zeta - d [E(\gamma_i) - V(\gamma_i)  + C(\gamma_i) ]
  = i(2\zeta - \frac{d}{2}) + d \sum_{\rho =1}^{C(\gamma_i )}  \frac{n(\gamma_i^{\rho}) -4 }{4}  = \frac{\epsilon}{2} i + 
  d \sum_{\rho =1}^{C(\gamma_i )}  \frac{n(\gamma_i^{\rho}) -4 }{4}   \;,
\ee
where we used $\zeta  = \frac{d+\epsilon}{4}$. So far we have been quite general: we only assumed that there are no subgraphs with zero or negative convergence degree, that is no two-point subgraphs. For the kite graph:
\begin{itemize}
 \item every strict subgraph (i.e. subgraph different from itself) has at least six external half-edges but less that 20 half edges hence $8\epsilon + 4 d \ge \omega(\gamma_i) \ge \frac{d}{2}$.
 \item the wheel itself (corresponding to $t_E$) has $4$ external half edges and 8 edges $\omega(\gamma_E) = 4\epsilon$.
\end{itemize}
Thus (recalling that there are $E!$ sectors) upper and lower bounds are:
\[
\frac{1}{\epsilon} \left( \frac{1}{ 4 (4d+ 8\epsilon)^E }  \right) \le A^G_{\sigma} \le \frac{1}{\epsilon}  \left( \frac{4^{E-1}}{ d^E} \right) \;\;\Rightarrow
\; \;\frac{ k^{ \epsilon- 5 \epsilon } }{\epsilon} \left( \frac{8!}{ 4 (4d+ 8\epsilon)^8 }  \right) \le A^G \le \frac{  k^{ \epsilon- 5 \epsilon } }{\epsilon}  \left( \frac{8! \; 4^{7}}{ d^8}  \right) \; .
\]

\section{Regularized trace of conformal partial waves} 
\label{app:I_eps}
In this appendix we show how to compute $\cI_\eps(J)$ \eqref{Iepsilon}. By homogeneity of the sphere, we can set $z=0$, and factor out  the volume of the d-sphere  $V_d=\int \dd z\, \Omega(z)^d$, given in \eqref{eq:Vol-Sd}. The most generic form of a three-point function $\la \phi \phi O \ra$ is fixed by conformal symmetry as in \eqref{eq:3pt} and we obtain:
\begin{equation}
	\cI_\eps(J)=V_d\int \dd x_1 \dd x_2 \frac{\left(Z^{\m_1} \cdots Z^{\m_J} - \text{``traces"}\right)\left(Z^{\m_1} \cdots Z^{\m_J} - \text{``traces"}\right)}{(1+x_1^2)^{\eps} (1+x_2^2)^{\eps}|x_1|^{d-\eps}|x_2|^{d-\eps}|x_1-x_2|^{d-\eps}} \, ,
\end{equation}
where $Z^{\m} = \f{|x_{1}||x_{2}|}{|x_1-x_2|} \left(\f{x_{1}^{\m}}{|x_{1}|^2} - \f{x_{2}^{\m}}{|x_{2}|^2}\right)$ has unit norm.
In order to take care of the spin structure we use the following identity (e.g.\ \cite{Dolan:2011dv,Costa:2016hju}), which follows from the addition theorem \eqref{eq:additionTh}:
\begin{equation}
\left(Z^{\m_1} \cdots Z^{\m_J} - \text{``traces"}\right)\left(Z^{\m_1} \cdots Z^{\m_J} - \text{``traces"}\right)=\frac{(d-2)_J}{2^l (\frac{d-2}{2})_J}=\frac{\Gamma(d-2+J)\Gamma(\tfrac{d-2}{2})}{2^J\Gamma(d-2)\Gamma(\tfrac{d-2}{2}+J)} \,.
\end{equation}
Then the relation $\mathcal{I}_\eps(J)=\frac{\Gamma(d-2+J)\Gamma(\tfrac{d-2}{2})}{2^J\Gamma(d-2)\Gamma(\tfrac{d-2}{2}+J)} \mathcal{I}_\eps(0)$ holds and it is sufficient to compute the integral at $J=0$.
Now we are ready to deal with the integral 
\begin{equation}
	\frac{\mathcal{I}_\eps(0)}{V_d}=(2a)^{3\epsilon-d}\int \dd x_1 \dd x_2 \frac{1}{(1+x_1^2)^{\eps} (1+x_2^2)^{\eps}|x_1|^{d-\eps}|x_2|^{d-\eps}|x_1-x_2|^{d-\eps}} \, .
\end{equation}
It is convenient to perform the change of variable: $x^\mu=\frac{x_1^\mu}{x_1^2}$ and $y^\mu=\frac{x_2^\mu}{x_2^2}$. The Jacobian determinant of the transformation is $x^{-2d}y^{-2d}$ and the integral simplifies to:
\begin{equation}
\frac{\mathcal{I}_\eps(0)}{V_d}=(2a)^{3\epsilon-d}\int \dd x \dd y \frac{1}{(1+x^2)^{\eps}(1+y^2)^{\eps}(|x-y|^2+\mu^2)^{\frac{d}{2}-\frac{\eps}{2}}} \,,
\end{equation}
where we have introduced another regulator $\mu^2$ that is needed to make the integral convergent.
Assuming $0<\eps<d$, we introduce three Schwinger parameters:
\begin{equation}
	\begin{split}
\frac{\mathcal{I}_\eps(0)}{V_d}=(2a)^{3\epsilon-d}\int \dd x \dd y \frac{1}{\Gamma(\eps)^2\Gamma(\tfrac{d}{2}-\tfrac{\eps}{2})}\int d \alpha_1 d \alpha_2 d \alpha_3  (\alpha_1 \alpha_2)^{\eps-1}\alpha_3^{\tfrac{d}{2}-1-\tfrac{\eps}{2}} \\
 \times e^{-\alpha_1(1+x^2)-\alpha_2(1+y^2)-\alpha_3(x^2+y^2-2xy)-\alpha_3 \mu^2} \, .
	\end{split}
\end{equation}
We can now perform the Gaussian integrals in x and y:
\begin{equation}
\frac{\mathcal{I}_\eps(0)}{V_d}=(2a)^{3\epsilon-d}\frac{\pi^d}{\Gamma(\eps)^2\Gamma(\frac{d}{2}-\frac{\eps}{2})}\int d \alpha_1 d  \alpha_2 d \alpha_3 \frac{(\alpha_1 \alpha_2)^{\eps-1}\alpha_3^{\frac{d}{2}-1-\frac{\eps}{2}}}{(\alpha_1\alpha_2+\alpha_3(\alpha_1+\alpha_2))^{d/2}}e^{-\alpha_1-\alpha_2-\mu^2\alpha_3}\,.
\end{equation}
Using the Mellin-Barnes representation \cite{Benedetti:2020rrq} we find:
\begin{equation}
	\begin{split}
\frac{\mathcal{I}_\eps(0)}{V_d}=\frac{(2a)^{3\epsilon-d}\pi^d}{\Gamma(\eps)^2\Gamma(\frac{d}{2}-\frac{\eps}{2})}\int_{x_0- i \infty}^{x_0+i\infty} \left( \frac{dz}{2\pi i} \right) \int d\alpha_1d \alpha_2 d\alpha_3 e^{-\alpha_1-\alpha_2-\mu^2\alpha_3}\\ \times
\frac{\Gamma(-z)\Gamma(z+\frac{d}{2})(\alpha_1\alpha_2)^{\epsilon+z-1} \alpha_3^{-z-\frac{\eps}{2}-1}}{\Gamma(\tfrac{d}{2})(\alpha_1+\alpha_2)^{z+d/2}}  \,.
	\end{split}
\end{equation}
Next, for  ${\rm Re}(z)<-\eps/2$ we can use the integral representation of the Gamma function to integrate $\alpha_3$.
And if we require also ${\rm Re} (z)>-\eps$ and ${\rm Re} (z)+2\eps>d/2$ the integration over $\alpha_1$ and $\alpha_2$ can be found in \cite{Benedetti:2020rrq}. We finally obtain:
\begin{equation}
\frac{\mathcal{I}_\eps(0)}{V_d}=\frac{(2a)^{3\epsilon-d} \pi^d}{\Gamma(\epsilon)^2\Gamma(\frac{d}{2}-\frac{\eps}{2})} \int_{x_0- i \infty}^{x_0+i\infty} \left( \frac{dz}{2\pi i} \right) \mu^{2z+\epsilon}  \frac{\Gamma (-z) \Gamma \left(\frac{d+2z}{2}\right) \Gamma \left(-\frac{2z+\epsilon }{2}\right) \Gamma \left(z+\epsilon\right)^2 \Gamma \left(\frac{-d+2z+4\epsilon}{2}\right)}{\Gamma \left(\frac{d}{2}\right) \Gamma \left(2 z+2\epsilon\right)} \,.
\end{equation}

Since we have the two conditions ${\rm Re}(z)<-\eps/2$ and ${\rm Re} (z)>-\eps$, we are free to choose $x_0 \in (-\eps,-\eps/2))$. 
Moreover we must take $\eps>d/2$ in order to avoid poles of the last gamma function in this range, but of course we will then analytically continue the result to small $\eps$. Closing the  contour to the right we pick only the poles at $z=n$ and $z=-\eps/2+n$ with $n\in\mathbb{N}_0$.  
Computing the residues at the poles we find the result:
\begin{equation}
	\begin{split}
\frac{\mathcal{I}_\eps(0)}{V_d}= (2a)^{3\epsilon-d}\left(\frac{\pi ^d \Gamma \left(\frac{\epsilon }{2}\right)^3 \Gamma \left(\frac{3 \epsilon }{2}-\frac{d}{2}\right)}{\Gamma \left(\frac{d}{2}\right) \Gamma \left(\epsilon\right)^3}\right)\left(1+ \mathcal{O}(\mu^2)\right) \\
+ \mu^{\eps} (2a)^{3\epsilon-d}\left( \frac{\pi ^d \Gamma \left(-\frac{\epsilon }{2}\right) \Gamma \left(\frac{4\epsilon -d}{2}\right)}{\Gamma \left(2\epsilon\right) \Gamma \left(\frac{d}{2}-\frac{\epsilon }{2}\right)}\right)\left(1+ \mathcal{O}(\mu^2)\right) \, .
	\end{split}
\end{equation}
When we remove the $\mu$ regulator, while keeping $\eps$ finite, only the first term coming from $z=-\eps/2$ survives and we find the final result \eqref{eq:I_eps}.

\section{Large-$J$ expansion} 
\label{app:NumericsLargeJ}

In this appendix we detail how the finite part from \eqref{eq:F-derviative} is extracted. As explained in the main text, in order to regularize the sum over $J$ it is important to consistently shift $\wtD\rightarrow\wtD-\eps$ and $\htilde\rightarrow\htilde-\eps$ everywhere. In particular, it is crucial to introduce $\eps$ in the product of normalization factors \eqref{eq:prod-cN}. At large $J$ this product reduces to:
\be
\cN^{\D}_{h,J}  \cN^{\wtD-\eps}_{\htilde-\eps,J} \sim \f{2^{3(d+\eps)/2+J}}{(2\pi)^{d}} \, J^{-3\eps} \left( 1 + \mathcal{O}(1/J) \right) \,,
\ee
and the factor $J^{-3\eps}$ turns out to suffice to render the sum convergent.
After setting $\Delta=\frac{d}{4}$,  the regularized version of \eqref{eq:F-derviative} reads:
\be 
- g\f{\p}{\p g}F_{\rm NNLO}^{\eps}  =N^2 \sum_{J\in \mathbb{N}_0}  \int_{\f{d}{2}}^{\f{d}{2}+\im\infty}  \f{{\rm d}h}{2\pi\im} \r(h,J) \,\f{k(h,J)^2}{1-k(h,J)}\, \cN^{\frac{d}{4}}_{h,J}  \cN^{\frac{3d}{4}-\eps}_{\htilde-\eps,J} \, \mathcal{I}_\eps(J)  \,,
\ee
where $\rho(h,J)$, $k(h,J)$, $\cN^{\D}_{h,J}$ and $\mathcal{I}_\eps(J)$ are all ratios of gamma functions.  

Integrating numerically each term at fixed $J$, and using standard convergence tests, one finds that the resulting series is divergent at $\eps=0$. 
In order to isolate and subtract the divergence, we need to identify the asymptotic behavior at large $J$. A naive expansion of the integrand in $1/J$ leads however to a divergent integral over $h$, due to the exchange of limit and integral, and a more careful analysis is needed.
It turns out to be convenient to make the change of variable $h=d/2 + \im \a J$ in the integral, after which we can use Stirling formula on the gamma functions with $J$ in the argument. 
With this procedure we find the following asymptotic behavior in $J$: 
\begin{equation} \label{eq:J-asympt}
	- \f{1}{N^2} g\f{\p}{\p g}F_{\rm NNLO}^{\eps} \to \sum_{J\in \mathbb{N}_+} \frac{1}{J^{1+3\eps}} \int_{0}^{+\infty} {\rm d}\a\, F(\a,\eps)=\zeta(1+\eps)f(\eps) \,,
\end{equation}
hence the series has a simple pole at $\eps=0$. 
The precise expressions of $F(\a,\eps)$ and $f(\eps)$ are:
\begin{align}
	&F(\a,\eps)=\frac{\pi ^{-\frac{d}{2}} \, g^4 \, 2^{-2 d+\frac{9 \epsilon }{2}-1} \,a^{3 \epsilon} \, \Gamma \left(\frac{d-2}{2}\right) \Gamma \left(\frac{\epsilon }{2}\right)^3   \Gamma \left(\frac{3 \epsilon }{2}-\frac{d}{2}\right)}{\Gamma (d-2) \Gamma \left(d\right) \Gamma (\epsilon )^3} \, \a^{d-2} \left(\a^2+1\right)^{1-d-\frac{3 \epsilon }{2}}\,,\\
	&f(\eps)=\frac{\pi ^{2-\frac{d}{2}} \, g^4 \, 2^{-3 d+\frac{3 \epsilon }{2}+4} \, a^{3 \epsilon}\, \Gamma \left(\frac{3 \epsilon }{2}-\frac{d}{2}\right) \Gamma \left(\frac{1}{2} (d+3 \epsilon -1)\right)}{\Gamma \left(d\right) \Gamma \left(\frac{\epsilon +1}{2}\right)^3 \Gamma \left(d+\frac{3 \epsilon }{2}-1\right)} \,,
\end{align}
and they are both analytic functions at $\eps=0$ (for $2<d<4$).
For numerical computation, it is convenient to add and subtract the asymptotic contribution \eqref{eq:J-asympt} to the original series, and write:
\be
\begin{split} \label{RegularizedSum}
	- g\f{\p}{\p g}F_{\rm NNLO}^{\eps}  =  &\,  N^2 \int_{\f{d}{2}}^{\f{d}{2}+\im\infty}  \f{{\rm d}h}{2\pi\im} \r(h,0) \,\f{k(h,0)^2}{1-k(h,0)}\, \cN^{\D}_{h,0}  \cN^{\wtD-\eps}_{\htilde-\eps,0} \, \cI_\eps (0) \\
	&+ N^2  \sum_{J\in \mathbb{N}_+} \Bigg( \int_{\f{d}{2}}^{\f{d}{2}+\im\infty}  \f{{\rm d}h}{2\pi\im}  \r(h,J) \,\f{k(h,J)^2}{1-k(h,J)}\, \cN^{\D}_{h,J}  \cN^{\wtD-\eps}_{\htilde-\eps,J} \, \cI_\eps (J) -\frac{f(\eps)}{J^{1+3\epsilon}}\Bigg) \\
	&  +N^2 \zeta(1+\eps)f(\eps)  \,,
\end{split}
\ee
where in the first term we isolated the $J=0$ contribution. The sum in the second line is now convergent for $\eps=0$, hence it can be computed numerically.
The last term expands in $\epsilon$ as:
\begin{equation}
	N^2f(\epsilon)\zeta(1+3\epsilon)=N^2\Big(\frac{f(0)}{3\epsilon}+\f13 f'(0)+\gamma f(0) + \mathcal{O}(\epsilon) \Big) \, ,
\end{equation}
and after subtracting the pure pole part, it yields an additional finite contribution. Overall we get the derivative of the sphere free energy:
\be
- g\f{\p}{\p g}F_{\rm NNLO} = \lim_{\eps\to 0} \left(- g\f{\p}{\p g}F_{\rm NNLO}^{\eps} - \frac{N^2 f(0)}{3\epsilon} \right) \, ,
\ee
which can be evaluated numerically. 
For example, for $d=3$, $g=1$ and $a=1$, we find:
\be
- g\f{\p}{\p g}F_{\rm NNLO}   = 7.57 \times 10^{-4} \, N^2 \, .
\ee




\bibliographystyle{JHEP}
\bibliography{Refs-TMV,Refs-CFT,Refs-QFT} 

\providecommand{\href}[2]{#2}\begingroup\raggedright\begin{thebibliography}{10}

\bibitem{Zamolodchikov:1986gt}
A.~B. Zamolodchikov, \emph{{Irreversibility of the Flux of the Renormalization
  Group in a 2D Field Theory}}, {\emph{JETP Lett.} {\bfseries 43} (1986) 730}.

\bibitem{Komargodski:2011vj}
Z.~Komargodski and A.~Schwimmer, \emph{{On Renormalization Group Flows in Four
  Dimensions}}, \href{https://doi.org/10.1007/JHEP12(2011)099}{\emph{JHEP}
  {\bfseries 12} (2011) 099} [\href{https://arxiv.org/abs/1107.3987}{{\ttfamily
  1107.3987}}].

\bibitem{Jafferis:2011zi}
D.~L. Jafferis, I.~R. Klebanov, S.~S. Pufu and B.~R. Safdi, \emph{{Towards the
  F-Theorem: N=2 Field Theories on the Three-Sphere}},
  \href{https://doi.org/10.1007/JHEP06(2011)102}{\emph{JHEP} {\bfseries 06}
  (2011) 102} [\href{https://arxiv.org/abs/1103.1181}{{\ttfamily 1103.1181}}].

\bibitem{Klebanov:2011gs}
I.~R. Klebanov, S.~S. Pufu and B.~R. Safdi, \emph{{F-Theorem without
  Supersymmetry}}, \href{https://doi.org/10.1007/JHEP10(2011)038}{\emph{JHEP}
  {\bfseries 10} (2011) 038} [\href{https://arxiv.org/abs/1105.4598}{{\ttfamily
  1105.4598}}].

\bibitem{Cardy:1988cwa}
J.~L. Cardy, \emph{{Is There a c Theorem in Four-Dimensions?}},
  \href{https://doi.org/10.1016/0370-2693(88)90054-8}{\emph{Phys. Lett. B}
  {\bfseries 215} (1988) 749}.

\bibitem{Komargodski:2011xv}
Z.~Komargodski, \emph{{The Constraints of Conformal Symmetry on RG Flows}},
  \href{https://doi.org/10.1007/JHEP07(2012)069}{\emph{JHEP} {\bfseries 07}
  (2012) 069} [\href{https://arxiv.org/abs/1112.4538}{{\ttfamily 1112.4538}}].

\bibitem{Jafferis:2010un}
D.~L. Jafferis, \emph{{The Exact Superconformal R-Symmetry Extremizes Z}},
  \href{https://doi.org/10.1007/JHEP05(2012)159}{\emph{JHEP} {\bfseries 05}
  (2012) 159} [\href{https://arxiv.org/abs/1012.3210}{{\ttfamily 1012.3210}}].

\bibitem{Pufu:2016zxm}
S.~S. Pufu, \emph{{The F-Theorem and F-Maximization}},
  \href{https://doi.org/10.1088/1751-8121/aa6765}{\emph{J. Phys. A} {\bfseries
  50} (2017) 443008} [\href{https://arxiv.org/abs/1608.02960}{{\ttfamily
  1608.02960}}].

\bibitem{Casini:2012ei}
H.~Casini and M.~Huerta, \emph{{On the RG running of the entanglement entropy
  of a circle}}, \href{https://doi.org/10.1103/PhysRevD.85.125016}{\emph{Phys.
  Rev. D} {\bfseries 85} (2012) 125016}
  [\href{https://arxiv.org/abs/1202.5650}{{\ttfamily 1202.5650}}].

\bibitem{Casini:2011kv}
H.~Casini, M.~Huerta and R.~C. Myers, \emph{{Towards a derivation of
  holographic entanglement entropy}},
  \href{https://doi.org/10.1007/JHEP05(2011)036}{\emph{JHEP} {\bfseries 05}
  (2011) 036} [\href{https://arxiv.org/abs/1102.0440}{{\ttfamily 1102.0440}}].

\bibitem{Casini:2004bw}
H.~Casini and M.~Huerta, \emph{{A Finite entanglement entropy and the
  c-theorem}},
  \href{https://doi.org/10.1016/j.physletb.2004.08.072}{\emph{Phys. Lett. B}
  {\bfseries 600} (2004) 142}
  [\href{https://arxiv.org/abs/hep-th/0405111}{{\ttfamily hep-th/0405111}}].

\bibitem{Casini:2017vbe}
H.~Casini, E.~Test\'e and G.~Torroba, \emph{{Markov Property of the Conformal
  Field Theory Vacuum and the a Theorem}},
  \href{https://doi.org/10.1103/PhysRevLett.118.261602}{\emph{Phys. Rev. Lett.}
  {\bfseries 118} (2017) 261602}
  [\href{https://arxiv.org/abs/1704.01870}{{\ttfamily 1704.01870}}].

\bibitem{Giombi:2014xxa}
S.~Giombi and I.~R. Klebanov, \emph{{Interpolating between $a$ and $F$}},
  \href{https://doi.org/10.1007/JHEP03(2015)117}{\emph{JHEP} {\bfseries 03}
  (2015) 117} [\href{https://arxiv.org/abs/1409.1937}{{\ttfamily 1409.1937}}].

\bibitem{Fei:2015oha}
L.~Fei, S.~Giombi, I.~R. Klebanov and G.~Tarnopolsky, \emph{{Generalized
  $F$-Theorem and the $\epsilon$ Expansion}},
  \href{https://doi.org/10.1007/JHEP12(2015)155}{\emph{JHEP} {\bfseries 12}
  (2015) 155} [\href{https://arxiv.org/abs/1507.01960}{{\ttfamily
  1507.01960}}].

\bibitem{Giombi:2015haa}
S.~Giombi, I.~R. Klebanov and G.~Tarnopolsky, \emph{{Conformal QED$_d$,
  $F$-Theorem and the $\epsilon$ Expansion}},
  \href{https://doi.org/10.1088/1751-8113/49/13/135403}{\emph{J. Phys. A}
  {\bfseries 49} (2016) 135403}
  [\href{https://arxiv.org/abs/1508.06354}{{\ttfamily 1508.06354}}].

\bibitem{Hogervorst:2015akt}
M.~Hogervorst, S.~Rychkov and B.~C. van Rees, \emph{{Unitarity violation at the
  Wilson-Fisher fixed point in 4-$\epsilon$ dimensions}},
  \href{https://doi.org/10.1103/PhysRevD.93.125025}{\emph{Phys. Rev. D}
  {\bfseries 93} (2016) 125025}
  [\href{https://arxiv.org/abs/1512.00013}{{\ttfamily 1512.00013}}].

\bibitem{Paulos:2015jfa}
M.~F. Paulos, S.~Rychkov, B.~C. van Rees and B.~Zan, \emph{Conformal invariance
  in the long-range {Ising} model},
  \href{https://doi.org/10.1016/j.nuclphysb.2015.10.018}{\emph{Nucl.Phys.B}
  {\bfseries 902} (2016) 246}
  [\href{https://arxiv.org/abs/1509.00008}{{\ttfamily 1509.00008}}].

\bibitem{Gaiotto:2014gha}
D.~Gaiotto, \emph{{Boundary F-maximization}},
  \href{https://arxiv.org/abs/1403.8052}{{\ttfamily 1403.8052}}.

\bibitem{Kobayashi:2018lil}
N.~Kobayashi, T.~Nishioka, Y.~Sato and K.~Watanabe, \emph{{Towards a
  $C$-theorem in defect CFT}},
  \href{https://doi.org/10.1007/JHEP01(2019)039}{\emph{JHEP} {\bfseries 01}
  (2019) 039} [\href{https://arxiv.org/abs/1810.06995}{{\ttfamily
  1810.06995}}].

\bibitem{Maldacena:2011jn}
J.~Maldacena and A.~Zhiboedov, \emph{{Constraining Conformal Field Theories
  with A Higher Spin Symmetry}},
  \href{https://doi.org/10.1088/1751-8113/46/21/214011}{\emph{J. Phys. A}
  {\bfseries 46} (2013) 214011}
  [\href{https://arxiv.org/abs/1112.1016}{{\ttfamily 1112.1016}}].

\bibitem{Binder:2019zqc}
D.~J. Binder and S.~Rychkov, \emph{{Deligne Categories in Lattice Models and
  Quantum Field Theory, or Making Sense of $O(N)$ Symmetry with Non-integer
  $N$}}, \href{https://doi.org/10.1007/JHEP04(2020)117}{\emph{JHEP} {\bfseries
  04} (2020) 117} [\href{https://arxiv.org/abs/1911.07895}{{\ttfamily
  1911.07895}}].

\bibitem{Benedetti:2019eyl}
D.~Benedetti, R.~Gurau and S.~Harribey, \emph{{Line of fixed points in a
  bosonic tensor model}},
  \href{https://doi.org/10.1007/JHEP06(2019)053}{\emph{JHEP} {\bfseries 06}
  (2019) 053} [\href{https://arxiv.org/abs/1903.03578}{{\ttfamily
  1903.03578}}].

\bibitem{Klebanov:2016xxf}
I.~R. Klebanov and G.~Tarnopolsky, \emph{Uncolored random tensors, melon
  diagrams, and the {SYK} models},
  \href{https://doi.org/10.1103/PhysRevD.95.046004}{\emph{Phys. Rev.}
  {\bfseries D95} (2017) 046004}
  [\href{https://arxiv.org/abs/1611.08915}{{\ttfamily 1611.08915}}].

\bibitem{Gubser:2017qed}
S.~S. Gubser, M.~Heydeman, C.~Jepsen, S.~Parikh, I.~Saberi, B.~Stoica et~al.,
  \emph{{Melonic theories over diverse number systems}},
  \href{https://doi.org/10.1103/PhysRevD.98.126007}{\emph{Phys.\ Rev.\ D}
  {\bfseries 98} (2018) 126007}
  [\href{https://arxiv.org/abs/1707.01087}{{\ttfamily 1707.01087}}].

\bibitem{Giombi:2017dtl}
S.~Giombi, I.~R. Klebanov and G.~Tarnopolsky, \emph{{Bosonic tensor models at
  large {$N$} and small {$\epsilon$}}},
  \href{https://doi.org/10.1103/PhysRevD.96.106014}{\emph{Phys. Rev.}
  {\bfseries D96} (2017) 106014}
  [\href{https://arxiv.org/abs/1707.03866}{{\ttfamily 1707.03866}}].

\bibitem{Bulycheva:2017ilt}
K.~Bulycheva, I.~R. Klebanov, A.~Milekhin and G.~Tarnopolsky, \emph{Spectra of
  operators in large {$N$} tensor models},
  \href{https://doi.org/10.1103/PhysRevD.97.026016}{\emph{Phys.\ Rev.\ D}
  {\bfseries 97} (2018) 026016}
  [\href{https://arxiv.org/abs/1707.09347}{{\ttfamily 1707.09347}}].

\bibitem{Prakash:2017hwq}
S.~Prakash and R.~Sinha, \emph{A complex fermionic tensor model in $d$
  dimensions}, \href{https://doi.org/10.1007/JHEP02(2018)086}{\emph{JHEP}
  {\bfseries 02} (2018) 086}
  [\href{https://arxiv.org/abs/1710.09357}{{\ttfamily 1710.09357}}].

\bibitem{Giombi:2018qgp}
S.~Giombi, I.~R. Klebanov, F.~Popov, S.~Prakash and G.~Tarnopolsky,
  \emph{Prismatic large {$N$} models for bosonic tensors},
  \href{https://doi.org/10.1103/PhysRevD.98.105005}{\emph{Phys. Rev.}
  {\bfseries D98} (2018) 105005}
  [\href{https://arxiv.org/abs/1808.04344}{{\ttfamily 1808.04344}}].

\bibitem{Benedetti:2019rja}
D.~Benedetti, N.~Delporte, S.~Harribey and R.~Sinha, \emph{{Sextic tensor field
  theories in rank $3$ and $5$}},
  \href{https://doi.org/10.1007/JHEP06(2020)065}{\emph{JHEP} {\bfseries 06}
  (2020) 065} [\href{https://arxiv.org/abs/1912.06641}{{\ttfamily
  1912.06641}}].

\bibitem{Chang:2018sve}
C.-M. Chang, S.~Colin-Ellerin and M.~Rangamani, \emph{{On Melonic Supertensor
  Models}}, \href{https://doi.org/10.1007/JHEP10(2018)157}{\emph{JHEP}
  {\bfseries 10} (2018) 157}
  [\href{https://arxiv.org/abs/1806.09903}{{\ttfamily 1806.09903}}].

\bibitem{Popov:2019nja}
F.~K. Popov, \emph{{Supersymmetric tensor model at large $N$ and small
  $\epsilon$}}, \href{https://doi.org/10.1103/PhysRevD.101.026020}{\emph{Phys.\
  Rev.\ D} {\bfseries 101} (2020) 026020}
  [\href{https://arxiv.org/abs/1907.02440}{{\ttfamily 1907.02440}}].

\bibitem{Lettera:2020uay}
D.~Lettera and A.~Vichi, \emph{{A large-$N$ tensor model with four
  supercharges}},  \href{https://arxiv.org/abs/2012.11600}{{\ttfamily
  2012.11600}}.

\bibitem{Benedetti:2020iku}
D.~Benedetti and N.~Delporte, \emph{{Remarks on a melonic field theory with
  cubic interaction}},
  \href{https://doi.org/10.1007/JHEP04(2021)197}{\emph{JHEP} {\bfseries 04}
  (2021) 197} [\href{https://arxiv.org/abs/2012.12238}{{\ttfamily
  2012.12238}}].

\bibitem{Klebanov:2018fzb}
I.~R. Klebanov, F.~Popov and G.~Tarnopolsky, \emph{{TASI} lectures on large
  {$N$} tensor models}, \href{https://doi.org/10.22323/1.305.0004}{\emph{PoS}
  {\bfseries TASI2017} (2018) 004}
  [\href{https://arxiv.org/abs/1808.09434}{{\ttfamily 1808.09434}}].

\bibitem{Benedetti:2020seh}
D.~Benedetti, \emph{{Melonic CFTs}},
  \href{https://doi.org/10.22323/1.376.0168}{\emph{PoS} {\bfseries CORFU2019}
  (2020) 168} [\href{https://arxiv.org/abs/2004.08616}{{\ttfamily
  2004.08616}}].

\bibitem{Benedetti:2020yvb}
D.~Benedetti, R.~Gurau and K.~Suzuki, \emph{{Conformal symmetry and composite
  operators in the $O(N)^{3}$ tensor field theory}},
  \href{https://doi.org/10.1007/JHEP06(2020)113}{\emph{JHEP} {\bfseries 06}
  (2020) 113} [\href{https://arxiv.org/abs/2002.07652}{{\ttfamily
  2002.07652}}].

\bibitem{Benedetti:2019ikb}
D.~Benedetti, R.~Gurau, S.~Harribey and K.~Suzuki, \emph{{Hints of unitarity at
  large $N$ in the $O(N)^3$ tensor field theory}},
  \href{https://doi.org/10.1007/JHEP02(2020)072}{\emph{JHEP} {\bfseries 02}
  (2020) 072} [\href{https://arxiv.org/abs/1909.07767}{{\ttfamily
  1909.07767}}].

\bibitem{Benedetti:2020sye}
D.~Benedetti, R.~Gurau and S.~Harribey, \emph{{The tri-fundamental quartic
  model}}, \href{https://doi.org/10.1103/PhysRevD.103.046018}{\emph{Phys. Rev.
  D} {\bfseries 103} (2021) 046018}
  [\href{https://arxiv.org/abs/2011.11276}{{\ttfamily 2011.11276}}].

\bibitem{Gubser:2002vv}
S.~S. Gubser and I.~R. Klebanov, \emph{{A Universal result on central charges
  in the presence of double trace deformations}},
  \href{https://doi.org/10.1016/S0550-3213(03)00056-7}{\emph{Nucl. Phys. B}
  {\bfseries 656} (2003) 23}
  [\href{https://arxiv.org/abs/hep-th/0212138}{{\ttfamily hep-th/0212138}}].

\bibitem{Sachdev:1992fk}
S.~Sachdev and J.~Ye, \emph{{Gapless spin fluid ground state in a random,
  quantum Heisenberg magnet}},
  \href{https://doi.org/10.1103/PhysRevLett.70.3339}{\emph{Phys. Rev. Lett.}
  {\bfseries 70} (1993) 3339}
  [\href{https://arxiv.org/abs/cond-mat/9212030}{{\ttfamily
  cond-mat/9212030}}].

\bibitem{Kitaev}
A.~Kitaev, \emph{{A simple model of quantum holography}}, {\emph{KITP strings
  seminar and Entanglement 2015} (Feb. 12, April 7, and May 27, 2015) }.

\bibitem{Maldacena:2016hyu}
J.~Maldacena and D.~Stanford, \emph{{Remarks on the Sachdev-Ye-Kitaev model}},
  \href{https://doi.org/10.1103/PhysRevD.94.106002}{\emph{Phys. Rev.}
  {\bfseries D94} (2016) 106002}
  [\href{https://arxiv.org/abs/1604.07818}{{\ttfamily 1604.07818}}].

\bibitem{Gross:2017vhb}
D.~J. Gross and V.~Rosenhaus, \emph{{A line of CFTs: from generalized free
  fields to SYK}}, \href{https://doi.org/10.1007/JHEP07(2017)086}{\emph{JHEP}
  {\bfseries 07} (2017) 086}
  [\href{https://arxiv.org/abs/1706.07015}{{\ttfamily 1706.07015}}].

\bibitem{Behan:2017emf}
C.~Behan, L.~Rastelli, S.~Rychkov and B.~Zan, \emph{A scaling theory for the
  long-range to short-range crossover and an infrared duality},
  \href{https://doi.org/10.1088/1751-8121/aa8099}{\emph{J.Phys.A} {\bfseries
  50} (2017) 354002} [\href{https://arxiv.org/abs/1703.05325}{{\ttfamily
  1703.05325}}].

\bibitem{Campa:2009rev}
A.~Campa, T.~Dauxois and S.~Ruffo, \emph{Statistical mechanics and dynamics of
  solvable models with long-range interactions},
  \href{https://doi.org/10.1016/j.physrep.2009.07.001}{\emph{Physics Reports}
  {\bfseries 480} (2009) 57} [\href{https://arxiv.org/abs/0907.0323}{{\ttfamily
  0907.0323}}].

\bibitem{Diaz:2007an}
D.~E. Diaz and H.~Dorn, \emph{{Partition functions and double-trace
  deformations in AdS/CFT}},
  \href{https://doi.org/10.1088/1126-6708/2007/05/046}{\emph{JHEP} {\bfseries
  05} (2007) 046} [\href{https://arxiv.org/abs/hep-th/0702163}{{\ttfamily
  hep-th/0702163}}].

\bibitem{Benedetti:2018goh}
D.~Benedetti and R.~Gurau, \emph{{2PI effective action for the SYK model and
  tensor field theories}},
  \href{https://doi.org/10.1007/JHEP05(2018)156}{\emph{JHEP} {\bfseries 05}
  (2018) 156} [\href{https://arxiv.org/abs/1802.05500}{{\ttfamily
  1802.05500}}].

\bibitem{Berges:2004yj}
J.~Berges, \emph{{Introduction to nonequilibrium quantum field theory}},
  \href{https://doi.org/10.1063/1.1843591}{\emph{AIP Conf. Proc.} {\bfseries
  739} (2005) 3} [\href{https://arxiv.org/abs/hep-ph/0409233}{{\ttfamily
  hep-ph/0409233}}].

\bibitem{Bonzom:2011zz}
V.~Bonzom, R.~Gurau, A.~Riello and V.~Rivasseau, \emph{{Critical behavior of
  colored tensor models in the large N limit}},
  \href{https://doi.org/10.1016/j.nuclphysb.2011.07.022}{\emph{Nucl.\ Phys.\ B}
  {\bfseries 853} (2011) 174}
  [\href{https://arxiv.org/abs/1105.3122}{{\ttfamily 1105.3122}}].

\bibitem{Carrozza:2015adg}
S.~Carrozza and A.~Tanasa, \emph{{$O(N)$} random tensor models},
  \href{https://doi.org/10.1007/s11005-016-0879-x}{\emph{Lett. Math. Phys.}
  {\bfseries 106} (2016) 1531}
  [\href{https://arxiv.org/abs/1512.06718}{{\ttfamily 1512.06718}}].

\bibitem{Bonzom:2019yik}
V.~Bonzom, V.~Nador and A.~Tanasa, \emph{{Diagrammatics of the quartic
  $O(N)^3$-invariant Sachdev-Ye-Kitaev-like tensor model}},
  \href{https://doi.org/10.1063/1.5095248}{\emph{J. Math. Phys.} {\bfseries 60}
  (2019) 072302} [\href{https://arxiv.org/abs/1903.01723}{{\ttfamily
  1903.01723}}].

\bibitem{Grabner:2017pgm}
D.~Grabner, N.~Gromov, V.~Kazakov and G.~Korchemsky, \emph{Strongly
  $\gamma$-deformed $\mathcal{N}=4$ supersymmetric yang-mills theory as an
  integrable conformal field theory},
  \href{https://doi.org/10.1103/PhysRevLett.120.111601}{\emph{Phys.Rev.Lett.}
  {\bfseries 120} (2018) 111601}
  [\href{https://arxiv.org/abs/1711.04786}{{\ttfamily 1711.04786}}].

\bibitem{Kazakov:2018qbr}
V.~Kazakov and E.~Olivucci, \emph{Biscalar integrable conformal field theories
  in any dimension},
  \href{https://doi.org/10.1103/PhysRevLett.121.131601}{\emph{Phys.Rev.Lett.}
  {\bfseries 121} (2018) 131601}
  [\href{https://arxiv.org/abs/1801.09844}{{\ttfamily 1801.09844}}].

\bibitem{Karateev:2018oml}
D.~Karateev, P.~Kravchuk and D.~Simmons-Duffin, \emph{{Harmonic Analysis and
  Mean Field Theory}},
  \href{https://doi.org/10.1007/JHEP10(2019)217}{\emph{JHEP} {\bfseries 10}
  (2019) 217} [\href{https://arxiv.org/abs/1809.05111}{{\ttfamily
  1809.05111}}].

\bibitem{Rubin:1984}
M.~A. Rubin and C.~R. Ordonez, \emph{{Eigenvalues and degeneracies for
  $n$-dimensional tensor spherical harmonics}},
  \href{https://doi.org/10.1063/1.526034}{\emph{Journal of Mathematical
  Physics} {\bfseries 25} (1984) 2888}.

\bibitem{Samko-book}
S.~Samko, \emph{Hypersingular Integrals and Their Applications}. CRC Press,
  London, 1~ed., 2002,
  \href{https://doi.org/10.1201/9781482264968}{10.1201/9781482264968}.

\bibitem{Farnsworth:2017tbz}
K.~Farnsworth, M.~A. Luty and V.~Prilepina, \emph{{Weyl versus Conformal
  Invariance in Quantum Field Theory}},
  \href{https://doi.org/10.1007/JHEP10(2017)170}{\emph{JHEP} {\bfseries 10}
  (2017) 170} [\href{https://arxiv.org/abs/1702.07079}{{\ttfamily
  1702.07079}}].

\bibitem{Allen:1985wd}
B.~Allen and T.~Jacobson, \emph{{Vector Two Point Functions in Maximally
  Symmetric Spaces}}, \href{https://doi.org/10.1007/BF01211169}{\emph{Commun.
  Math. Phys.} {\bfseries 103} (1986) 669}.

\bibitem{Dowker:1975tf}
J.~S. Dowker and R.~Critchley, \emph{{Effective Lagrangian and Energy Momentum
  Tensor in de Sitter Space}},
  \href{https://doi.org/10.1103/PhysRevD.13.3224}{\emph{Phys. Rev. D}
  {\bfseries 13} (1976) 3224}.

\bibitem{Kwasnicki:2017}
M.~Kwasnicki, \emph{{Ten equivalent definitions of the fractional Laplace
  operator}}, {\emph{Fractional Calculus and Applied Analysis} {\bfseries 20}
  (2017) 51 } [\href{https://arxiv.org/abs/1507.07356}{{\ttfamily
  1507.07356}}].

\bibitem{Stinga:2009}
P.~Stinga and J.~Torrea, \emph{Extension problem and harnack's inequality for
  some fractional operators}, {\emph{Communications in Partial Differential
  Equations} {\bfseries 35} (2009) 2092 }
  [\href{https://arxiv.org/abs/0910.2569}{{\ttfamily 0910.2569}}].

\bibitem{Ferrara:1972uq}
S.~Ferrara, A.~Grillo, G.~Parisi and R.~Gatto, \emph{{The shadow operator
  formalism for conformal algebra. Vacuum expectation values and operator
  products}}, \href{https://doi.org/10.1007/BF02907130}{\emph{Lett. Nuovo Cim.}
  {\bfseries 4S2} (1972) 115}.

\bibitem{Samko:2003}
S.~Samko, \emph{On inversion of fractional spherical potentials by spherical
  hypersingular operators},  in \emph{Singular Integral Operators,
  Factorization and Applications}, A.~B{\"o}ttcher, M.~A. Kaashoek, A.~B.
  Lebre, A.~F. dos Santos and F.-O. Speck, eds., (Basel), pp.~357--368,
  Birkh{\"a}user Basel, 2003,
  \href{https://doi.org/10.1007/978-3-0348-8007-7_19}{DOI}.

\bibitem{Branson:1997}
T.~Branson, \emph{Spectral theory of invariant operators, sharp inequalities,
  and representation theory},  in \emph{Proceedings of the 16th Winter School
  "Geometry and Physics"}, pp.~[29]--54, Circolo Matematico di Palermo, 1997,
  \href{http://eudml.org/doc/220067}{http://eudml.org/doc/220067}.

\bibitem{gonzalez2016recent}
M.~del Mar~Gonzalez, \emph{{Recent progress on the fractional Laplacian in
  conformal geometry}},  \href{https://arxiv.org/abs/1609.08988}{{\ttfamily
  1609.08988}}.

\bibitem{Branson:1995}
T.~P. Branson, \emph{Sharp inequalities, the functional determinant, and the
  complementary series}, {\emph{Transactions of the American Mathematical
  Society} {\bfseries 347} (1995) 3671}.

\bibitem{PavSam84}
P.~M. Pavlov and S.~G. Samko, \emph{Description of spaces $l^\alpha_p(s_{n-1})$
  in terms of spherical hypersingular integrals}, {\emph{Dokl. Akad. Nauk SSSR}
  {\bfseries 276} (1984) 546}.

\bibitem{Dobrev:1976vr}
V.~Dobrev, G.~Mack, I.~Todorov, V.~Petkova and S.~Petrova, \emph{{On the
  Clebsch-Gordan Expansion for the Lorentz Group in n Dimensions}},
  \href{https://doi.org/10.1016/0034-4877(76)90057-4}{\emph{Rept. Math. Phys.}
  {\bfseries 9} (1976) 219}.

\bibitem{Dobrev:1975ru}
V.~Dobrev, V.~Petkova, S.~Petrova and I.~Todorov, \emph{{Dynamical Derivation
  of Vacuum Operator Product Expansion in Euclidean Conformal Quantum Field
  Theory}}, \href{https://doi.org/10.1103/PhysRevD.13.887}{\emph{Phys. Rev. D}
  {\bfseries 13} (1976) 887}.

\bibitem{Dobrev:1977qv}
V.~Dobrev, G.~Mack, V.~Petkova, S.~Petrova and I.~Todorov, \emph{{Harmonic
  Analysis on the n-Dimensional Lorentz Group and Its Application to Conformal
  Quantum Field Theory}}, vol.~63. Springer, Berlin, Heidelberg, 1977,
  \href{https://doi.org/10.1007/BFb0009678}{10.1007/BFb0009678}.

\bibitem{Caron-Huot:2017vep}
S.~Caron-Huot, \emph{{Analyticity in Spin in Conformal Theories}},
  \href{https://doi.org/10.1007/JHEP09(2017)078}{\emph{JHEP} {\bfseries 09}
  (2017) 078} [\href{https://arxiv.org/abs/1703.00278}{{\ttfamily
  1703.00278}}].

\bibitem{Simmons-Duffin:2017nub}
D.~Simmons-Duffin, D.~Stanford and E.~Witten, \emph{{A spacetime derivation of
  the Lorentzian OPE inversion formula}},
  \href{https://doi.org/10.1007/JHEP07(2018)085}{\emph{JHEP} {\bfseries 07}
  (2018) 085} [\href{https://arxiv.org/abs/1711.03816}{{\ttfamily
  1711.03816}}].

\bibitem{Liu:2018jhs}
J.~Liu, E.~Perlmutter, V.~Rosenhaus and D.~Simmons-Duffin,
  \emph{{$d$-dimensional SYK, AdS Loops, and $6j$ Symbols}},
  \href{https://doi.org/10.1007/JHEP03(2019)052}{\emph{JHEP} {\bfseries 03}
  (2019) 052} [\href{https://arxiv.org/abs/1808.00612}{{\ttfamily
  1808.00612}}].

\bibitem{Mack:2009mi}
G.~Mack, \emph{{D-independent representation of Conformal Field Theories in D
  dimensions via transformation to auxiliary Dual Resonance Models. Scalar
  amplitudes}},  \href{https://arxiv.org/abs/0907.2407}{{\ttfamily 0907.2407}}.

\bibitem{Costa:2012cb}
M.~S. Costa, V.~Goncalves and J.~Penedones, \emph{{Conformal Regge theory}},
  \href{https://doi.org/10.1007/JHEP12(2012)091}{\emph{JHEP} {\bfseries 12}
  (2012) 091} [\href{https://arxiv.org/abs/1209.4355}{{\ttfamily 1209.4355}}].

\bibitem{Murugan:2017eto}
J.~Murugan, D.~Stanford and E.~Witten, \emph{{More on Supersymmetric and 2d
  Analogs of the SYK Model}},
  \href{https://doi.org/10.1007/JHEP08(2017)146}{\emph{JHEP} {\bfseries 08}
  (2017) 146} [\href{https://arxiv.org/abs/1706.05362}{{\ttfamily
  1706.05362}}].

\bibitem{Gadde:2017sjg}
A.~Gadde, \emph{{In search of conformal theories}},
  \href{https://arxiv.org/abs/1702.07362}{{\ttfamily 1702.07362}}.

\bibitem{Hogervorst:2017sfd}
M.~Hogervorst and B.~C. van Rees, \emph{{Crossing symmetry in alpha space}},
  \href{https://doi.org/10.1007/JHEP11(2017)193}{\emph{JHEP} {\bfseries 11}
  (2017) 193} [\href{https://arxiv.org/abs/1702.08471}{{\ttfamily
  1702.08471}}].

\bibitem{Sleight:2018ryu}
C.~Sleight and M.~Taronna, \emph{{Anomalous Dimensions from Crossing Kernels}},
  \href{https://doi.org/10.1007/JHEP11(2018)089}{\emph{JHEP} {\bfseries 11}
  (2018) 089} [\href{https://arxiv.org/abs/1807.05941}{{\ttfamily
  1807.05941}}].

\bibitem{Sleight:2018epi}
C.~Sleight and M.~Taronna, \emph{{Spinning Mellin Bootstrap: Conformal Partial
  Waves, Crossing Kernels and Applications}},
  \href{https://doi.org/10.1002/prop.201800038}{\emph{Fortsch. Phys.}
  {\bfseries 66} (2018) 1800038}
  [\href{https://arxiv.org/abs/1804.09334}{{\ttfamily 1804.09334}}].

\bibitem{deMelloKoch:2018ivk}
R.~de~Mello~Koch, A.~Jevicki, K.~Suzuki and J.~Yoon, \emph{{AdS Maps and
  Diagrams of Bi-local Holography}},
  \href{https://doi.org/10.1007/JHEP03(2019)133}{\emph{JHEP} {\bfseries 03}
  (2019) 133} [\href{https://arxiv.org/abs/1810.02332}{{\ttfamily
  1810.02332}}].

\bibitem{Aharony:2020omh}
O.~Aharony, S.~M. Chester and E.~Y. Urbach, \emph{{A Derivation of AdS/CFT for
  Vector Models}}, \href{https://doi.org/10.1007/JHEP03(2021)208}{\emph{JHEP}
  {\bfseries 03} (2021) 208}
  [\href{https://arxiv.org/abs/2011.06328}{{\ttfamily 2011.06328}}].

\bibitem{Benedetti:2021qyk}
D.~Benedetti, \emph{{Instability of complex CFTs with operators in the
  principal series}},
  \href{https://doi.org/10.1007/JHEP05(2021)004}{\emph{JHEP} {\bfseries 05}
  (2021) 004} [\href{https://arxiv.org/abs/2103.01813}{{\ttfamily
  2103.01813}}].

\bibitem{Dolan:2011dv}
F.~A. Dolan and H.~Osborn, \emph{{Conformal Partial Waves: Further Mathematical
  Results}},  \href{https://arxiv.org/abs/1108.6194}{{\ttfamily 1108.6194}}.

\bibitem{Hepp:1966eg}
K.~Hepp, \emph{{Proof of the Bogolyubov-Parasiuk theorem on renormalization}},
  \href{https://doi.org/10.1007/BF01773358}{\emph{Commun. Math. Phys.}
  {\bfseries 2} (1966) 301}.

\bibitem{Zimmermann:1969jj}
W.~Zimmermann, \emph{{Convergence of Bogolyubov's method of renormalization in
  momentum space}}, \href{https://doi.org/10.1007/BF01645676}{\emph{Commun.
  Math. Phys.} {\bfseries 15} (1969) 208}.

\bibitem{rivasseau2014perturbative}
V.~Rivasseau, \emph{From perturbative to constructive renormalization}.
  Princeton University Press, 2014.

\bibitem{Bergere:1974zh}
M.~C. Bergere and J.~B. Zuber, \emph{{Renormalization of feynman amplitudes and
  parametric integral representation}},
  \href{https://doi.org/10.1007/BF01646611}{\emph{Commun. Math. Phys.}
  {\bfseries 35} (1974) 113}.

\bibitem{Bergere:1980sm}
M.~C. Bergere and F.~David, \emph{{Integral Representation for the
  Dimensionally Renormalized Feynman Amplitude}},
  \href{https://doi.org/10.1007/BF01941797}{\emph{Commun. Math. Phys.}
  {\bfseries 81} (1981) 1}.

\bibitem{Costa:2016hju}
M.~S. Costa, T.~Hansen, J.~a. Penedones and E.~Trevisani, \emph{{Projectors and
  seed conformal blocks for traceless mixed-symmetry tensors}},
  \href{https://doi.org/10.1007/JHEP07(2016)018}{\emph{JHEP} {\bfseries 07}
  (2016) 018} [\href{https://arxiv.org/abs/1603.05551}{{\ttfamily
  1603.05551}}].

\bibitem{Benedetti:2020rrq}
D.~Benedetti, R.~Gurau, S.~Harribey and K.~Suzuki, \emph{{Long-range
  multi-scalar models at three loops}},
  \href{https://doi.org/10.1088/1751-8121/abb6ae}{\emph{J. Phys. A} {\bfseries
  53} (2020) 445008} [\href{https://arxiv.org/abs/2007.04603}{{\ttfamily
  2007.04603}}].

\end{thebibliography}\endgroup

\addcontentsline{toc}{section}{References}


\end{document}